\documentclass[reprint, nofootinbib]{revtex4-2}
\usepackage{graphicx}
\usepackage{dcolumn}
\usepackage{bm}
\usepackage[colorlinks=true, linkcolor=blue, citecolor=blue, urlcolor=blue]{hyperref}
\usepackage[mathlines]{lineno}
\usepackage{amsmath,amssymb}
\usepackage{mathrsfs}
\usepackage{tikz}
\usepackage{float}
\usepackage{xcolor}
\usepackage{subcaption} 
\usepackage{enumitem} 

\begin{document}

\preprint{APS/123-QED}

\title{Why collective behaviours self-organise to criticality: A primer on information-theoretic and thermodynamic utility measures}

\author{Qianyang Chen}
\email{qianyang.chen@sydney.edu.au}
\author{Mikhail Prokopenko}%
\affiliation{%
 Centre for Complex Systems, Faculty of Engineering,\\
 The University of Sydney, Sydney, NSW 2006, Australia
}%

\begin{abstract}
Collective behaviours are frequently observed to self-organise to criticality. Existing proposals to explain these phenomena are fragmented across disciplines and only partially answer the question. This primer compares the underlying, \textit{intrinsic}, utilities that may explain the self-organisation of collective behaviours near criticality. We focus on information-driven approaches (predictive information, empowerment, and active inference), as well as an approach incorporating both information theory and thermodynamics (thermodynamic efficiency). By interpreting the Ising model as a perception-action loop, we compare how different intrinsic utilities shape collective behaviour and analyse the distinct characteristics that arise when each is optimised. In particular, we highlight that thermodynamic efficiency --- measuring the ratio of predictability gained by the system to its energy costs --- reaches its maximum at the critical regime. Finally, we propose the \textit{Principle of Super-efficiency}, suggesting that collective behaviours self-organise to the critical regime where optimal efficiency is achieved with respect to the entropy reduction relative to the thermodynamic costs.
\end{abstract}

\maketitle


\section{\label{sec:level1}Introduction}
Self-organisation is a process where a system spontaneously develops new structured patterns or functions, without being explicitly controlled by an external force. This process is observed in a wide range of natural and artificial systems, where local interactions among components generate global order. As a fundamental concept in complexity science, self-organisation is extensively studied in various disciplines, including systems theory, condensed matter physics, systems biology, as well as social sciences.

From a physics perspective, self-organisation is generally viewed as entropy reduction or increase in order in an open system ``without specific interference from outside'' \cite{haken_synergetics_1983, haken_information_2006}. From a biological perspective, self-organisation is typically defined as a pattern-formation process that relies entirely on interactions among the lower-level components of the system \cite{camazine_self-organization_2003}. 
There are three key aspects to self-organisation \cite{haken_synergetics_1983,kauffman_investigations_2000,camazine_self-organization_2003,haken_information_2006,prokopenko_information-theoretic_2009}:
\begin{enumerate}[label=(\roman*)]
    \item Spontaneous order: the system evolves into a more organised state without external control; 
    \item Emergence of coherent global behaviour: there is an observable transition to a more coherent collective behaviour; 
    \item Local interactions and long-range correlations: system components operate on local information but exhibit long-range interaction and connectivity.
\end{enumerate}

One of the underlying principles for the spontaneous order created in self-organisation, as suggested by Kauffman \cite{kauffman_investigations_2000}, is the ``constraint closure'', which means that the system carries out some work to create constrains on the release of energy, and those constraints, in turn, channel the energy to perform more useful work. Thus, a successful framework describing self-organisation needs to account for thermodynamic characteristics of the spontaneous order, capturing the corresponding energy flows and costs.

\begin{figure*}[!ht]
    \begin{subfigure}[b]{0.23\textwidth}
        \includegraphics[width=\textwidth]{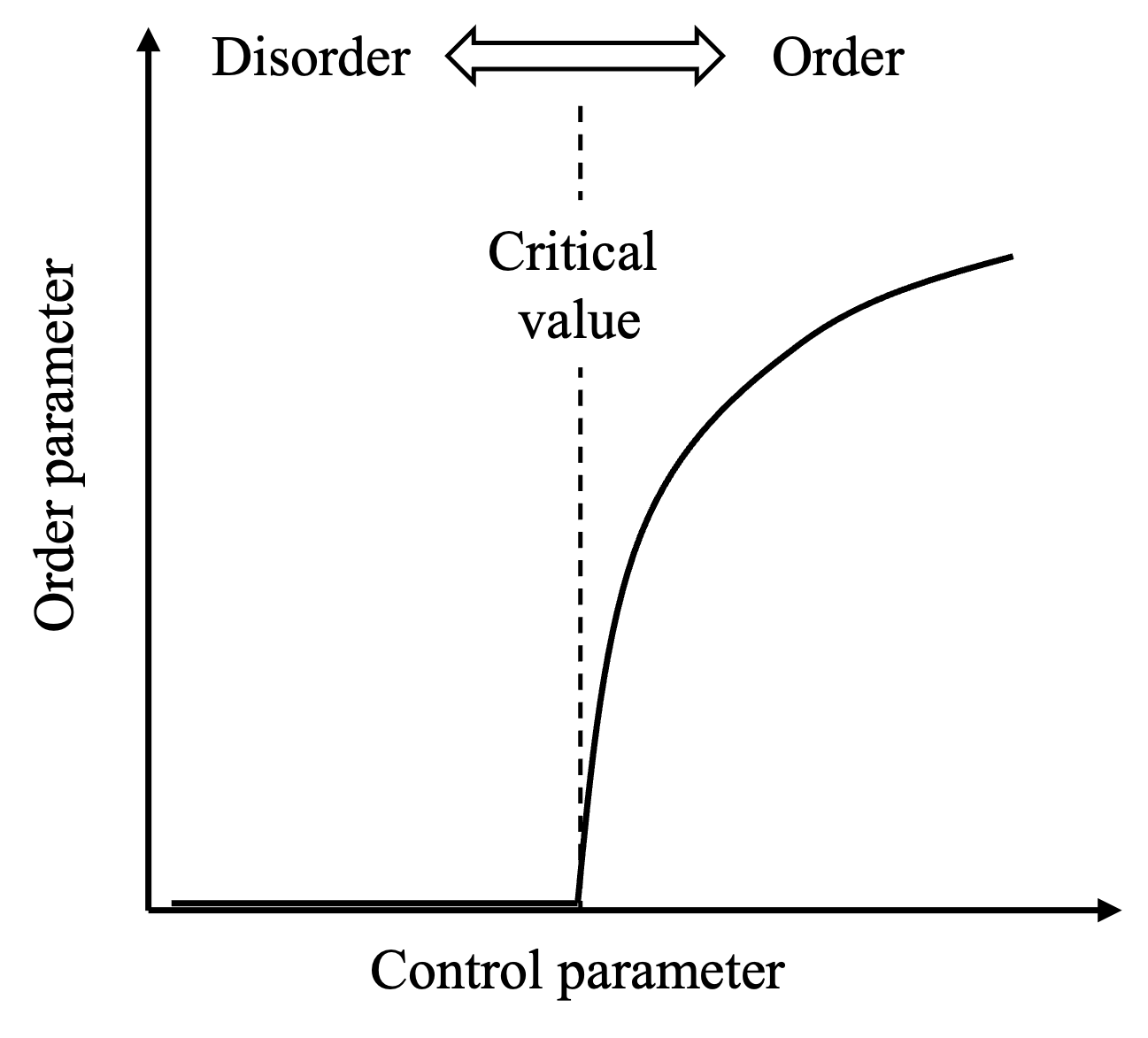}
        \caption{Order-disorder phase transition}
        \label{fig:crit_phasetransition}
    \end{subfigure}
    \hfill
    \begin{subfigure}[b]{0.23\textwidth}
        \includegraphics[width=\textwidth]{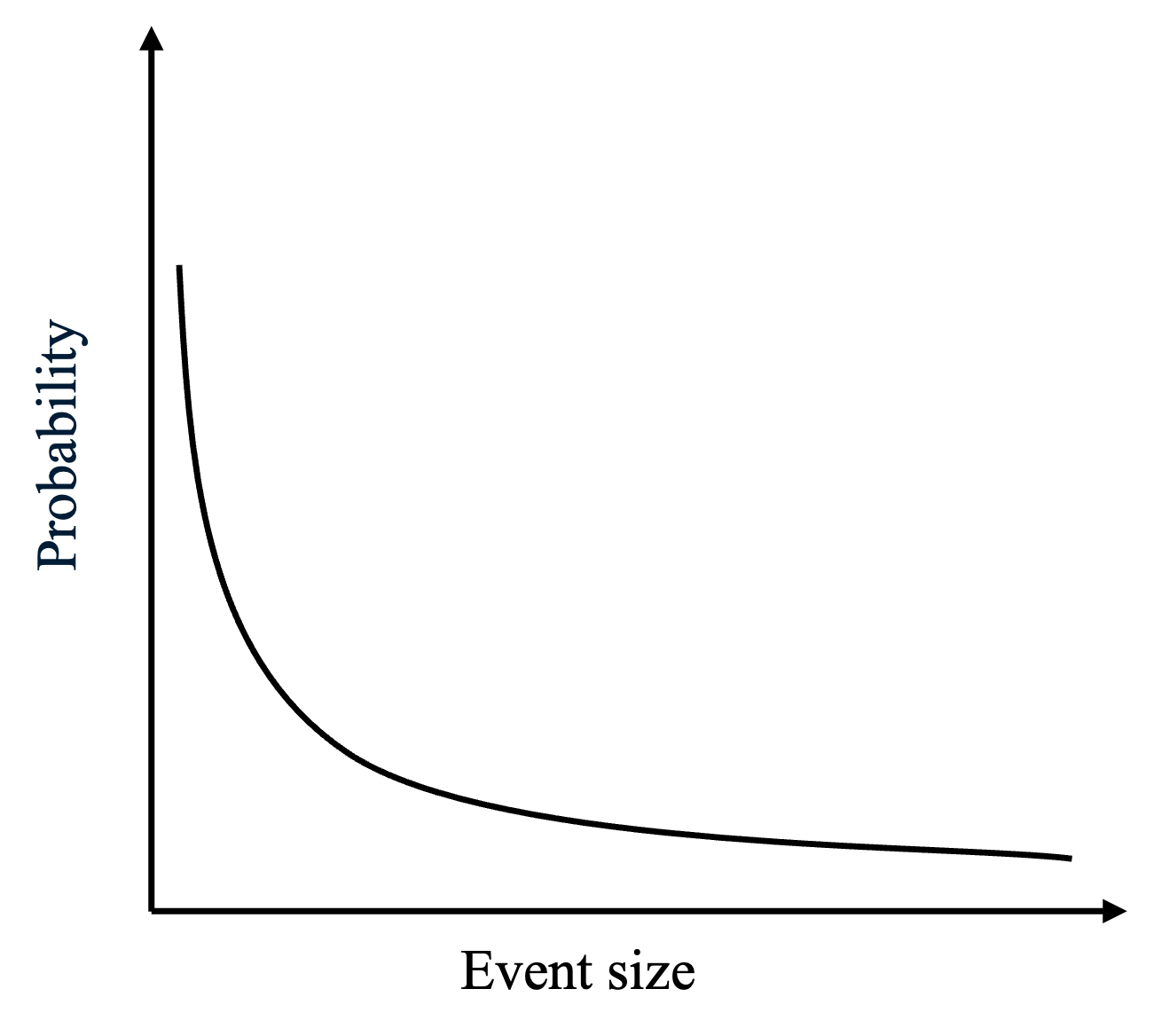}
        \caption{Power-law distribution of event sizes}
        \label{fig:crit_scaleinvariance}
    \end{subfigure}
    \hfill
    \begin{subfigure}[b]{0.23\textwidth}
        \includegraphics[width=\textwidth]{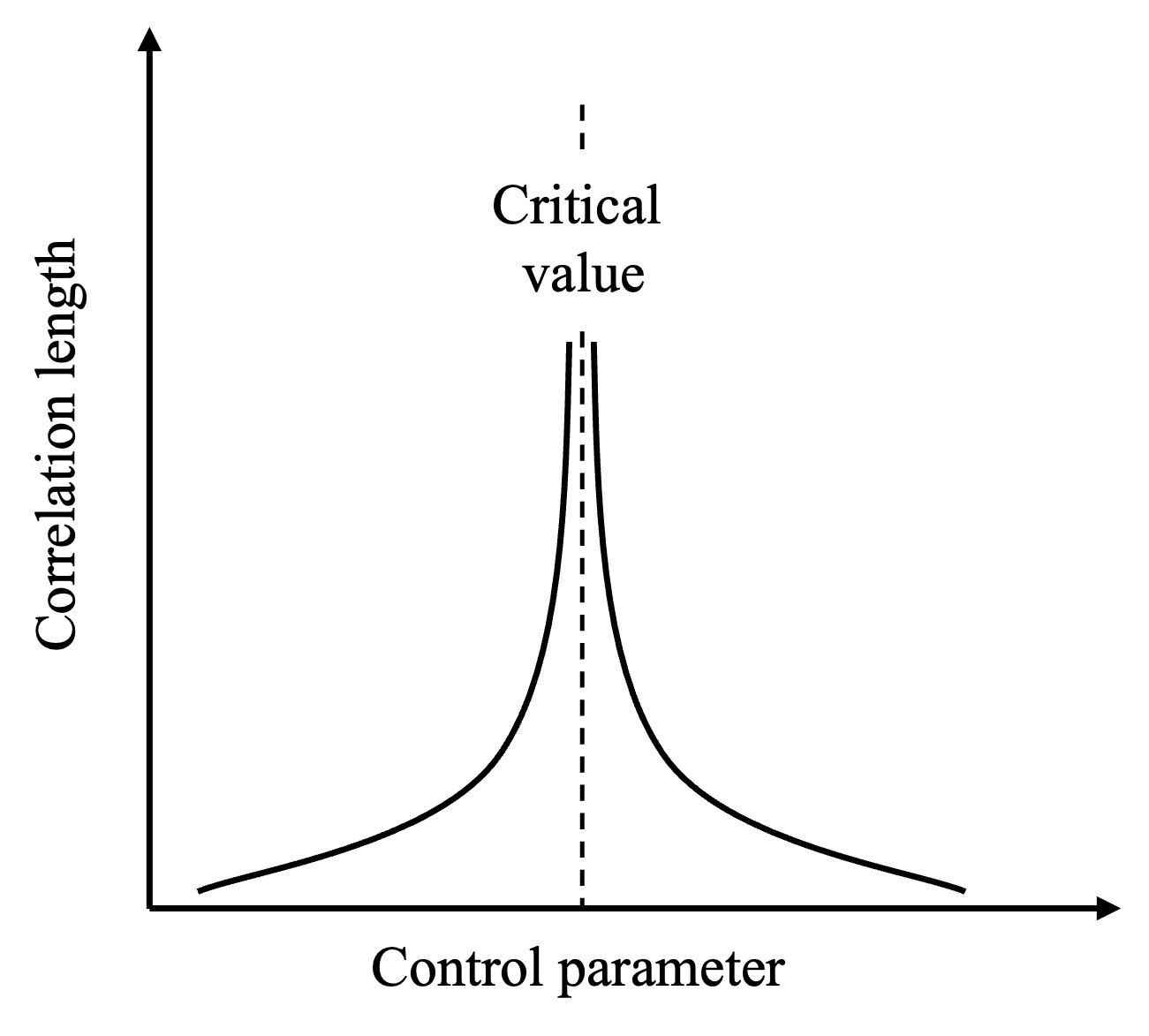}
        \caption{Divergence of correlation length}
        \label{fig:crit_divcorrlength}
    \end{subfigure}
    \hfill
    \begin{subfigure}[b]{0.23\textwidth}
        \includegraphics[width=\textwidth]{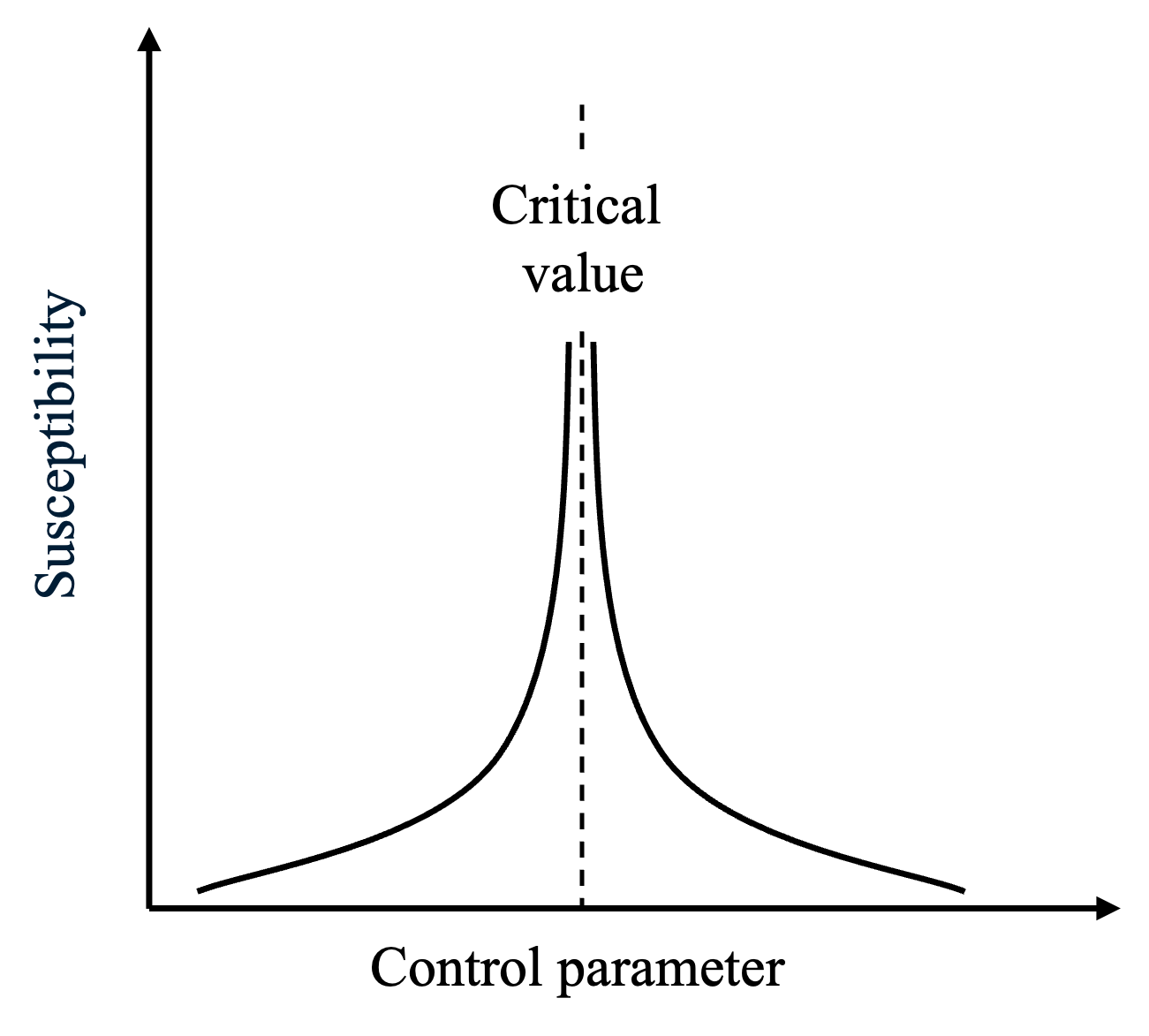}
        \caption{Divergence of susceptibility}
        \label{fig:crit_divresponse}
    \end{subfigure}    
    \caption{Schematic representations for second order phase transition and critical phenomena. The control parameter is a variable that influences the state of the system, such as temperature or pressure, while the order parameter quantifies the degree of order within the system, with non-zero value only in the ordered phase. (a) Order parameter changes continuously  in response to changes in the control parameter during a second-order phase transition. (b) Scale-invariance observed at the critical regime indicates a power-law distribution of event sizes, e.g. avalanche or earthquake magnitudes. (c) Correlation length diverges at the critical point, facilitating long-range correlation of fluctuations between constituent parts of the system. (d) Susceptibility diverges at the critical point, reflecting the system's increased sensitivity to external perturbations.}
    \label{fig:schema_criticality}
\end{figure*}

Typically, self-organised collective behaviours, such as magnetisation, ant colony foraging, swarming, slime mould aggregation, flocking of birds, and neural processing in the brain, exhibit critical phenomena \cite{beekman_phase_2001,beggs_neuronal_2003, shmulevich_eukaryotic_2005,ramo_perturbation_2006, beggs_criticality_2008, cavagna_scale-free_2010, fessel_physarum_2012,krotov_morphogenesis_2014, cavagna_dynamic_2017}. These phenomena occur at the critical point of a continuous phase transition, and include scale-invariance \cite{newman_power_2005}, divergence of correlation length, and divergence of the response function \cite{stanley_introduction_1971}. These hallmarks are observed in physical \cite{halperin_scaling_1969, binney_theory_1992}, biological \cite{beggs_neuronal_2003, cavagna_scale-free_2010, bialek_social_2014}, social\cite{stanley_scale_2000, bettencourt_growth_2007} and hybrid systems \cite{scheffer_critical_2009}. 

Scale-invariance means that the system near the critical regime does not exhibit a typical length scale, i.e., patterns appear similar on many magnification levels. Consequently, the size of events at criticality follows a power-law distribution. The correlation length measures the scale on which fluctuations or changes at one point in the system affect those at another point, and the divergence of this quantity implies the long-range interaction between constituent components of the system. In the context of collective systems such as groups of biological organisms, long-range interactions may generate more coherent global behaviour for the group. The response function characterises the system's response to perturbations. For example, magnetic susceptibility represents the change of magnetisation of a material in response to an applied magnetic field, and is known to diverge at criticality, as even a small field can induce large changes in magnetisation. At the critical regime, systems typically become highly sensitive to small changes in parameters, showing large responses to minor perturbations. Another implication to collective behaviour in biological systems is that the groups may become more sensitive to stimuli from the external environment, such as detection of predators. 

Physical systems, such as fluids or magnets, can be driven to criticality by adjusting a control parameter, e.g., temperature or pressure, that influences the state of the system. As the control parameter reaches a critical value, the order parameter, which measures the degree of order or organisation within the system, undergoes a transition, e.g., from zero (disordered phase) to non-zero (ordered phase). However, for biological systems, there are typically no well-defined protocols to adjust the control parameters. Nevertheless, nature somehow finds its way to poise the system at or near criticality. 

A canonical framework describing the mechanism behind such dynamics is the theory of Self-Organised Criticality (SOC), initially introduced by Bak et al. \cite{Bak1987} based on a mathematical model known as the Bak–Tang–Wiesenfeld (BTW) Sandpile Model. Central to this theory is the interplay of two opposing forces that push a system to criticality. The first force is the driving force, characterised by gradual, incremental changes that increase the system’s energy, disorder, or stress (e.g., adding sand to a sand pile). When the accumulated stress or energy reaches a certain threshold, a stabilising force comes into play, triggering a response that dissipates or redistributes the energy, typically in a sudden and possibly widespread manner (e.g., sand avalanche). Under the influence of these two opposing forces, the system ``evolves'' to criticality and remains there. The concept of SOC inspired a series of studies that applied it to develop an understanding of the underlying mechanism that generates critical phenomena in various complex systems such as forest fires \cite{drossel_self-organized_1992, clar_forest_1996, Malamud1998}, earthquakes \cite{chen_self-organized_1991} and brain activities \cite{beggs_neuronal_2003, beggs_criticality_2008, hesse_self-organized_2014, Shew2015}. However, it can be argued that SOC provides a possible explanation for \textit{how} criticality occurs, rather than \textit{why} it benefits the system. 

We are interested in exploring the intrinsic utility for a self-organising system approaching and operating at the critical regime. Here, an intrinsic utility is understood broadly, as the inherent benefit or value gained by the system from its own organisation, independently of external rewards and objectives~\cite{schmidhuber_formal_2010, oudeyer_what_2007}. Recent research on intrinsic utilities shaping self-organising behaviours primarily examined how systems, especially autonomous robots and biological entities, utilise task-independent objectives in order to optimise and adapt their behaviours. Notable strategies include predictive information maximisation \cite{Ay2008, der_predictive_2008, zahedi_higher_2010, ay_information-driven_2012, martius_information_2013, Der2013, der_role_2014}, empowerment maximisation \cite{klyubin_empowerment_2005, capdepuy_maximization_2007, Klyubin2008, capdepuy_perception-action_2012, Salge2014a, tiomkin_intrinsic_2024}, and free energy minimisation \cite{friston_free_2006, friston_free-energy_2010,friston_active_2015, friston_active_2017, friston_free_2023-1} (which encompasses both intrinsic and extrinsic utilities). A consistent feature of these approaches is their employment of information theory in quantifying the intrinsic motivation for the spontaneous order and emergence of collective behaviours. Informally, one identifies a change in suitably defined entropic quantities with relevant pattern formation at macro-level. Although these approaches can sometimes induce critical behaviours \cite{martius_information_2013, Der2013, der_role_2014, heins_collective_2024}, this outcome is not invariably guaranteed. Thus, to understand the fundamental drivers of critical phenomena in collective behaviours, it is essential to give a thermodynamic account of the intrinsic motivation.

The three frameworks mentioned above predominantly focus on the informational benefit (e.g., increase in predictability, order or potential influence, reduction in uncertainty or surprise) without explicitly addressing the associated energy costs. 
Although the free energy minimisation incorporates the term ``energy'' in its name, the utilised concept is an information-theoretic construct which does not align with the thermodynamic free energy (more details provided in Section \ref{sec:infodriven selforganisation}). As a result, the trade-offs between informational benefits and thermodynamic costs are not captured explicitly. Furthermore, the lack of a common example that could be used to directly compare these intrinsic utilities makes understanding these trade-offs more challenging.

In this work, we compare several intrinsic utilities applied to the same system and explore whether they attain optimality near the critical regime. In doing so, we highlight the salient features of these approaches, providing a concise primer on information-theoretic and thermodynamic utility
measures in the context of self-organisation. We then formulate a unifying principle connecting (i) the intrinsic functional benefits of collective behaviour (measured as entropy reduction or gained predictability) and (ii) the associated thermodynamic costs. Studies of various complex dynamical systems such as urban growth \cite{crosato_critical_2018}, self-propelled particles \cite{crosato_thermodynamics_2018}, contagion network \cite{harding_thermodynamic_2018}, and the canonical Curie-Weiss model for magnetisation \cite{nigmatullin_thermodynamic_2021}, have shown that systems at critical points exhibit maximum thermodynamic efficiency defined as a ratio of the gained predictability (i.e., reduction in uncertainty, or the increase in the internal order) to the amount of work required to change the underlying control parameter. These studies strongly suggested that the rate of entropy reduction relative to the carried-out work diverges (peaks in finite systems) at critical points. We further strengthen the argument by analytically demonstrating that the thermodynamic efficiency diverges at the critical point of the canonical two-dimensional Ising model.

We argue that these studies exemplify a general principle of \textit{super-efficiency}: at the critical regime, a self-organising system of interacting agents achieves the optimal thermodynamic efficiency by gaining maximal predictability of collective behaviour per unit of the expended work. Informally, one can say that a complex system finds the regime where the cost of ``keeping it together'' is justified. On one hand, given some available energy to change the control parameter, the system identifies the control parameter value where the gain in predictability maximises. On the other hand, given a required predictability gain, the system finds the point (the value of the control parameter) where the energy cost associated with the change would be minimal. The principle of super-efficiency suggests that this point aligns with the critical point.

This principle encapsulates the intrinsic utility of self-organising collective behaviour, elucidating why some systems gravitate towards criticality. Our discussion will begin with a background on criticality and phase transitions provided in Section \ref{sec:SOC}. Section \ref{sec:infodriven selforganisation} provides an overview of established intrinsic utility approaches, while Section \ref{sec:thermodynamic efficiency} describes thermodynamic efficiency defined at the intersection of information theory and thermodynamics. A common example using the two-dimensional Ising model is presented in Section \ref{sec:example} to compare the self-organising behaviours driven by different intrinsic utilities. Section \ref{sec:principle} offers a more in-depth discussion on the principle of super-efficiency, and Section \ref{sec:conclusion} summarises the findings and presents the final discussion. This study offers insights into the distinct characteristics of collective behaviour derived within each framework, and emphasises how the principle of super-efficiency captures the intrinsic utility for collective behaviour at the critical regime. 

\section{\label{sec:SOC}Phase transitions and criticality}
Generally, there are two types of phase transitions: first-order phase transition, characterised by discontinuity in the system's order parameter during the transition, and second-order phase transition, where the order parameter changes smoothly and continuously. Critical phenomena are observed in second-order phase transitions. 

The percolation model serves as a canonical example for understanding phase transition and criticality \cite{stauffer_introduction_1992, newman_power_2005}. Let us consider an infinite-size lattice where each site can either be vacant or occupied. In the simplest version, each site is independently occupied with a probability $p$, leading to the formation of clusters of connected occupied sites. A percolating cluster is a group of connected occupied sites that span across the lattice from one side to the opposite. The percolation phase transition is often studied by tracing percolating cluster sizes with respect to the control parameter $p$ (the site occupancy probability), and one can distinguish between: 
\begin{itemize}
    \item Largest cluster size (LCS): a low LCS value indicates a more disordered state where most clusters are small, while a high LCS value indicates a more ordered state where majority of the sites belong to the same cluster;   
    \item Average cluster size (ACS) measures the average cluster size excluding the percolating cluster. This quantity corresponds to the initial susceptibility in the scaling theory for magnetic systems \cite{stauffer_scaling_1979, deckmyn_properties_1995}. 
\end{itemize}

The divergence of ACS at the critical point can be understood as follows: at low $p$, predominantly small clusters form, resulting in a small ACS. It increases with $p$ until the percolation threshold $p_c$ is reached. At $p=p_c$, a percolating cluster forms for the first time, spanning the entire lattice, and the ACS diverges. Beyond $p_c$, smaller clusters are progressively absorbed by the giant percolating cluster as more sites become occupied. The absorption reduces the average size of the remaining finite clusters, causing the ACS to decrease.

\begin{figure}[ht]
    \begin{subfigure}[b]{0.5\textwidth}
        \includegraphics[width=\textwidth]{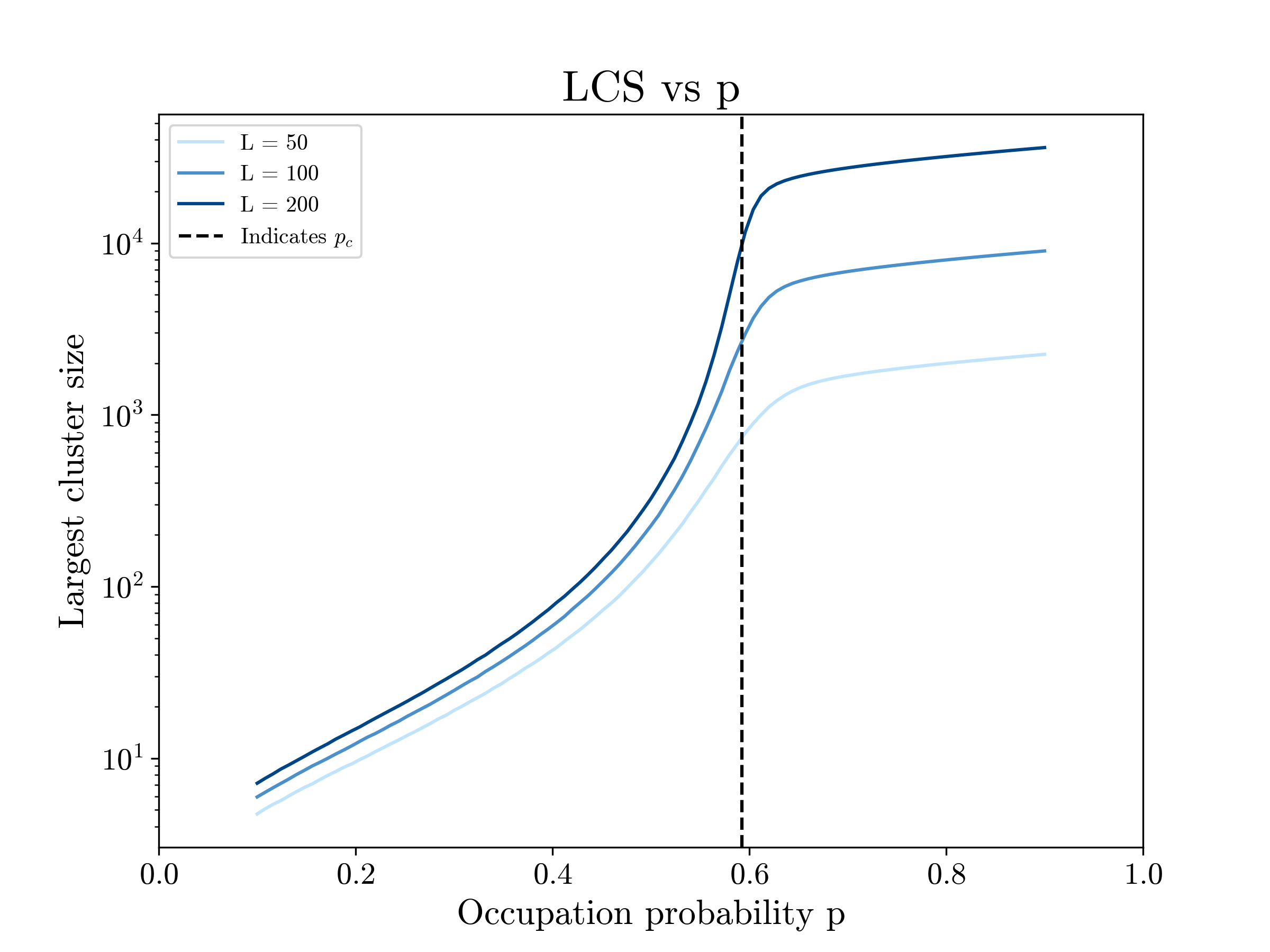}
        \caption{Order-disorder phase transition (largest cluster size)}
        \label{fig:perco_LCS}
    \end{subfigure}
    \hfill
    \begin{subfigure}[b]{0.5\textwidth}
        \includegraphics[width=\textwidth]{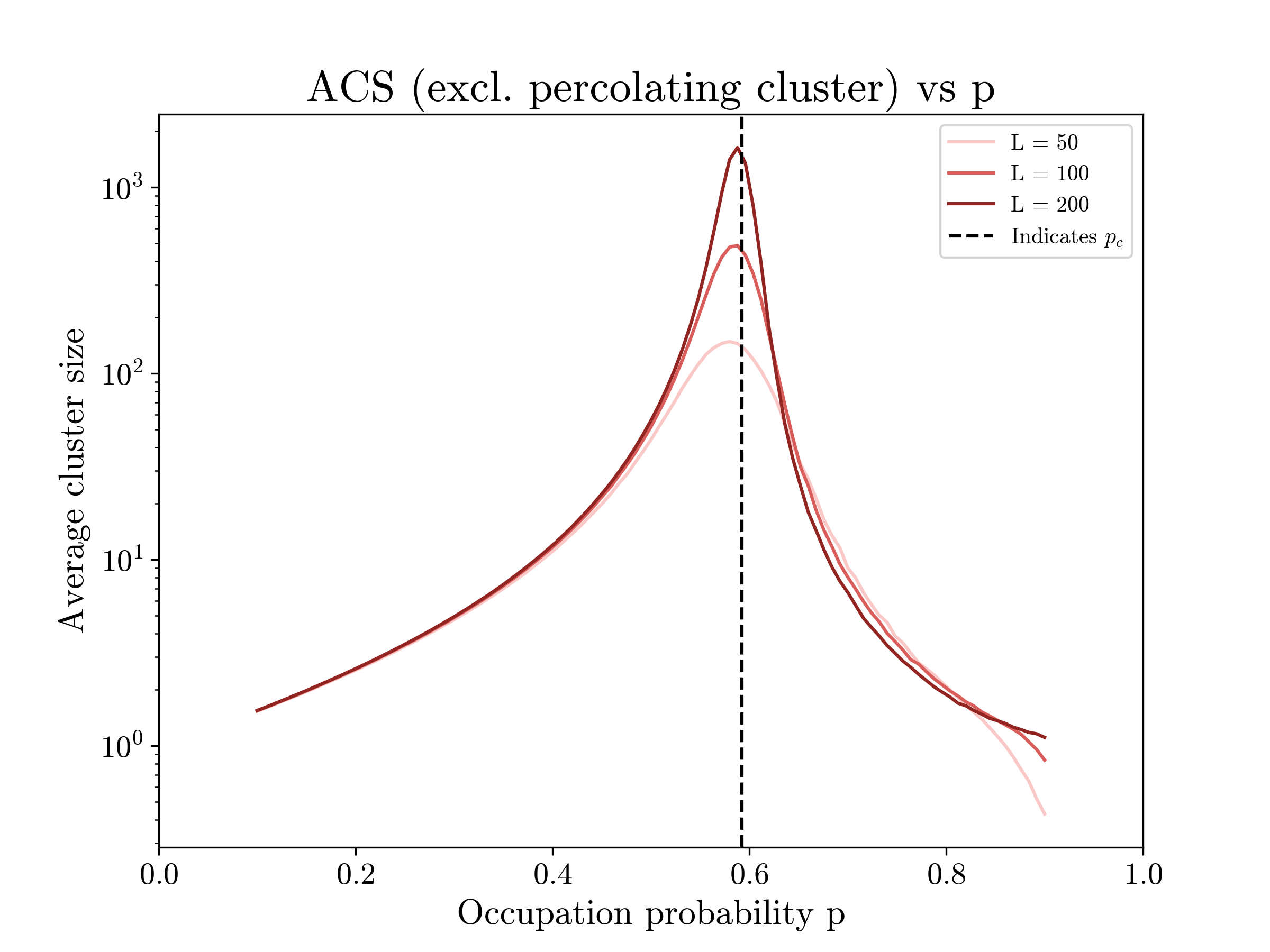}
        \caption{Divergence of susceptibility (average cluster size)}
        \label{fig:perco_ACS}
    \end{subfigure}   
    \caption{Site percolation model: largest cluster size and average cluster size (excluding percolating cluster) simulated for different lattice sizes $L$ \cite{qianyang_chen_2024_13784627}. Vertical axis in log-scale.
    }
    \label{fig:percolation}
\end{figure}


Similarly, the correlation length $\xi$ also diverges at the critical point. The divergence of $\xi$ indicates that fluctuations at one single site can propagate infinitely far across the system, reflecting extensive long-range interactions throughout the lattice at the critical regime. 

The relationship between average cluster size $\langle S \rangle$ or correlation length $\xi$ and the control parameter $p$ at the critical regime can be expressed as:
\begin{align}
    \label{eq:crit_divergence1}
    \langle S \rangle &\propto |p - p_c|^{-\gamma} \\
    \label{eq:crit_divergence2}
    \xi &\propto |p - p_c|^{-\nu}
\end{align}
where $\gamma, \nu$ are critical exponents, positive numbers whose value depends on the dimension of the system. Put simply, at criticality, the system is very sensitive to small perturbations.

The critical phenomena observed in the lattice model provide insights into the dynamics of collective behaviour near the critical point. Structurally, a percolating cluster spanning the entire lattice forms when the control parameter reaches its critical value. Functionally, the system demonstrates a heightened responsiveness to changes in control parameters, exhibiting long-range correlations between its constituents, characterised by the divergence of average cluster size and correlation length. The ability to propagate information over long distances and maximise responsiveness enables more coherent global behaviour and enhances the collective group's sensitivity to external changes, thus offering advantages for the system to operate at or near the critical regime.

The ubiquity of critical phenomena in nature led to vigorous research into potential mechanisms generating criticality. In the 1980s, the theory of Self-Organised Criticality (SOC) was proposed by Bak, Tang and Wiesenfeld \cite{Bak1987} as a possible explanation. Under SOC, specific dissipative dynamical systems naturally evolve towards criticality regardless of their initial states. At the core of the SOC models is the presence of a slow driving force that pushes a subcritical system towards critical or supercritical state, and a fast regulating force that brings the system back from supercritical states. The driving force permits the build-up of energy that is later released by the regulating force, propagating throughout the system via localised interactions. 

One of the classical SOC models is the Drossel and Schwabl forest-fire model \cite{drossel_self-organized_1992, clar_forest_1996}, which enhances the percolation model. Let the occupied sites represent trees grown in the forest. Additionally, the model includes another dynamics: lightning strikes a random site with a probability of $f$. At each time step, the lattice sites are updated based on four rules:
\begin{enumerate}
    \item on an empty site, a tree grows with probability $p$;
    \item lightning strikes a random site in the forest with probability $f$, turning a tree into a burning tree;
    \item a tree burns if at least one of its adjacent neighbours is burning;
    \item a burning tree turns into an empty site, and the model runs indefinitely.
\end{enumerate}

Following the above rules, as trees grow, they form clusters that contribute to propagation of forest fires. The tree growth probability at any given site mirrors the site occupancy probability in the percolation model. Lightning acts as an external factor that regulates tree growth. While the percolation model requires a manual adjustment of $p$ to reach the critical point, the dynamics of the DS forest-fire model self-regulates to criticality. Thus, critical phenomena, such as long-range correlations and scale-invariance, are considered to be self-organising in response to the model's inherent dynamics. 

\begin{figure}[ht]
    \begin{subfigure}[b]{0.5\textwidth}
        \includegraphics[width=\textwidth]{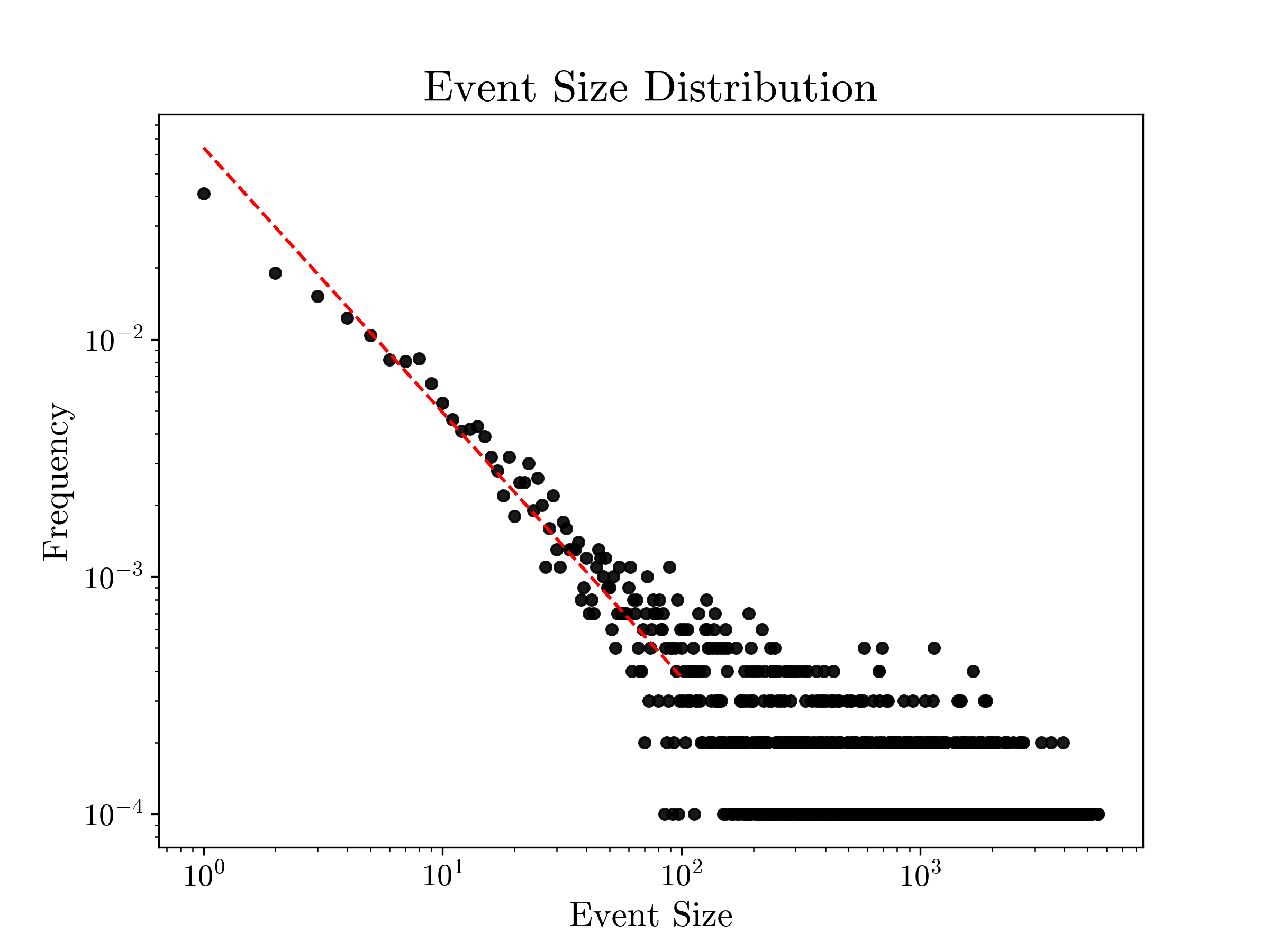}
        \caption{Power-law distribution of event sizes}
        \label{fig:DSFFM_eventSize}
    \end{subfigure}
    \vfill
    \begin{subfigure}[b]{0.5\textwidth}
        \includegraphics[width=\textwidth]{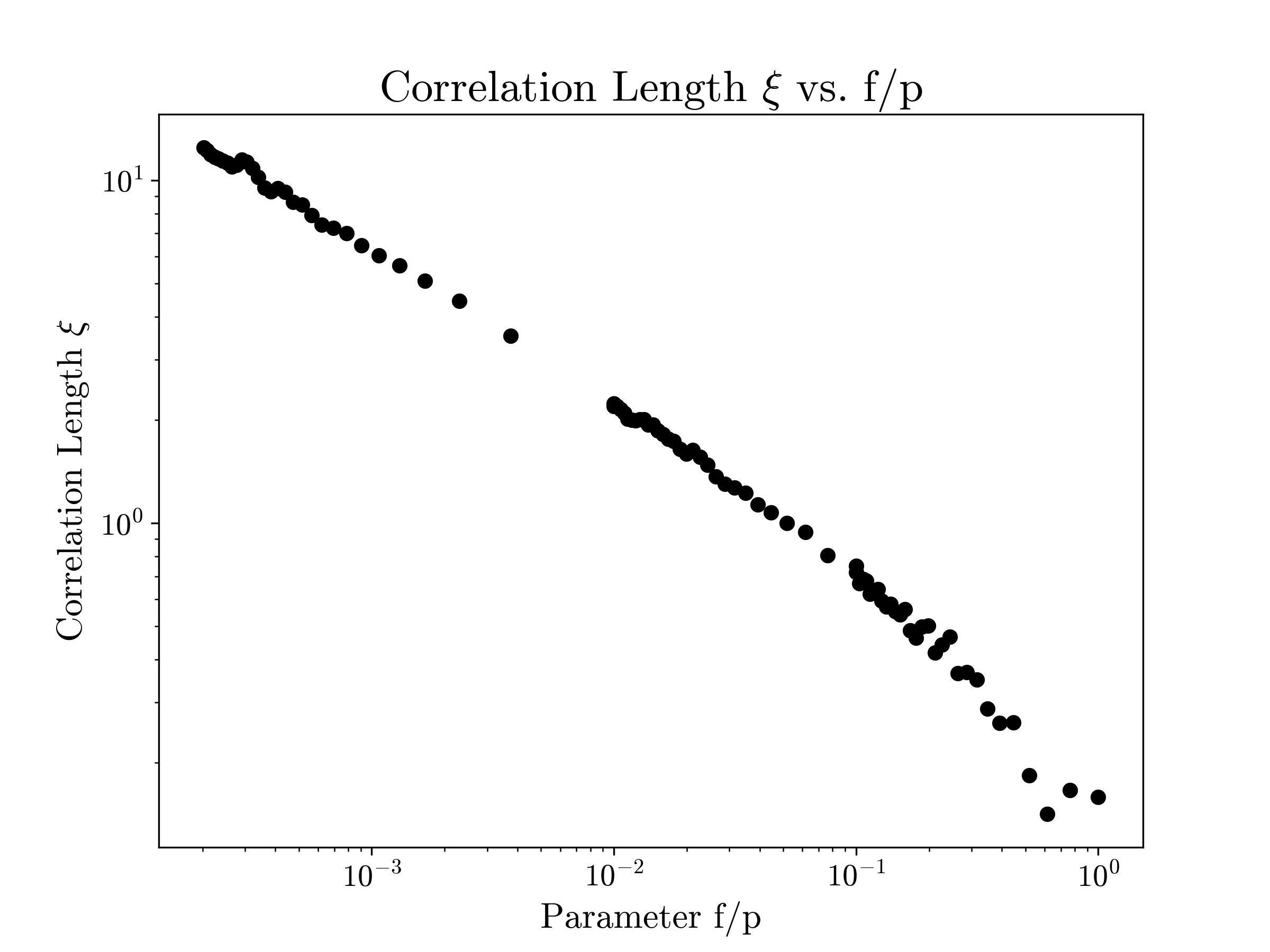}
        \caption{Divergence of correlation length}
        \label{fig:DSFFM_corrLen}
    \end{subfigure}   
    \caption{DS forest-fire model: event size distribution and correlation length plots (log-log scale), simulated on a 100$\times$100 square lattice \cite{qianyang_chen_2024_13784627}. Critical point occurs in the limit $f/p \rightarrow 0$ \cite{clar_forest_1996}.}
    \label{fig:DSFFM}
\end{figure}

The system's behaviour is governed by two opposing forces: lightning strikes and tree growth. When trees are sparse, the chance of them getting hit by a lightning strike is low; therefore, trees continue to grow. When tree density goes beyond the critical density, giant clusters form, spanning the entire lattice. A lightning strike on such a cluster results in a fire that consumes the entire cluster, returning the system to a subcritical state. The system approaches a steady state at the critical point under the influence of these two forces, where the fire event sizes follow a power-law distribution and the correlation length $\xi$ diverges (Figure \ref{fig:DSFFM}).

The dynamic process that drives the system to the critical point hinges on the condition of ``separation of time scales''. This condition specifies that the time required to burn down the whole cluster is much shorter than the time it takes to grow a tree, which in turn is much shorter than the time between two lightning strikes at the same site. This condition translates to a slow dynamic of the external driving force (the lightning) and a fast dynamic of local interaction (fire spreading and tree burning). It ensures that (1) no trees grow during a burn, and (2) there is a substantial energy accumulation between lightning events, leading to the destruction of many trees by a single lightning strike \cite{drossel_self-organized_1992}.

There is an ongoing debate regarding the characteristics of the power-law distribution and criticality generated by the forest fire model or SOC models in general. For example, numerical results indicate that the scaling law observed in the forest-fire model is transient and does not hold at larger scales \cite{grassberger_critical_2002}. Findings of \cite{palmieri_emergence_2018} challenge the view that SOC systems inherently exhibit exact critical scaling, but suggest that the forest-fire model demonstrates weak criticality --- a concept that is based only on the way that correlation length diverges, without the requirement for an exact power law distribution. 

The theory of Self-Organised Criticality elucidates the mechanism behind self-regulation of the dynamics towards a critical point where scale-invariance and long-range correlations may induce collective behaviours. However, SOC does not explain what intrinsic utility is attained by the collective behaviour of the system operating at the critical regime. The subsequent sections will review several approaches that explore the issue of intrinsic motivations.

\section{\label{sec:infodriven selforganisation}Information-driven self-organisation} 
Information-driven self-organisation is an active area of research that applies information theory to study the behaviours of an agent or a group of agents. The information-theoretic utility functions used to derive the behaviours have the advantage of being universal and domain-invariant. These measures are considered strong candidates for capturing the informational benefit of increased order in collective systems. Two broad categories can be identified: purely intrinsic, and ``hybrid'' measures which incorporate, in addition, an extrinsic target or preference.

In this work, we adopt the following notation:
\begin{itemize}
    \item Information theoretic quantities (definitions provided in Appendix \ref{appendix:prelim_info}):
    \begin{itemize}
        \item entropy $H(.)$
        \item conditional entropy $H(.|.)$
        \item mutual information $I(.;.)$
        \item Kullback-Leiber divergence $D(.||.)$
    \end{itemize}
    \item Capital letters $W, S, A, M$... for random variables;
    \item Small letters $w, s, a, m$... for a realisation of the corresponding random variable;
    \item Letters $\mathcal{I}, \mathfrak{E},\mathcal{F}$ for quantities computed using the information-theoretic approaches, corresponding to predictive information, empowerment and variational free energy, respectively;
    \item Blackboard bold font $\mathbb{E}, \mathbb{S}, \mathbb{F}, \mathbb{W}, \mathbb{Q}, \mathbb{I}$ for thermodynamic quantities or statistical quantities, corresponding to energy, entropy, thermodynamic free energy, work, heat and Fisher information respectively. Technical preliminaries are provided in Appendix \ref{appendix:prelim_therm}.
\end{itemize}

Additionally, we adopt the notation where subscript $x_t$ denotes the state at time $t$ in a time series, and superscript $x^{(i)}$ indicates the $i^{th}$ instance in the population.

\subsection{The perception-action loop}
\begin{figure}[!h]
\centering
\includegraphics[scale=0.5]{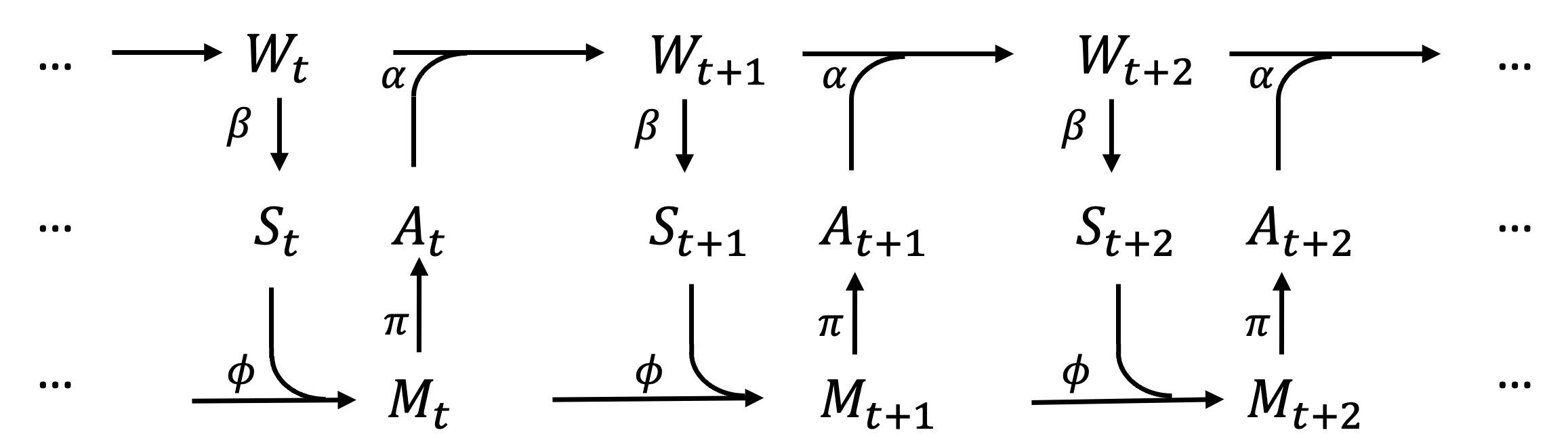}
\caption{\label{fig:pa_loop_mem} Causal structure of perception-action loop of an agent with memory, traced over time.}
\end{figure}

Approaches formalising information-driven self-organisation typically assume an underlying model of agent-world interaction. This interaction is generally modelled with a perception-action (or sensorimotor) loop, using random variables to reflect the probabilistic nature of the dynamic. Figure \ref{fig:pa_loop_mem} illustrates the causal network of the perception-action loop traced over time, where $W_t$, $S_t$, $A_t$, $M_t$ represent the state of the world, the sensor, the actuator and memory (the controller) at time $t$. The perception-action loop captures the following dynamics:
\begin{itemize}
    \item At any given time $t$, the world state $W_t$ leads to an update of the agent's sensory state $S_t$. The mapping from $W$ to $S$ is specified by kernel $\beta:W\rightarrow S$, representing the agent's sensory mechanism;
    \item The agent's memory (or controller) $M_t$ is influenced by both memory from the previous time step $M_{t-1}$ and the current sensory $S_t$, a relationship represented by kernel $\phi:M\times S \rightarrow M$;
    \item Depending on the memory state, the agent updates its action $A_t$ according to the policy $\pi: M \rightarrow A$. $M_t$ also carries through to the future $M_{t+1}$. 
    \item The action $A_t$ and the world state $W_t$ jointly update the next world state $W_{t+1}$. The mapping is specified by kernel $\alpha: W \times A \rightarrow W$, representing the agent's actuation mechanism.
\end{itemize}
It is worth noting that $\alpha$ and $\beta$ are kernels that capture the agent's embodiment in terms of the agent's sensor and actuator capabilities. They set constraints to how the agent may explore the environment, act and learn \cite{ay_causal_2013,ay_causal_2014}. 

In the next subsections, we will elaborate on the variations of causal network representations within different intrinsic utility frameworks. A common example will be presented in section \ref{sec:example}.

\subsection{Intrinsic utility approaches}
The notion of ``intrinsic utility'' suggests that the utility provided to the agent is internal and task-independent \cite{schmidhuber_curious_1991, schmidhuber_formal_2010}. Predictive information maximisation \cite{Ay2008, der_predictive_2008, zahedi_higher_2010, ay_information-driven_2012, martius_information_2013, Der2013, der_role_2014} and empowerment maximisation \cite{klyubin_empowerment_2005, capdepuy_maximization_2007, Klyubin2008, capdepuy_perception-action_2012, Salge2014a, tiomkin_intrinsic_2024} are two important approaches that utilise information-theoretic measures as intrinsic motivation for inducing self-organising behaviours.  

\subsubsection{Predictive information} \label{sec:pi_theory}
\begin{figure}[h]
\centering
\includegraphics[scale=0.45]{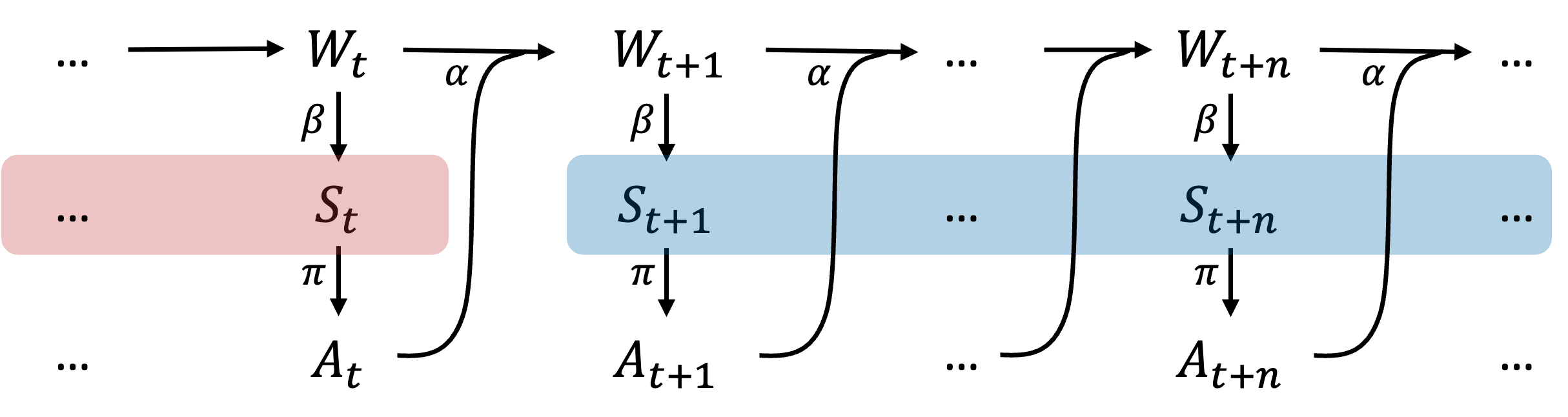}
\caption{\label{fig:pa_loop} Causal structure of perception-action loop of a memory-less agent traced over time. Coloured blocks represent the two components that mutual information is calculated for predictive information.}
\end{figure}

Predictive information \cite{bialek_predictability_2001}, also known as effective measure complexity \cite{grassberger_toward_1986} or excess entropy \cite{crutchfield_inferring_1989}, measures how much the observed history reduces uncertainty about the future. In the context of robotic behaviour development, predictive information in the sensor space may serve as an objective function for behaviour learning. Predictive information maximisation has been implemented for memory-less agents \cite{Ay2008, der_predictive_2008, zahedi_higher_2010, ay_information-driven_2012, martius_information_2013, Der2013, der_role_2014} but potentially can be adapted to incorporate external memory. Figure \ref{fig:pa_loop} illustrates a reduced causal network for a simple memory-less agent (reactive control) on which predictive information is applied.

Predictive information (used as intrinsic utility) is defined as the mutual information between past and future sensory states \cite{Ay2008}. This can be further decomposed into components that represent the diversity and predictability \cite{Ay2008, prokopenko_information-theoretic_2009}:
\begin{equation} \label{eq:pred_info}
    \mathcal{I} := \quad I(S_{past}; S_{future}) \quad = \underbrace{H(S_{future})}_\text{Diversity of future states} - \underbrace{H(S_{future}|S_{past})}_\text{Unpredictability of future} 
\end{equation}

Considering only one time step into the future, predictive information is defined as: 
\begin{equation} \label{eq: pi_one_step}
    \mathcal{I} := I(S_t; S_{t+1})
\end{equation}
and measured in bits. Equation (\ref{eq:pred_info}) indicates that predictive information is large when the entropy of the future sensory states $H(S_{future})$ is large, corresponding to a rich future experience, and/or when conditional entropy $H(S_{future}|S_{past})$ is small, representing a more predictable future. In both extremes, where there is complete order (no diversity) or complete randomness (no predictability), the predictive information will be zero. 

\begin{figure}[!h]
\centering
\includegraphics[scale=0.5]{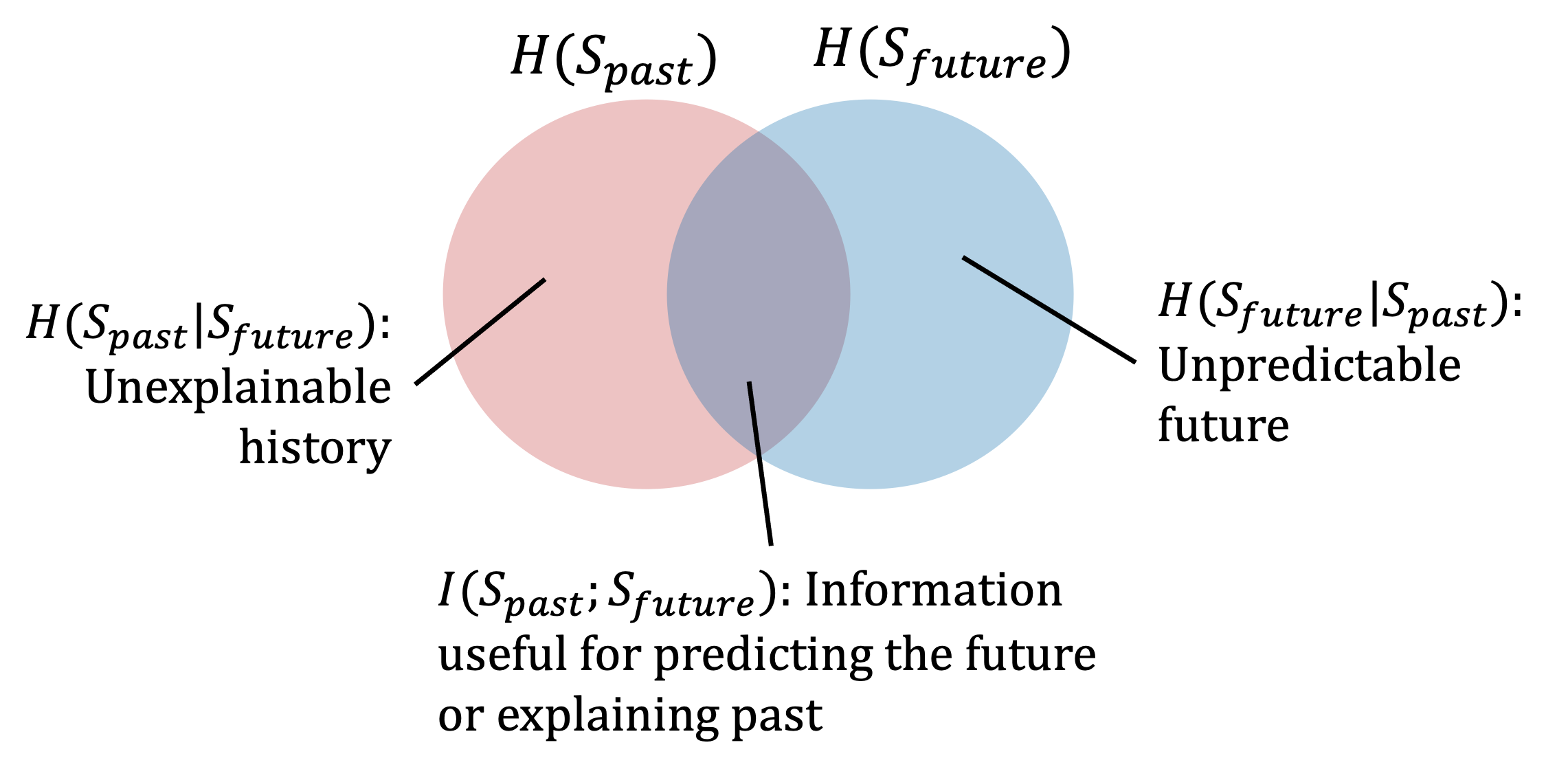}
\caption{\label{fig:venn_predInfo} The Venn diagram of predictive information shown as the mutual information (the overlap area) between past and future sensory states: it represents how useful the past is for predicting the future.}
\end{figure}

The Venn diagram in Figure \ref{fig:venn_predInfo} illustrates the relationship between the time series of past and future sensory states. We note that conditional entropy $H(S_{past}|S_{future})$ represents the remaining entropy of historical sensory states given the future states, which is the part of history that we are unable to reconstruct using information from the future. For example, reconstructing the question given the answer to that question.

Behavioural rules can be derived using the predictive information maximisation approach, forming a policy $\pi$ mapping sensory states to actions. Policy $\pi$ can be either deterministic, such as a simple mapping from sensor values to actions, or stochastic, which is represented by a conditional probability distribution $\pi\equiv p(a|s)$. The general form of the objective function for a one-step predictive information-driven agent is then expressed as:
\begin{equation}
        \pi^*(a_t, s_t) \: = \: \underset{\pi(a_t, s_t)}{\arg\max} \: \{\mathcal{I}\} \: = \: \underset{\pi(a_t, s_t)}{\arg\max} \: \{I(S_{t+1}; S_t)\}
\end{equation}
where $\pi^*$ denotes the optimal policy.

An agent motivated to maximise predictive information chooses policies that result in more diverse and, at the same time, predictable outcomes. In the context of collective behaviour, maximising predictive information has also been shown to induce cooperative behaviour under decentralised control \cite{prokopenko_evolving_2006, zahedi_higher_2010, martius_information_2013}. The increase of predictive information differs from merely reducing randomness in the system; it enhances the richness of structure in the collective system. As shown in Section \ref{sec:example}, collective behaviour resulting from maximising predictive information for each individual may appear random, but locally, it maintains a high level of diversity, aligned with the predictability of an individual's sensory states.

\subsubsection{Empowerment}

\begin{figure}[h]
\centering
\includegraphics[scale=0.45]{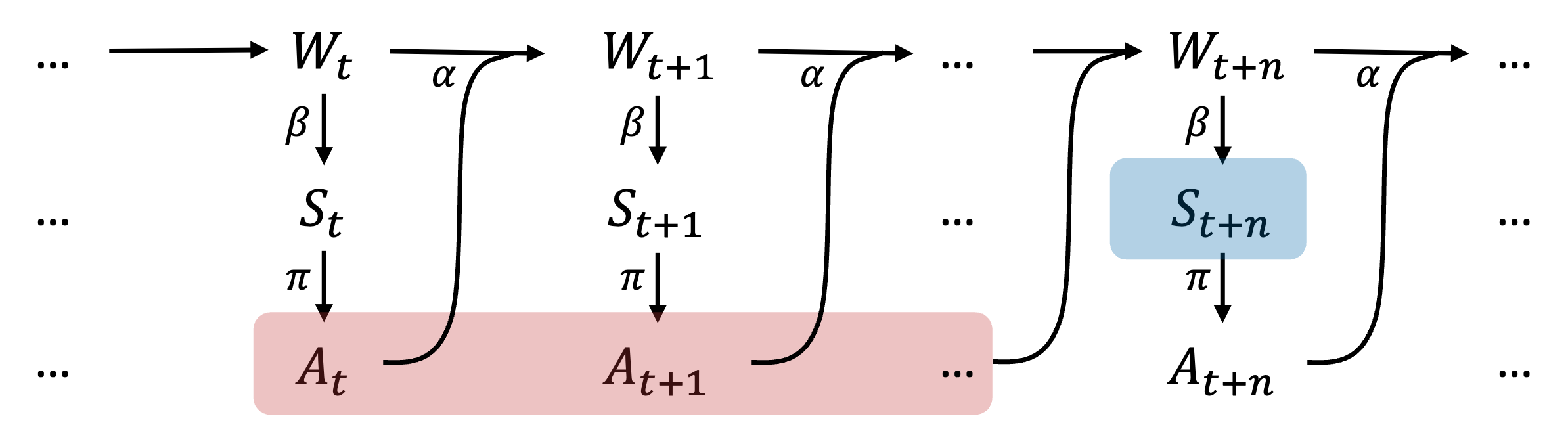}
\caption{\label{fig:pa_loop_emp} The causal structure of the perception-action loop of a memory-less agent traced over time. Coloured blocks represent the components of mutual information which determines empowerment.
}
\end{figure}

Alternatively, we can focus on a specific segment of the causal network that captures the influence of actions on subsequent sensory states through the external world (Figure \ref{fig:pa_loop_emp}). \textit{Empowerment} measures this influence as the maximum amount of information an agent can inject from its actuators ($A$) to its sensors ($S$) at a future time via the environment. 

The n-step empowerment is defined as the Shannon channel capacity $C$ or maximum mutual information between the current sequence of actions $A_t^n = \{A_t, A_{t+1}, ..., A_{t+n-1}\}$ and future sensor value $S_{t+n}$ \cite{klyubin_empowerment_2005}:
\begin{equation} \label{eq:emp}
\begin{split}
    \mathfrak{E} :&= C(A_t^n \rightarrow S_{t+n})\\
    &\equiv \max_{p(a_t^n)} \: I(S_{t+n}; A_t^n)
\end{split}
\end{equation}

Empowerment is measured in bits. Considering only the most immediate future, one-step empowerment can be computed by: 
\begin{equation} \label{eq: emp_one_step}
    \mathfrak{E} := \max_{p(a_t)} \: I(S_{t+1}; A_t)
\end{equation}
The definition~(\ref{eq: emp_one_step}) is referred to as \textit{general} or \textit{context-free} empowerment since it measures only the agent's general ability to inject information into its future sensory states. In order to use empowerment as the driver for an agent's action, one needs to distinguish between different states of the environment, so that the agent can make decisions accordingly. This is achieved by  \textit{context-dependent} empowerment. The context refers to the state of the environment $w$ that affects the perception-action loop characteristic $p(s|a)$. More specifically, the future sensory state of the agent is affected by both its past actions and the historical states of the world. In other words, the same actions can lead to different distributions of future sensory states when the external environment has changed. Instead of considering a general action-perception characteristic $p(s|a)$, empowerment should be considered for a specific world state or context \cite{klyubin_empowerment_2005, capdepuy_maximization_2007, Salge2014a, salge_empowerment_2017}:
\begin{equation} \label{eq: emp_context}
    \mathfrak{E}(w_{t}):= \max_{p(a_t^n|w_{t})} \: I(S_{t+n}; A_{t}^{n}|w_{t})
\end{equation}

Given that an action $a_t$ stochastically leads to a collection of possible future world states $\Gamma$, the resulting average context-dependent empowerment is computed as:
\begin{equation} \label{eq: emp_contextAve}
    \mathfrak{E}(W_t):= \sum_{w_t \in \Gamma}p(w_t)\mathfrak{E}(w_t)
\end{equation}
This quantity can be used as an objective function for an empowerment-driven agent to make decisions on which action to take. More commonly, the state of the world $W$ would be replaced by some context $K$ that approximates it if the full world information is not available.

The general empowerment, as defined in equations (\ref{eq:emp}) and (\ref{eq: emp_one_step}), is different from the average context-dependent empowerment \cite{Salge2014a}. General empowerment does not consider the varying influence of actions in different states, since the channel capacity is computed using only $p(s|a) = \sum_{w}p(s|a,w)p(w)$. In contrast, the average context-dependent empowerment, as defined in equation~(\ref{eq: emp_contextAve}), captures the nuanced ways in which different states can affect the actuation-sensing channel by computing $\max_{p(a|w)} \: I(S; A|w)$, and then average over all possible states.

The objective function for an n-step empowerment-driven agent is:
\begin{widetext}
\begin{equation} \label{eq: emp_of}
\begin{split}
        a_t^* &= \underset{a_t}{\arg\max} \{ \mathfrak{E}(W_{t+1}) \}\\
        &=  \underbrace{\underset{a_t}{\arg\max}}_\text{empw-driven} \Bigl\{ \sum_{w}p(w_{t+1})\underbrace{ \underbrace{\max_{p(a_{t+1}^n|w_{t+1})}}_\text{free to act}\: I(S_{t+n+1}; A_{t+1}^n|w_{t+1})}_\text{potential empowerment} \Bigr\}
\end{split}
\end{equation}
\end{widetext}
where $a^*$ denotes the optimal action under which average context-dependent empowerment is maximised.

Referring to the maximisation expression in equations (\ref{eq:emp}) -- (\ref{eq: emp_context}) and (\ref{eq: emp_of}), we emphasise that $p(a)$ is assumed to be chosen without constraints, that is, an empowerment-driven agent is free to act, so that the channel capacity can potentially be achieved. This needs to be distinguished from predictive information maximisation, where the agent's action is mapped to the sensory input via a policy $\pi$ and hence, is constrained. 

Furthermore, equation (\ref{eq: emp_of}) indicates that the action selected at time $t$ is such that the potential empowerment is maximised at time $t+1$. Therefore, the chosen action $a_t^*$ is different from the action distribution $p^*(a_{t+1}^n|w_{t+1})$ that maximises the mutual information \cite{Salge2014a, salge_empowerment_2017}. As pointed out in \cite{Salge2014a}, \textit{``Empowerment considers only the potential information flow, so the agent will only calculate how it could affect the world, rather than actually carry out its potential."}. 

\begin{figure}[h!]
\centering
\includegraphics[scale=0.5]{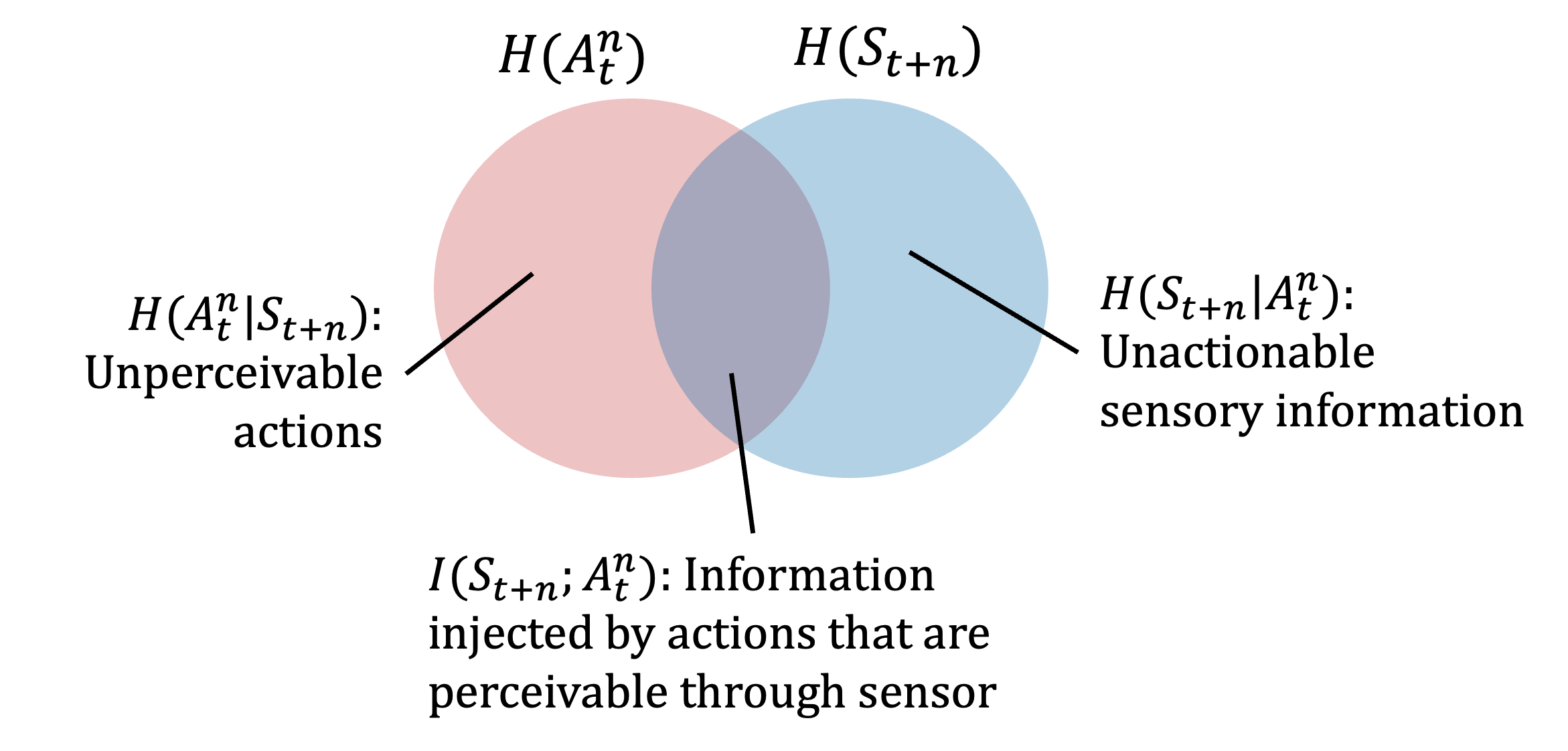}
\caption{\label{fig:venn_emp} The Venn diagram of mutual information $I(S_{t+n};A^n_t)$. Empowerment is the maximum of this mutual information for a given action channel.}
\end{figure} 

Equation (\ref{eq: emp_of}) also implies dependence on the current state of the world $w_t$, as the context of the empowerment ($w_{t+1}$) is a combined result of the current world state $w_t$ and the evaluated action $a_t$.

Similar to predictive information, a decomposition of the mutual information in equation (\ref{eq:emp}) is shown in Figure \ref{fig:venn_emp}. To intuitively understand two conditional entropies, we utilise the box-pushing example presented in \cite{klyubin_empowerment_2005}: a grid world with a robot that can move anywhere except where the box is. If the box is pushable but the robot's sensors cannot capture the box's location, then the robot cannot perceive its box-pushing actions. This is captured in $H(A_t^n|S_{t+n})$, the unperceivable actions. On the other hand, if the robot can see where the box is but cannot move it, then this information is reflected in $H(S_{t+n}|A_t^n)$, the unactionable sensory information. Only the amount of information that is both actionable and perceivable contributes to empowerment.

In summary, an empowerment-driven agent takes actions that maximise its ability to influence the external world in ways that are perceivable by its own sensors. In multi-agent settings, it has been shown that empowerment-maximisation for individual agents leads to spontaneous coordination among the collective \cite{capdepuy_maximization_2007,ronzhin_empowerment_2017, grasso_empowered_2022}. This coordination arises because shared information enhances an individual's empowerment, or informally, its ability to make an influence.

Examples of predictive information and empowerment in collective systems, along with their comparisons, are presented in Section \ref{sec:example}.

\subsection{Beyond intrinsic motivation}
Another prominent approach to derive behaviours based on fundamental principles is the free-energy principle which offers a formal account for the representational capacities of physical systems \cite{friston_free-energy_2010}. The free-energy principle was initially proposed by Friston et al.\cite{friston_free_2006} as an attempt to explain embodied perception-action loops in neuroscience, thus providing an understanding of the dynamics of the brain and decision-making.  Adoption of this principle led to wide applications in the study of learning \cite{friston_active_2015, friston_active_2017, proietti_active_2023}, evolutionary dynamics \cite{ramstead_answering_2018}, social interactions \cite{veissiere2020} and collective intelligence \cite{kaufmann_active_2021, heins_collective_2024}. The principle centres on the idea that self-organising biological agents have a natural inclination to resist disorder. It is argued that, as a result, the brain attempts to minimise uncertainty or surprise.

\begin{figure}[h]
\centering
\includegraphics[scale=0.5]{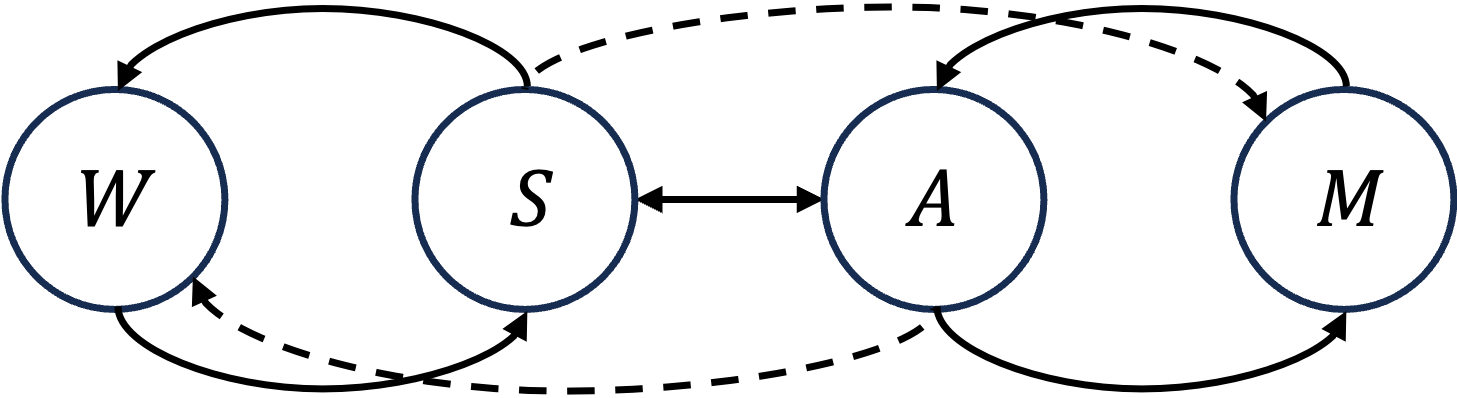}
\centering
\caption{\label{fig:pa_loop_fep2} The diagram illustrates interactions between elements in the active inference framework. Solid lines represent influences between components. Dash lines represent directed influence from sensory to internal or from action to external, which correspond to the two stages of active inference. Figure is adapted from \cite{friston_free_2023-1}.}
\end{figure}

The mechanism derived from the free-energy principle is commonly referred to as active inference. Similar to predictive information and empowerment, active inference can be conceptualised under the perception-action loop representation, although based on different relationships between state variables (Figure \ref{fig:pa_loop_fep2}). 

An underlying assumption in active inference is that the brain makes Bayesian inference over the external (world) states. Bayesian inference relies on some prior probability distribution over the unknown world and updates the distribution when more information is available. The main ingredients in the formulation of active inference are the generative model $p$ and the approximate posterior distribution $q$. The generative model $p$ maps causes (external states $W$) to consequences (sensory $S$, action $A$ and internal state $M$). It encodes the dynamics of the external world and integrates the agent's prior preferences of behaviour \cite{friston_free_2023-1}. While Bayesian inference relies on updating the prior distribution $p$ to the posterior $p(\cdot|observations)$ given the observed data, the posterior is notoriously costly to compute. To reduce the computational difficulty, a parameterised distribution $q$ is employed as an approximation to the true posterior $p(\cdot|observations)$. The approximation distribution $q$ is parameterised by the internal states $M$, which supplies the sufficient statistics of the conditional distribution. The Bayesian inference process with an approximated posterior distribution is referred to as \textit{variational Bayesian inference}. It is worth noting that the integration of external goals in the generative model sets active inference apart from the other pure intrinsic motivation approaches.

The variational free energy is defined as the Kullback–Leibler divergence between the approximate posterior distribution $q$ and the generative model $p$. The expression can be expanded in terms of the difference between a term that resembles expected energy and an entropy term, hence the name ``free energy''\cite{friston_free_2023-1}:

\begin{eqnarray} \label{eq:def_fep}
    \mathcal{F} &:=& \: \mathbb{E}_q\left[\log\frac{ q(w)}{p(s, a, m,w)}\right] \notag \\ 
    & = & \underbrace{\mathbb{E}_q[- \log p(s, a, m, w)]}_\text{Expected energy} - \underbrace{\mathbb{E}_q[-\log q(w)]}_\text{Entropy} 
\end{eqnarray}

However, the quantity called in this approach ``free energy'' is different from the thermodynamic free energy. In active inference, it is instead the variational free energy formulated in terms of information-theoretic quantities, relating to the Bayesian inference process \cite{friston_free_2006,friston_free_2023-1}. Informally, anything that can be represented in the form:
\begin{equation}
    \text{free energy} = \text{energy} \pm \text{const.}\times\text{entropy}
\end{equation}
can be interpreted as ``free energy'' \cite{gottwald_two_2020}. 

Active inference involves two alternating stages: belief update and action selection. During belief update, the agent optimises the internal representation of the generative model $p$ given the sensory samples; in action selection, the agent's action ensures that it samples sensory data that aligns with its current representation. The belief update stage addresses uncertainty about the current generative model, while the action selection stage addresses uncertainty about the future (including future hidden states and future observable outcomes) \cite{friston_free-energy_2009, friston_active_2017}.

The active inference approach has been shown to generate collective behaviour in a group of individuals, each driven by the free-energy minimisation scheme \cite{heins_collective_2024}. Collective dynamics are influenced by the individual's belief about uncertainty and can also be tuned to the changing environment by parameter learning over a slower timescale.

Equation (\ref{eq:def_fep}) can be rearranged in terms of the complexity of the internal model and the accuracy of its representation. In this configuration, minimising free energy is equivalent to reducing complexity, consequently resulting in optimised energy consumption \cite{landauer_irreversibility_1961, friston_free_2023-1}: 
\begin{equation} \label{eq:def_fep2}  
\begin{split}
    \mathcal{F} &= \mathbb{E}_q[\log q(w) - \log p(w)] - \mathbb{E}_q[\log p(s,a,m | w)] \\
    &= \underbrace{\text{KL}[q(w) || p(w)]}_\text{Complexity} - \underbrace{\mathbb{E}_q[\log p(s,a,m | w)]}_\text{Accuracy}
\end{split}
\end{equation}

While equation (\ref{eq:def_fep2}) implies a connection between minimising free energy and reducing energy cost under Landauer's principle \cite{landauer_irreversibility_1961}, the relationship is not explicitly formulated as a ratio of informational gain to energetic costs. 

A more detailed example of free-energy minimisation and the comparison with other approaches is provided in Section \ref{sec:example}.

\section{\label{sec:thermodynamic efficiency}Thermodynamic efficiency}
At this stage we point out that the three information-theoretic approaches reviewed in the previous section do not explicitly account for the corresponding energy costs. Thermodynamic efficiency, on the other hand, takes into consideration both the benefits and the associated costs of maintaining order within the system. Before presenting a formal definition of thermodynamic efficiency, it is important to differentiate between thermal and thermodynamic efficiency. 

\subsection{Thermal vs thermodynamic efficiency}
\begin{figure}[!h]
\centering
\includegraphics[scale=0.2]{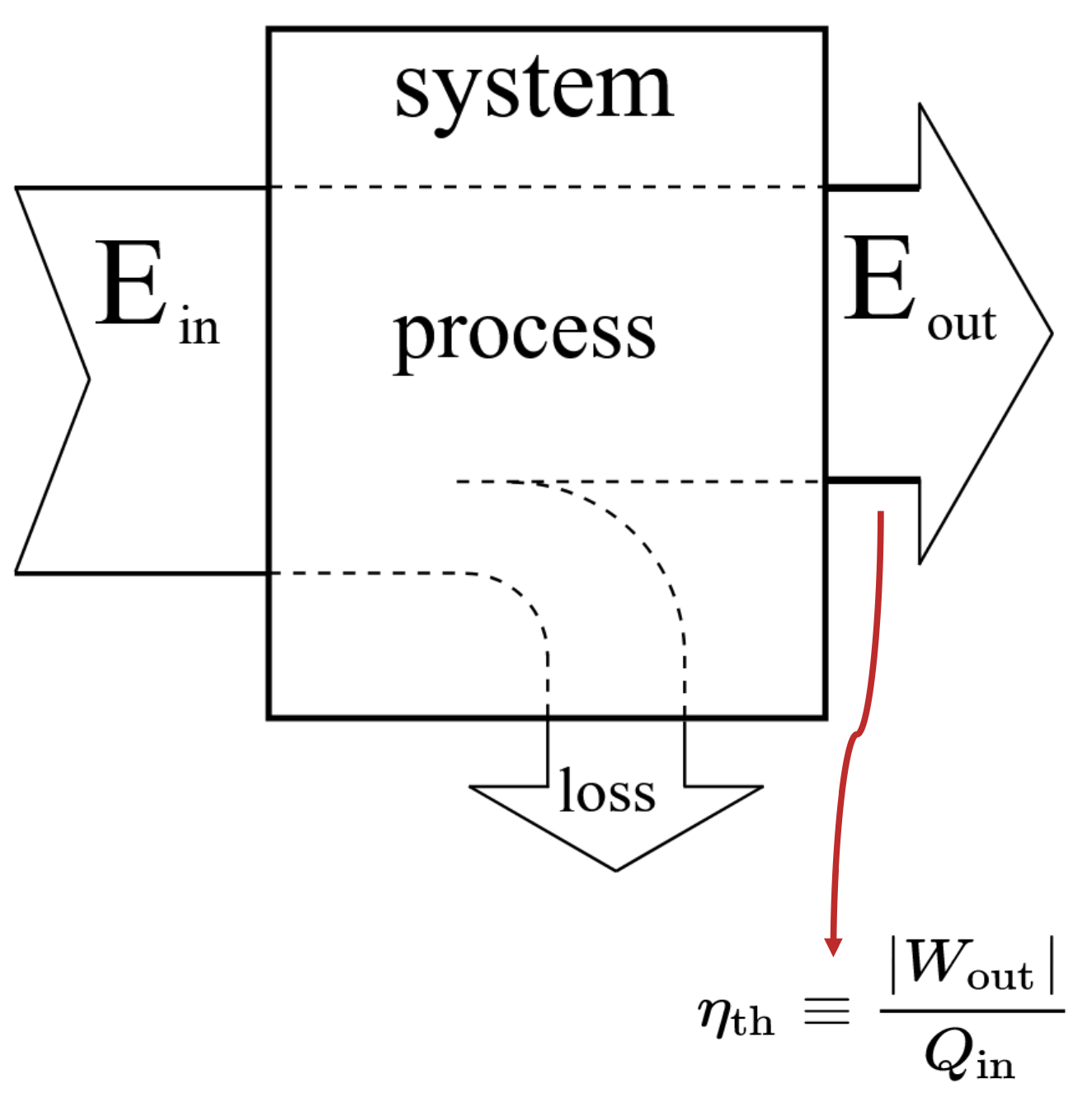}
\caption{\label{fig:thermalEff} Thermal efficiency for a system undergoing a specific process. It is generally defined as the dimensionless ratio between the total work output and the total energy input. Adapted from \cite{wiki_thermal_nodate}.}
\end{figure}

Let us consider a system undergoing a non-ideal process in which it receives energy and performs useful work. Not all the received energy is converted into work; some is inevitably lost as heat, which does not contribute to work output (Figure \ref{fig:thermalEff}). Thermal efficiency measures the system's efficiency of converting energy to work, and is defined as the ratio of useful work output to total energy input, both measured in joules, rendering it a dimensionless quantity. In a non-ideal process, the second law of thermodynamics implies that this ratio is less than one. 

\begin{figure}[!h]
\centering
\includegraphics[scale=0.2]{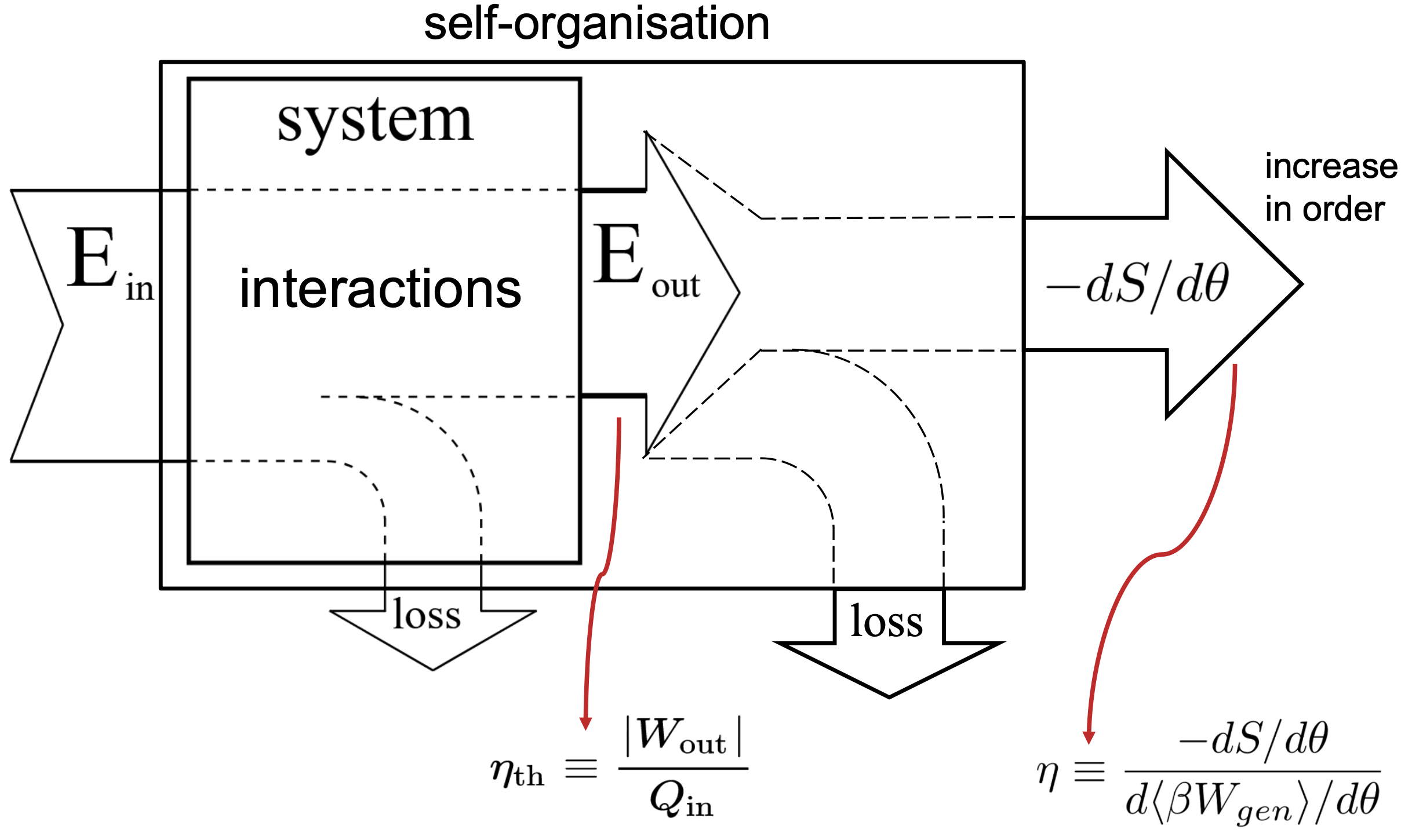}
\caption{\label{fig:thermalDymEff} Thermodynamic efficiency for a system. It is defined as the ratio between the increase in order and change in the generalized work carried out to generate the order.}
\end{figure}

In contrast, thermodynamic efficiency assesses the conversion of work into the system order, measured during a quasi-static change in the underlying control parameter. It pertains to systems involving interactions among multiple components and considers the benefit of increasing order within a collective system against the thermodynamic cost incurred. A system may transition from a disordered to an ordered state by altering a control parameter according to a specific protocol. Thermodynamic efficiency evaluates how efficiently the system converts the carried out work into order, at each specific value of the control parameter (Figure \ref{fig:thermalDymEff}). It is quantified as the ratio of the reduction in the system's configuration entropy (predictability gain) to the generalised work performed during the control parameter adjustment (subject to the unit of the Boltzmann constant $k_B$, e.g., see the expression for entropy, defined in the context of thermodynamics in \eqref{eq:eta_entropy}):
\begin{equation} \label{eq:def_eta}
    \eta(\theta) = \frac{-d \mathbb{S} / d \theta}{d \langle \beta\mathbb{W}_{gen}\rangle / d \theta}
\end{equation}
where $\theta$ is the control parameter, $\mathbb{S}$ denotes the configuration entropy of the system, and $\mathbb{W}_{gen}$ denotes the generalised work performed to change the control parameter. 

\subsection{Perspectives on thermodynamic efficiency}
The \textit{thermodynamic efficiency of interactions} can be expressed in two different ways~\cite{crosato_thermodynamics_2018,nigmatullin_thermodynamic_2021} (technical details are provided in Appendix \ref{appendix:prelim_therm}):
\begin{equation} \label{eq:def_eta2}
        \eta(\theta) = \underbrace{\frac{-d \mathbb{S} / d \theta}{d \langle \beta\mathbb{W}_{gen}\rangle / d \theta}}_\text{In thermodynamic terms}\\
        = \underbrace{\frac{-d \mathbb{S} / d \theta}{\int_{\theta}^{\theta^{*}}{\mathbb{I}(\theta')}d\theta'}}_\text{In computational terms}
\end{equation}
where $\mathbb{I}$ denotes Fisher information.

Thermodynamic efficiency offers a dual perspective on the energy dynamics within systems, encompassing both thermodynamic and computational dimensions. From the thermodynamic viewpoint, this quantity captures the gain in internal order within a collective system of interacting agents (e.g., a swarm) relative to the overall work required to adjust the agent interactions. From a computational viewpoint, thermodynamic efficiency measures the increase in predictability (reduction of uncertainty) of collective action gained by accumulating additional sensitivity to changes in the control parameter along the path $\theta \rightarrow \theta^*$. For example, a swarm may gain predictability of a collective response by adjusting the individual's alignment strength or the number of effective neighbours influencing an individual. This, however, may come at the expense of additional sensitivity to changes in these parameters, so that coherent motion may be disrupted by a change of alignment strength or a reduced number of effective neighbours.

\section{\label{sec:example}A common example}
In previous sections, we explored different approaches to quantifying the intrinsic utility of collective behaviour. In this section, we compare the four considered approaches  --- predictive information, empowerment, active inference, and thermodynamic efficiency --- using the canonical Ising model as a common example. This example considers a system at equilibrium, in order to provide a direct comparison of the collective behaviours resulting from optimising utility functions in the absence of external fluxes.

\subsection{The 2D Ising model}
The 2D Ising model offers a simplified representation of ferromagnetism in statistical mechanics. It models a collection of sites that can each exhibit either an up-spin or down-spin configuration while interacting with their neighbours to create a complex aggregate dynamic. The 2D Ising model is particularly relevant to our comparison due to its characteristic phase transition in the collective dynamics.

\begin{figure}[h]
\centering
\includegraphics[scale=0.5]{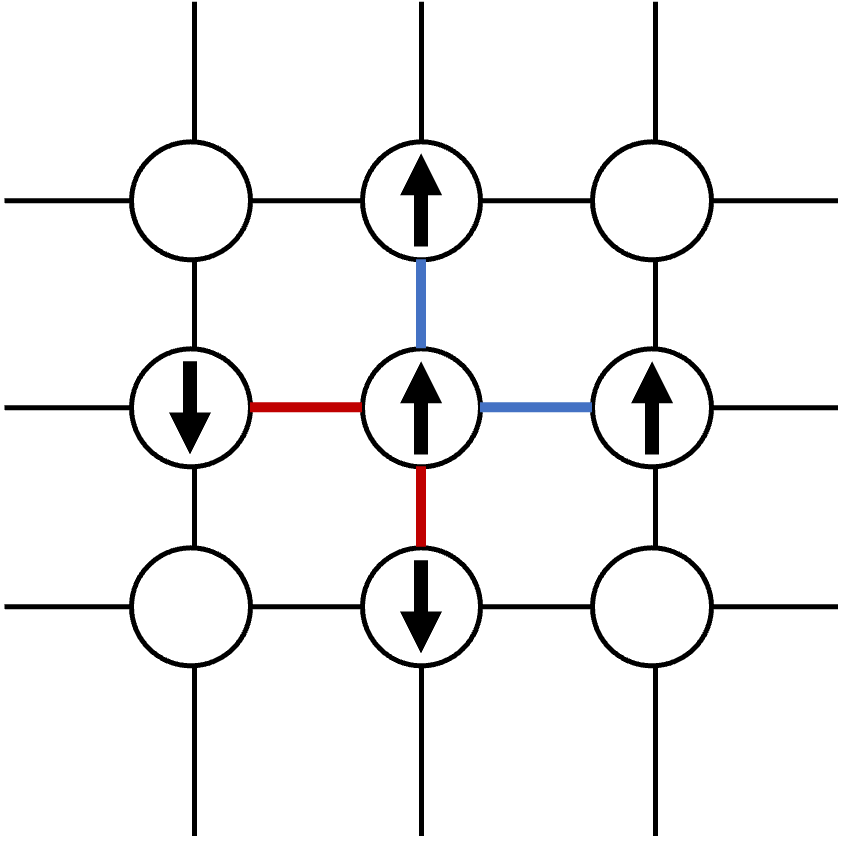}
\caption{\label{fig:ising} A lattice of atoms with dipole magnetic moments. Links in red represent higher energy bonds (where two adjacent atoms have opposite spins), and blue represents lower energy bonds (where two adjacent atoms are aligned).}
\end{figure}

The 2D-Ising model considers a lattice of atoms with magnetic spins oriented either up or down (Figure \ref{fig:ising}). The vertices of the lattice are referred to as ``sites'' and the edges as ``links''. Assuming the absence of an external magnetic field, the energy of a site is determined by the total energies in the links with its neighbours. Each site prefers to be in a lower energy state. For ferromagnetic materials maintaining a link between two sites with opposite spins requires additional energy, hence there is a natural tendency for a site to align its spin with those of its neighbours. The susceptibility of a site to neighbouring influences depends on the coupling strength $J$. A high value of $J$ indicates strong coupling between sites, leading to a greater tendency for spins to align with neighbouring sites.

In this example, the dynamics of the Ising model are interpreted from a perception-action loop perspective: each site acts as an agent that senses the energy of its neighbourhood and ``decides'' whether to flip its spin or maintain its current state. The agency of each site is determined by the choice of the coupling strength $J$. With a high $J$ value a site is more prone to align with its neighbours, and vice versa. This parameter governs the strength of the site's response to the neighbourhood's energy landscape, influencing its decision to align with neighbouring spins.

To draw a clearer connection between the Ising model and the perception-action loop, we consider that each of the four elements of the perception-action loop has a corresponding representation in the Ising model:
\begin{itemize}
    \item $W$ (world): the magnetisation (average spin) of the lattice;
    \item $S$ (sensory): the energy state of a site (defined in eq. (\ref{eq:energy}));
    \item $A$ (action): flipping the spin or remaining unchanged, with flip = $-1$ and no-flip = $+1$;
\end{itemize}

To quantify the energy, we define:
\begin{itemize}
    \item $N$: the number of sites in the lattice;
    \item $J$: the coupling strength between adjacent sites;
    \item $\sigma^{(i)}$: the spin of site $i$, with $+1$ representing up spin and $-1$ is down;
    \item $\underline{\sigma}$: the configuration of the lattice $\underline{\sigma} = \{\sigma^{(1)}, \sigma^{(2)}, ..., \sigma^{(N)}\}$.
\end{itemize}
Let $i$,$j$ be two sites connected by a link, then:
\begin{equation*}
  \sigma^{(i)}\sigma^{(j)} =
    \begin{cases}
      +1 & \text{if sites i, j aligned}\\
      -1 & \text{if sites i, j misaligned}
    \end{cases}       
\end{equation*}
Considering the interactions between a site and its nearest four neighbours only, the total energy of this site is:
\begin{equation} \label{eq:energy}
E^{(i)} = \sum_{j\in \nu^{(i)}}{-J\sigma^{(i)}\sigma^{(j)}} 
\end{equation}
where $\nu^{(i)}$ denotes the set of neighbouring sites of $i$.\\

We simulate the process for both Glauber dynamics \cite{Glauber1963} and Metropolis \cite{bhanot_metropolis_1988, janke_coarsening_2019} dynamics. At each time step, a site was selected uniformly at random. Under Glauber dynamics, the site's spin flips with the probability:

\begin{equation} \label{eq:glauber}
  p_G(\text{flip}) = 0.5\left[ 1- \tanh(0.5\beta dE^{(i)}) \right]
\end{equation}

Alternatively, using Metropolis dynamics, the probability to flip is:
\begin{equation} \label{eq:metropolis}
  p_M(\text{flip}) = \min \left[1, e^{-\beta dE^{(i)}}\right]
\end{equation}
where $\beta$ is the inverse of temperature and $dE^{(i)}$ is the change in energy after a flip. We assume $\beta=1$ for the purpose of this experiment. The energy change $dE^{(i)}$ is computed as:
\begin{equation}
\begin{split}
    dE^{(i)} &= E^{(i)}(\text{after flip}) - E^{(i)}(\text{before flip})\\
     &= \sum_{j\in\nu^{(i)}}2J\sigma^{(i)}\sigma^{(j)}
\end{split}
\end{equation}

The simulation setting is detailed in Appendix \ref{appendix:simulation}, and the source code is available in \cite{qianyang_chen_2024_13784627}. For the analysis, we computed all four intrinsic utility measures for systems at different values of of control parameter $J$. The goal is to compare the range of control parameter $J$ that optimises each utility and discuss the implications of the different optimal ranges for each utility measure and the associated characteristics of self-organised behaviours.

\subsection{Computational results}
During the simulations, we hold the coupling strength $J$ constant and run the simulation until the system reaches equilibrium. We then calculate the corresponding predictive information, empowerment, free energy (active inference), and thermodynamic efficiency. To eliminate the effects of initial conditions, we average these quantities across multiple simulations for each $J$. This process is repeated for a range of $J$ values. We aim to identify the optimal range of $J$ values under each approach in order to answer the question: ``If the coupling strength $J$ evolves independently using each of these quantities as the fitness function, what behaviour should we expect when fitness is optimised?''

We collect the following data for the selected site and the lattice at time $t$:
\begin{itemize}
    \item $a_t$: the action of flip (-1) or no-flip (+1) at time $t$;
    \item $\sigma_t, \sigma_{t+1}$: spin of the selected site before and after the action is performed;
    \item $w_{t}, w_{t+1}$: the magnetisation of the lattice before and after the action is performed;
    \item $s_{t}, s_{t+1}$: the selected site's sensory state before and after the action is performed;
\end{itemize}

The data form time series $\{a_t\}$, $\{\sigma_t\}$, $\{\sigma_{t+1}\}$, $\{w_t\}$, $\{w_{t+1}\}$, $\{s_t\}$, $\{s_{t+1}\}$, using which we compute the intrinsic utility measures. Leveraging the homogeneity of the lattice sites, we can aggregate the random samples from different sites to compute the measures. This approach ensures that the results represent the intrinsic utility values corresponding to the coupling strength $J$ as experienced by an average site within the lattice.

\subsubsection{Predictive information}
For a given coupling strength $J$, the corresponding one-step predictive information is the mutual information between the pre-action sensory state ($S_{t}$) and post-action sensory state ($S_{t+1}$):
\begin{equation}
\begin{split}
    \mathcal{I} &= I(S_{t+1}; S_t) \text{  [bits]}\\
    &= \sum_{s_t}\sum_{s_{t+1}}p(s_t, s_{t+1})\log\frac{p(s_t,s_{t+1})}{p(s_t)p(s_{t+1})}
\end{split}
\end{equation}
where the probability distributions are parameterised by $J$.

\begin{figure}[ht]
    \begin{subfigure}[b]{0.5\textwidth}
        \includegraphics[width=\textwidth]{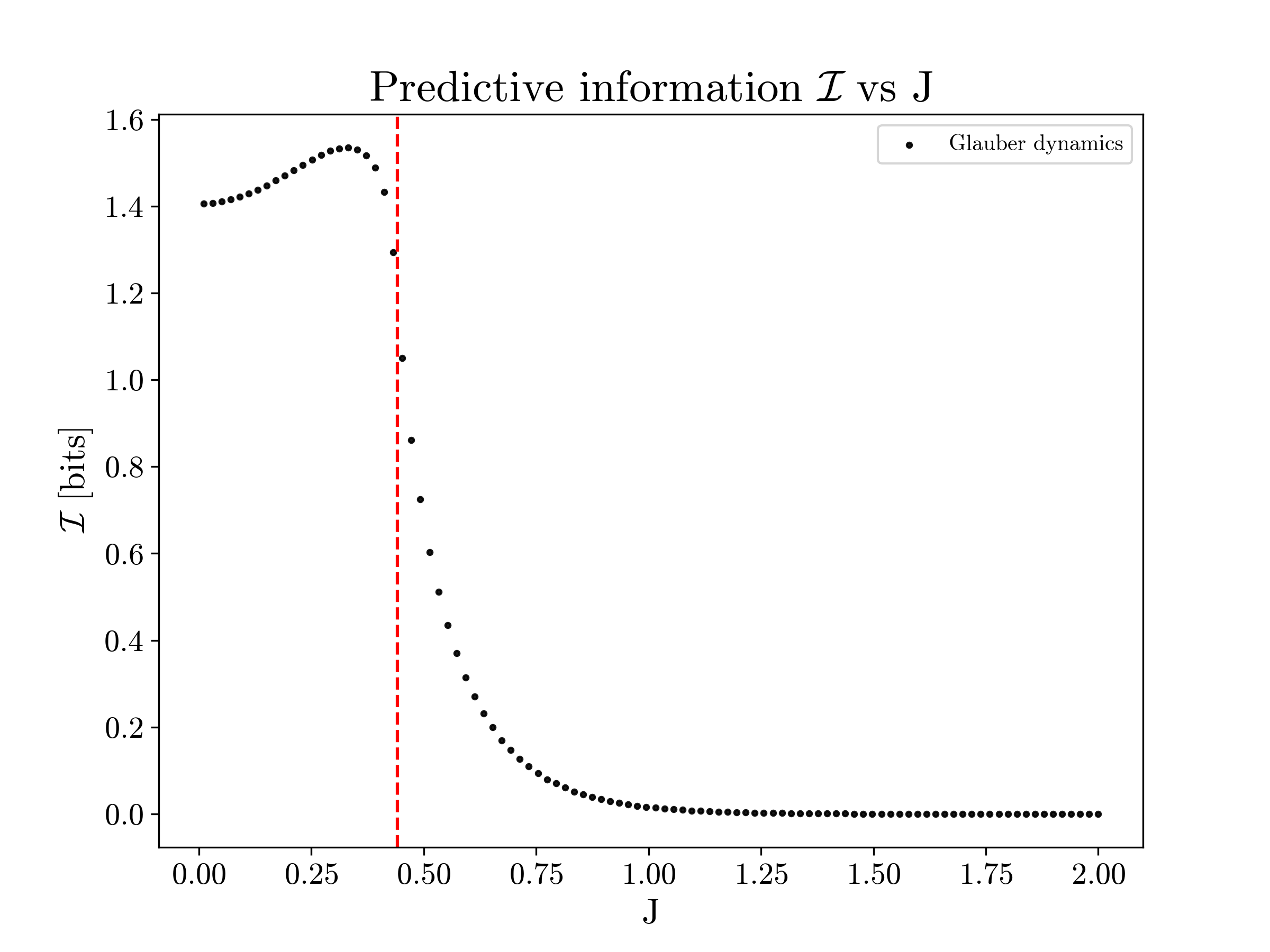}
        \caption{Glauber dynamics}
        \label{fig:result_pi_gla}
    \end{subfigure}  
    \hfill
     \begin{subfigure}[b]{0.5\textwidth}
        \includegraphics[width=\textwidth]{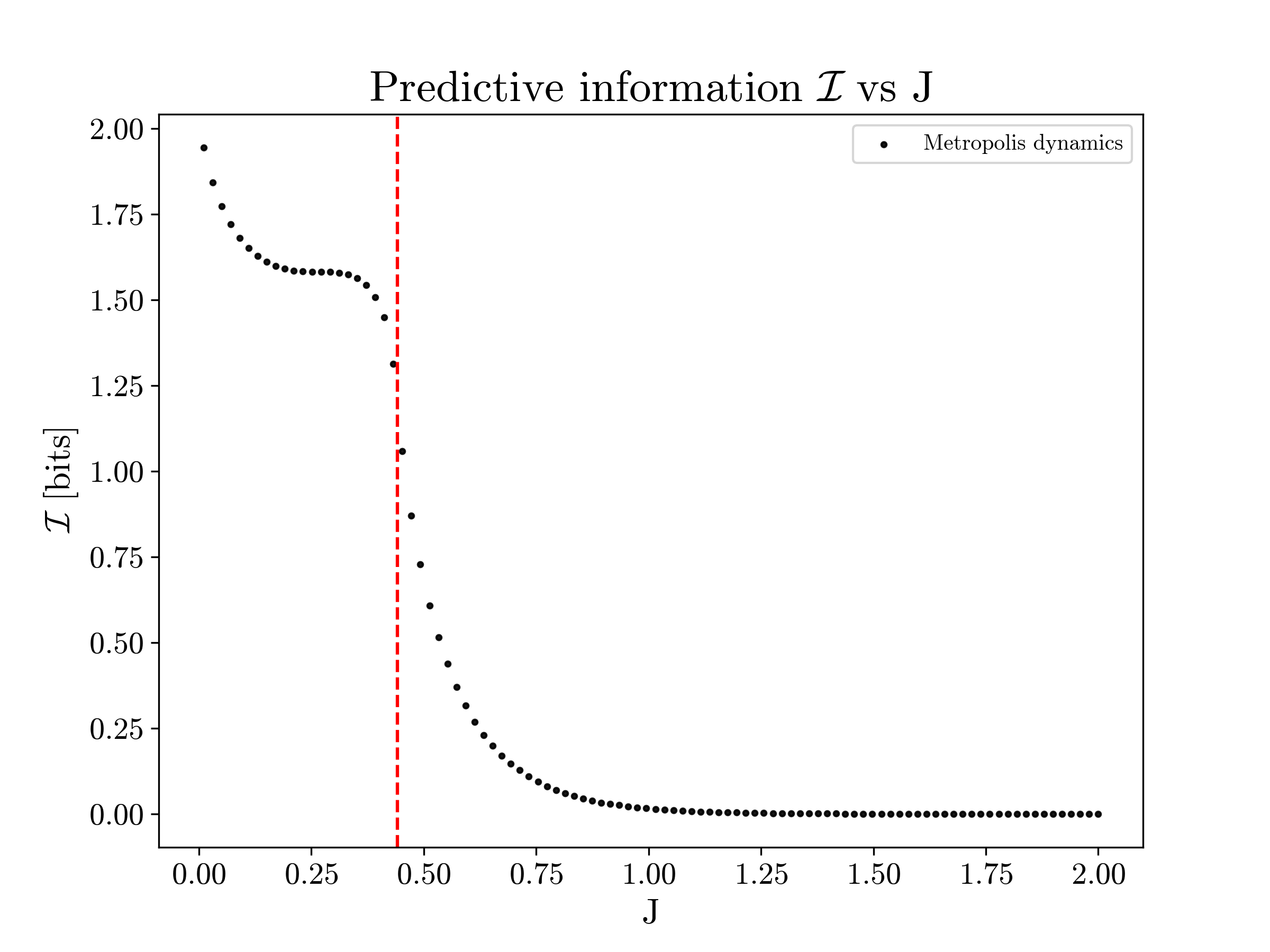}
        \caption{Metropolis dynamics}
        \label{fig:result_pi_met}
    \end{subfigure}
    \caption{\label{fig:result_pi} Average predictive information plotted against different values of $J$, computed from 100 simulations, each of 20 million time steps on a $50\times 50$ square lattice with periodic boundary conditions. Predictive information maximises when $J$ is small (weak coupling), where an average site exhibits explorative behaviour.}
\end{figure}

Figure \ref{fig:result_pi} shows predictive information $\mathcal{I}$ for each coupling strength $J$, with two main observations highlighted. Firstly, predictive information is higher under weak coupling and decreases to nearly zero under strong coupling; this observation applies to both Glauber and Metropolis dynamics. Recall that predictive information can be decomposed into two terms:
\begin{equation} \label{eq:pi_breakdown2}
    I(S_{t+1};S_t) = \underbrace{H(S_{future})}_\text{Diversity of future states} - \underbrace{H(S_{future}|S_{past})}_\text{Unpredictability of future} 
\end{equation}
The overall trend is driven by the diversity component of the equation. Weak coupling promotes more exploratory behaviours, leading to diverse sensory states. In contrast, strong coupling (as $J\rightarrow +\infty$) prevents sites from flipping after settling into the lower energy state, resulting in a predictable system with little sensory diversity.

\begin{figure}[ht]
    \begin{subfigure}[b]{0.5\textwidth}
        \includegraphics[width=\textwidth]{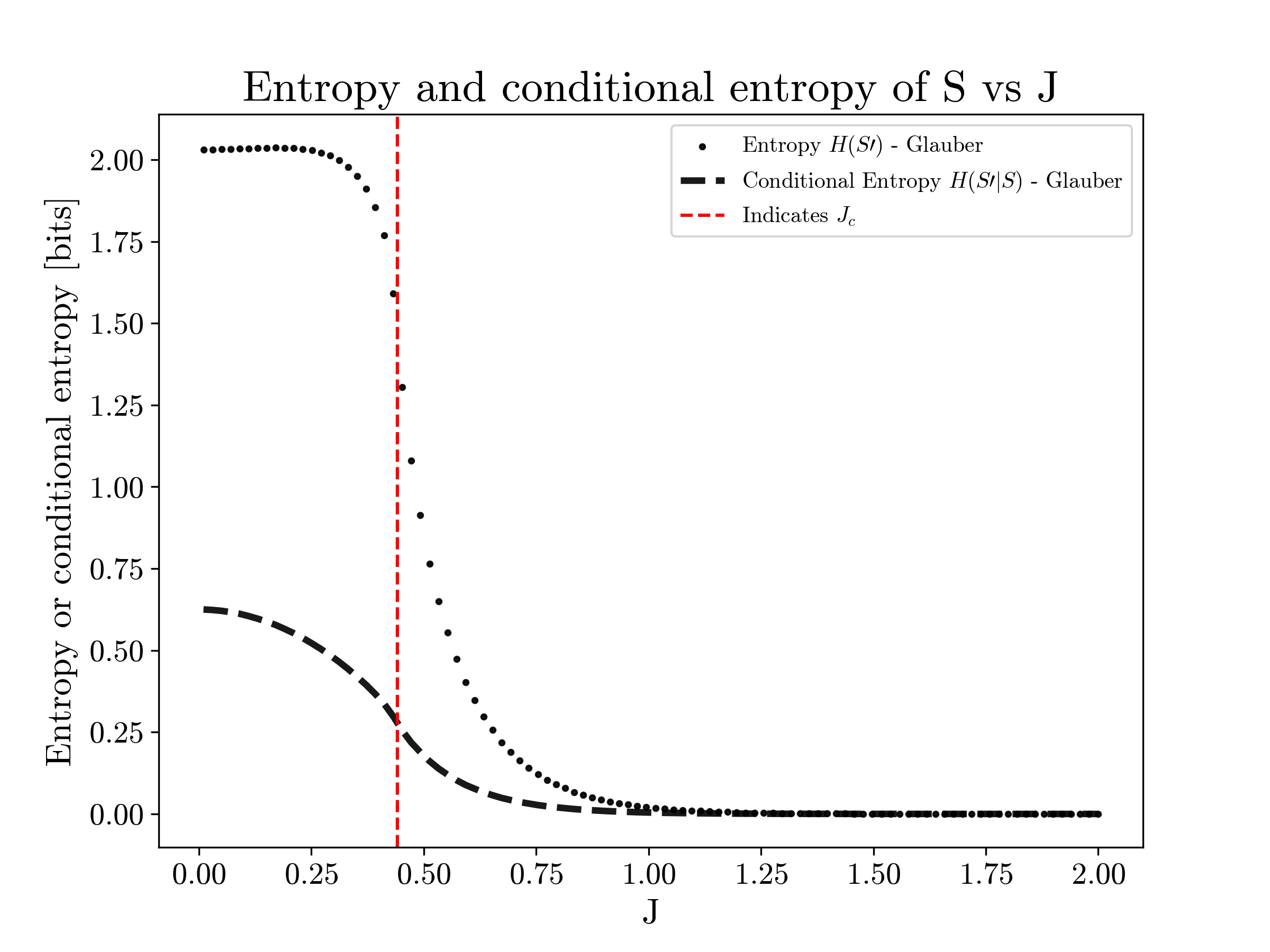}
        \caption{Glauber dynamics}
        \label{fig:result_hs_gla}
    \end{subfigure}  
    \hfill
     \begin{subfigure}[b]{0.5\textwidth}
        \includegraphics[width=\textwidth]{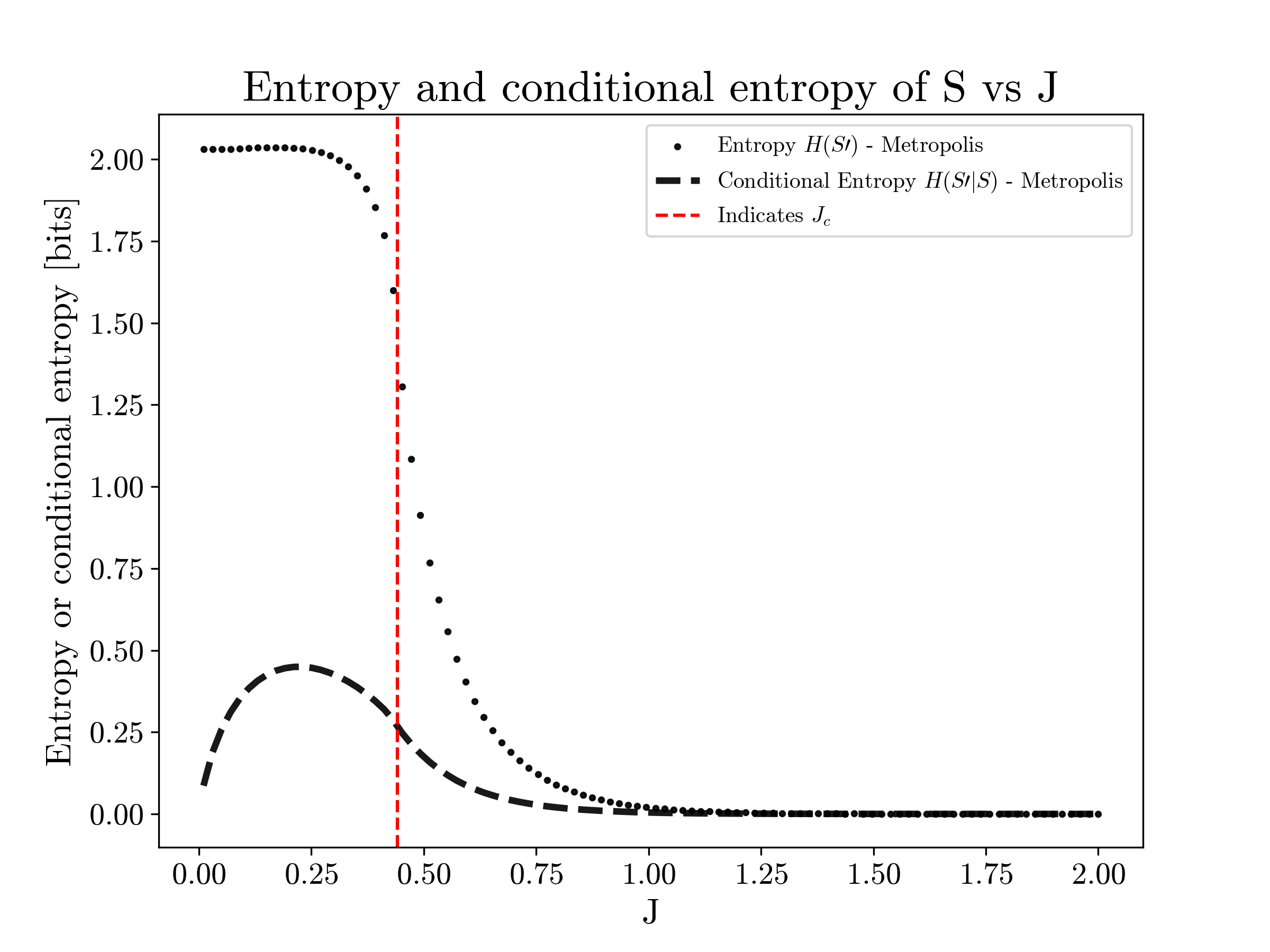}
        \caption{Metropolis dynamics}
        \label{fig:result_hs_met}
    \end{subfigure}
    \caption{\label{fig:pi_decompose} Decomposition of predictive information into richness (dotted line) and unpredictability (dash line) components; predictive information is the difference between the two curves. Results obtained from the average over 100 simulations, each of 20 million time steps on a $50\times 50$ square lattice with periodic boundary conditions.}
\end{figure}

Secondly, predictive information optimises at different coupling strengths for Glauber and Metropolis dynamics -- a phenomenon driven by the unpredictability term (Figure \ref{fig:pi_decompose}). Due to different behaviours of the conditional entropies in the sub-critical regime, predictive information maximises close to the critical point in Glauber dynamics, while peaks as $J \rightarrow 0$ under Metropolis dynamics. A detailed comparison of the two dynamics is provided in Appendix \ref{appendix:simulation}.

\subsubsection{Empowerment}
To compute the average empowerment of a site at equilibrium, we first analytically derive the channel capacity of the action channel. In this model, the action channel is defined by the conditional probability $p(s_{t+1}|a_t)$. The state of pre-action sensory $s_t$ determines this conditional probability, and thus, two cases must be considered separately: $p(s_{t+1}|a_t, s_t \neq 0)$ and $p(s_{t+1}|a_t, s_t = 0)$. 

\begin{figure}[ht]
\centering
\includegraphics[scale=0.7]{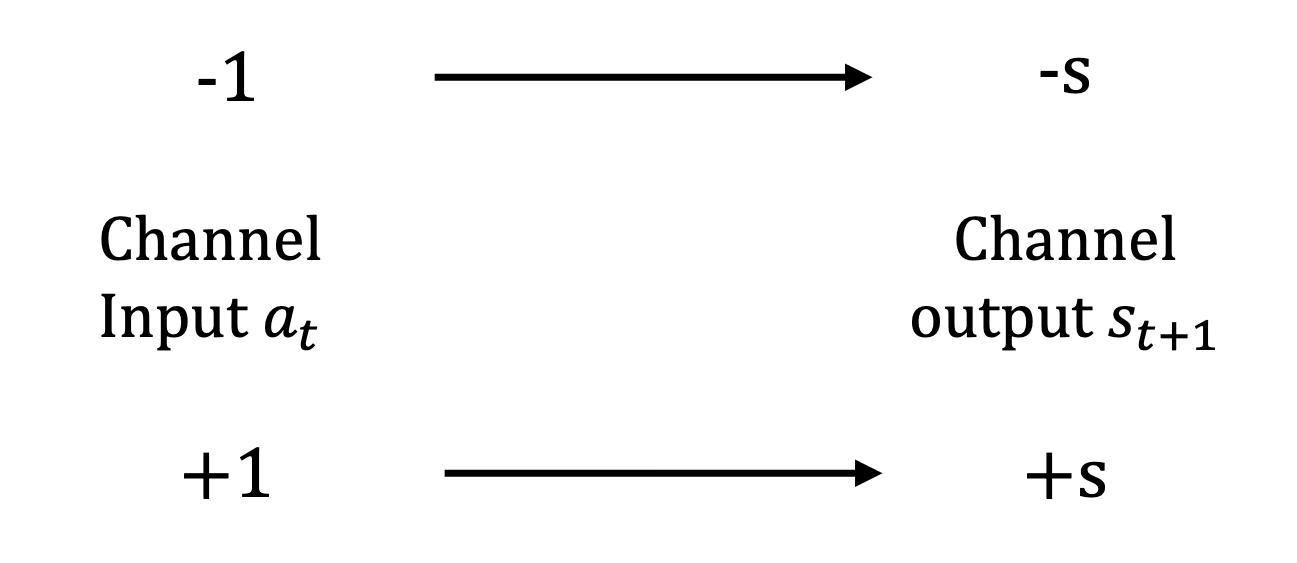}
\caption{\label{fig:channel0} A noiseless binary channel. Channel capacity C = 1 bit.}
\end{figure}

When $s_t \neq 0$, meaning that the up and down spins of the neighbours are not perfectly balanced, a flipping action will result in the next sensory state becoming the opposite of what it was before the flip. The action channel, in this case, resembles the one shown in Figure \ref{fig:channel0}. This is a noiseless binary channel and, by definition, has channel capacity $C(s_t) = 1$ bit. Full capacity is achieved when the site follows action distribution $p(a) = (\frac{1}{2}, \frac{1}{2})$.

\begin{figure}[ht]
\centering
\includegraphics[scale=0.7]{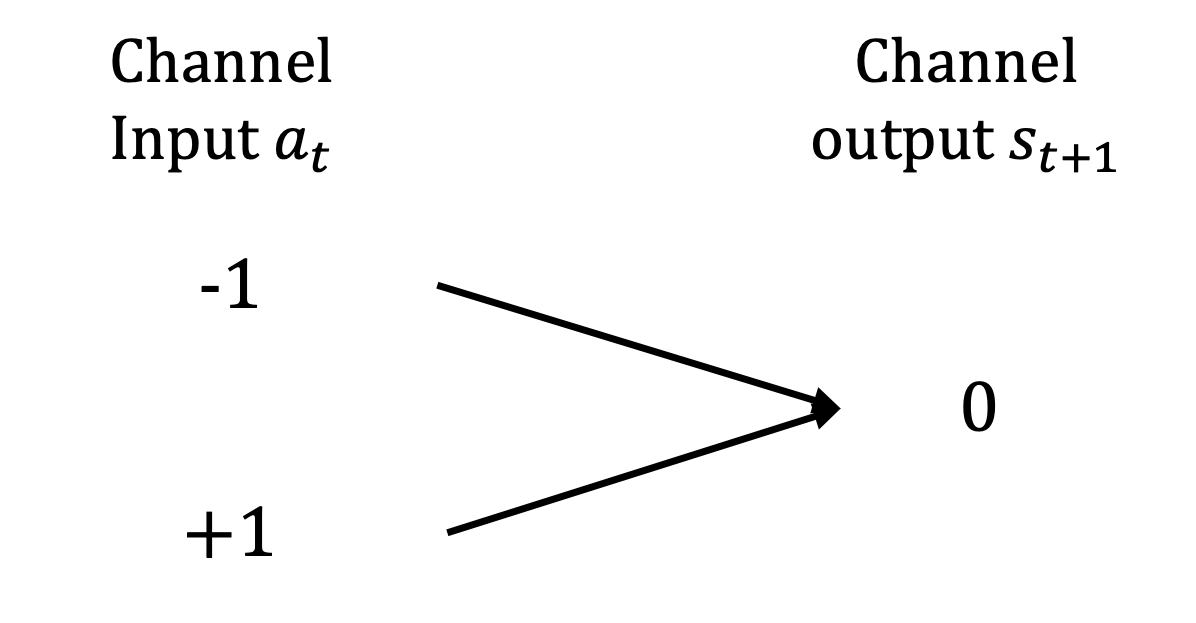}
\caption{\label{fig:channel1} A single output channel. Channel capacity C = 0 bit.}
\end{figure}

If $s_t = 0$, the channel simply reduces to the one shown in Figure \ref{fig:channel1}, that is, a channel that carries no information as the output is always the same. This means that the channel capacity is zero, $C(s_t) = 0$ bit.

\noindent Combining these two cases, we obtain:
\begin{equation}
  C(s_t) =
    \begin{cases}
      1 & \text{if } s_t \neq 0\\
      0 & \text{if } s_t =  0
    \end{cases}
\end{equation}

The average one-step empowerment, averaged over the distribution of channels, is computed as follows:
\begin{equation}
    \mathfrak{\bar{E}} = \sum_{s_t}{p(s_t)C(s_t)} \text{  [bits]}
\end{equation}

\begin{figure}[ht]
    \begin{subfigure}[b]{0.5\textwidth}
        \includegraphics[width=\textwidth]{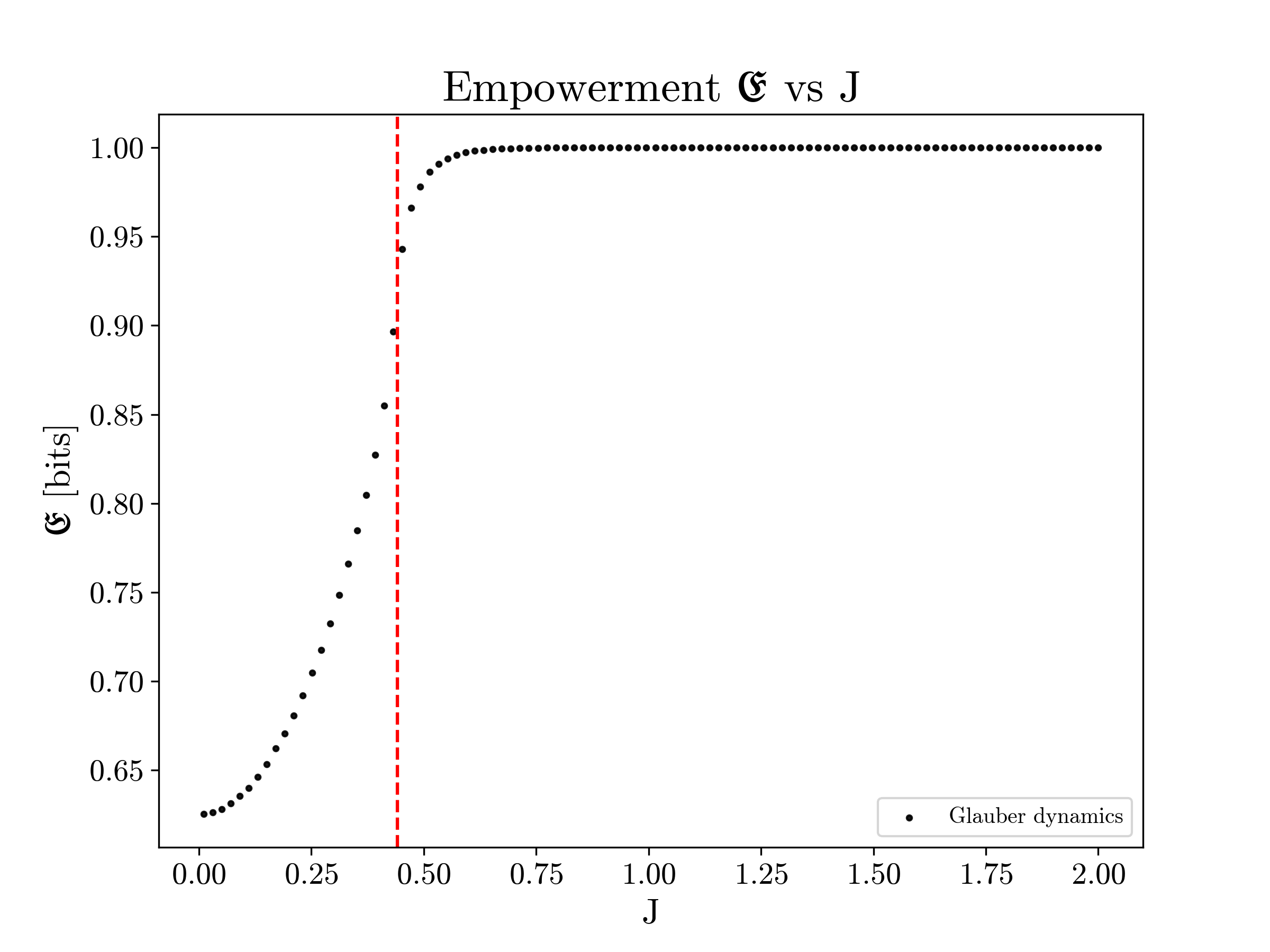}
        \caption{Glauber dynamics}
        \label{fig:result_emp_gla}
    \end{subfigure}  
    \hfill
     \begin{subfigure}[b]{0.5\textwidth}
        \includegraphics[width=\textwidth]{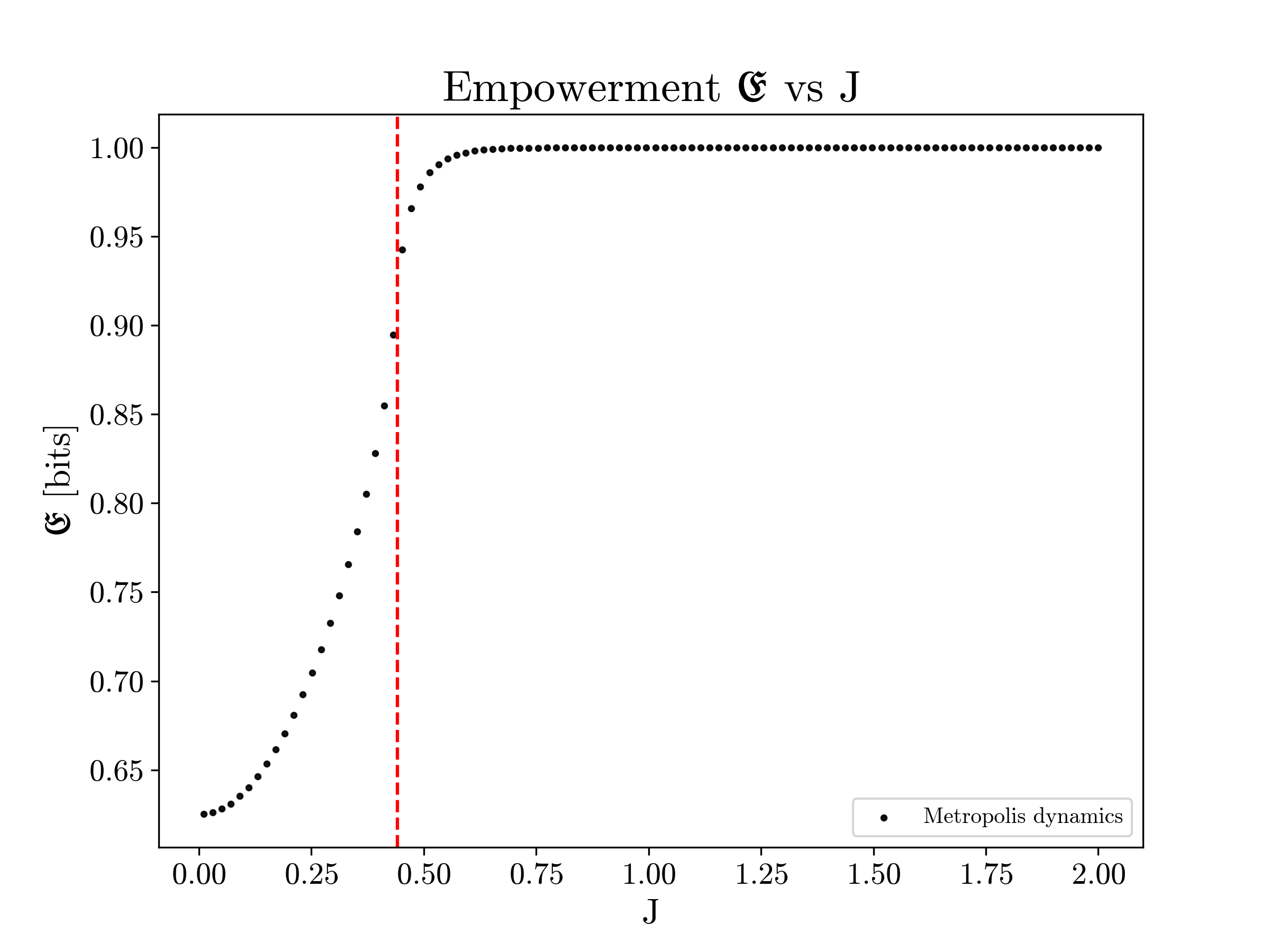}
        \caption{Metropolis dynamics}
        \label{fig:result_emp_met}
    \end{subfigure}
    \caption{\label{fig:result_emp} Average empowerment plotted against different values of $J$, computed from 100 simulations, each of 20 million time steps on a $50\times 50$ square lattice with periodic boundary conditions. Empowerment optimises at strong coupling, where the collective stabilises to a uniformly aligned configuration at equilibrium. Such configuration maximises an average site's ability to retrieve the impact of its action through future sensory states.}
\end{figure}

Empowerment optimises at a strong coupling (Figure \ref{fig:result_emp}), where the lattice stabilises with most atoms aligned at the equilibrium. Empowerment measures an agent's ability to inject information into the environment via current actions and later retrieve the information via its sensors. A site's action is most perceivable when all its neighbours align in the same direction, in which case the action of flip or no-flip leads to distinct sensory outputs ($s$ or $-s$, $s\neq 0$). Conversely, if four neighbours have an equal split between up and down spins, flipping the spin of a site does not change its sensory state. That is, the site will not be able to perceive the impact of its action. A large positive $J$ value increases the probability of an average site being at the configuration where all its neighbours have the same spin, thereby maximising the site's empowerment by ensuring its actions produce noticeable changes to its future sensory inputs. Empowerment is not affected by the choice of spin-flip dynamics (Glauber or Metropolis).

\subsubsection{Variational free energy (active inference)}
For the purpose of this study, the active inference framework is adopted from \cite{biehl_expanding_2018}, thus focusing solely on its intrinsic component. A negative sign is placed before the expression, effectively transforming the minimisation problem into maximising the action value $-\mathcal{F}$. For each possible action $a_t\in\{-1 \:(\textit{flip}), +1 \:(\textit{no-flip})\}$, we compute the one-step negative free energy following the derivation from \cite{biehl_expanding_2018}:
\begin{equation}
\begin{split}
    -\mathcal{F}{(a_t)} &= -H(S_{t+1}|W_{t+1}, a_t) \\
    &= \sum_{s_{t+1}}\sum_{w_{t+1}}p(s_{t+1},w_{t+1}|a_t)\log p(s_{t+1}|w_{t+1}, a_t)
\end{split}
\end{equation}
where the probability distribution $p(.)$ is parameterised by coupling strength $J$. 

\begin{figure}[ht]
    \begin{subfigure}[b]{0.5\textwidth}
        \includegraphics[width=\textwidth]{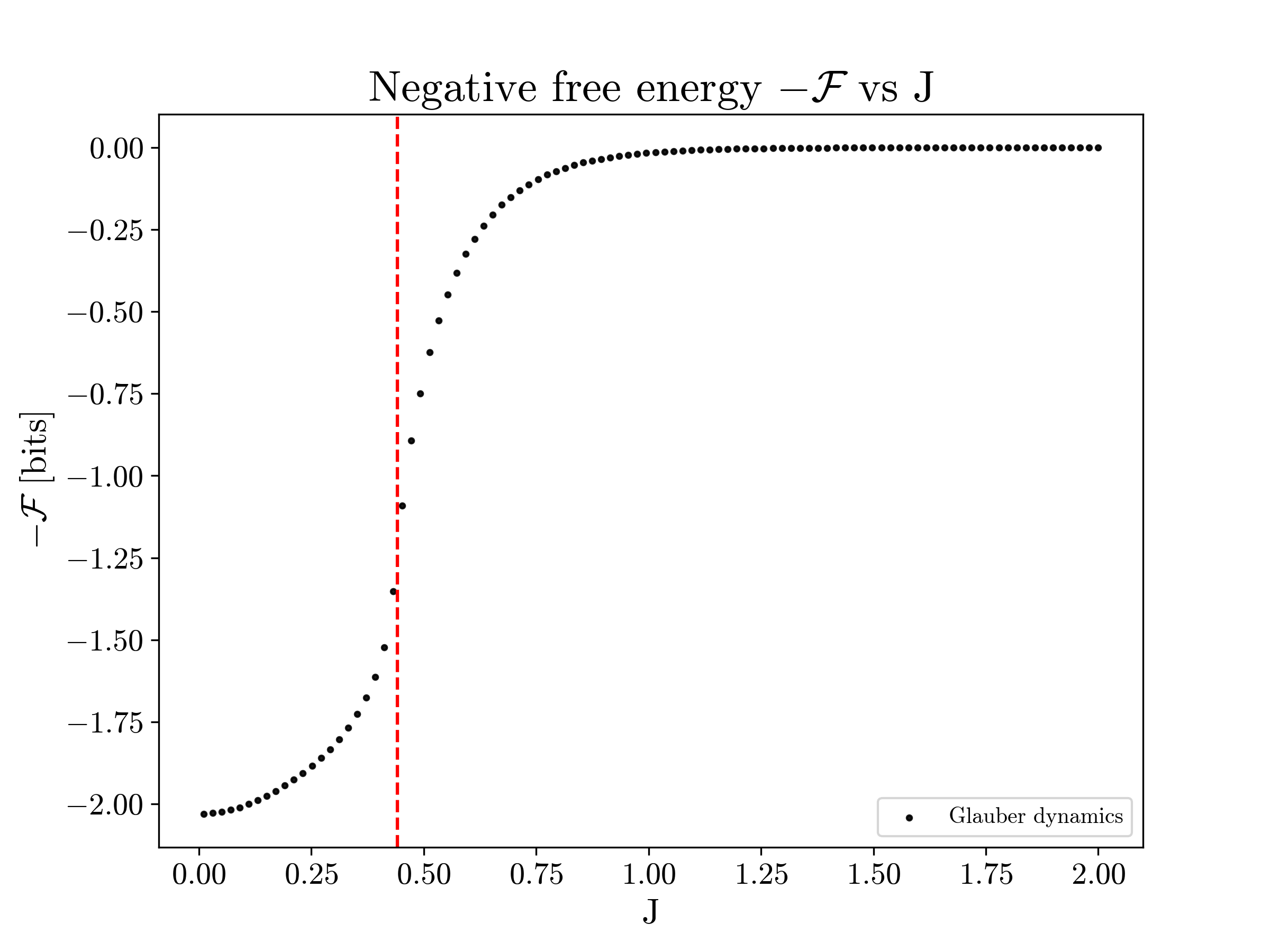}
        \caption{Glauber dynamics}
        \label{fig:result_nfe_gla}
    \end{subfigure}  
    \hfill
     \begin{subfigure}[b]{0.5\textwidth}
        \includegraphics[width=\textwidth]{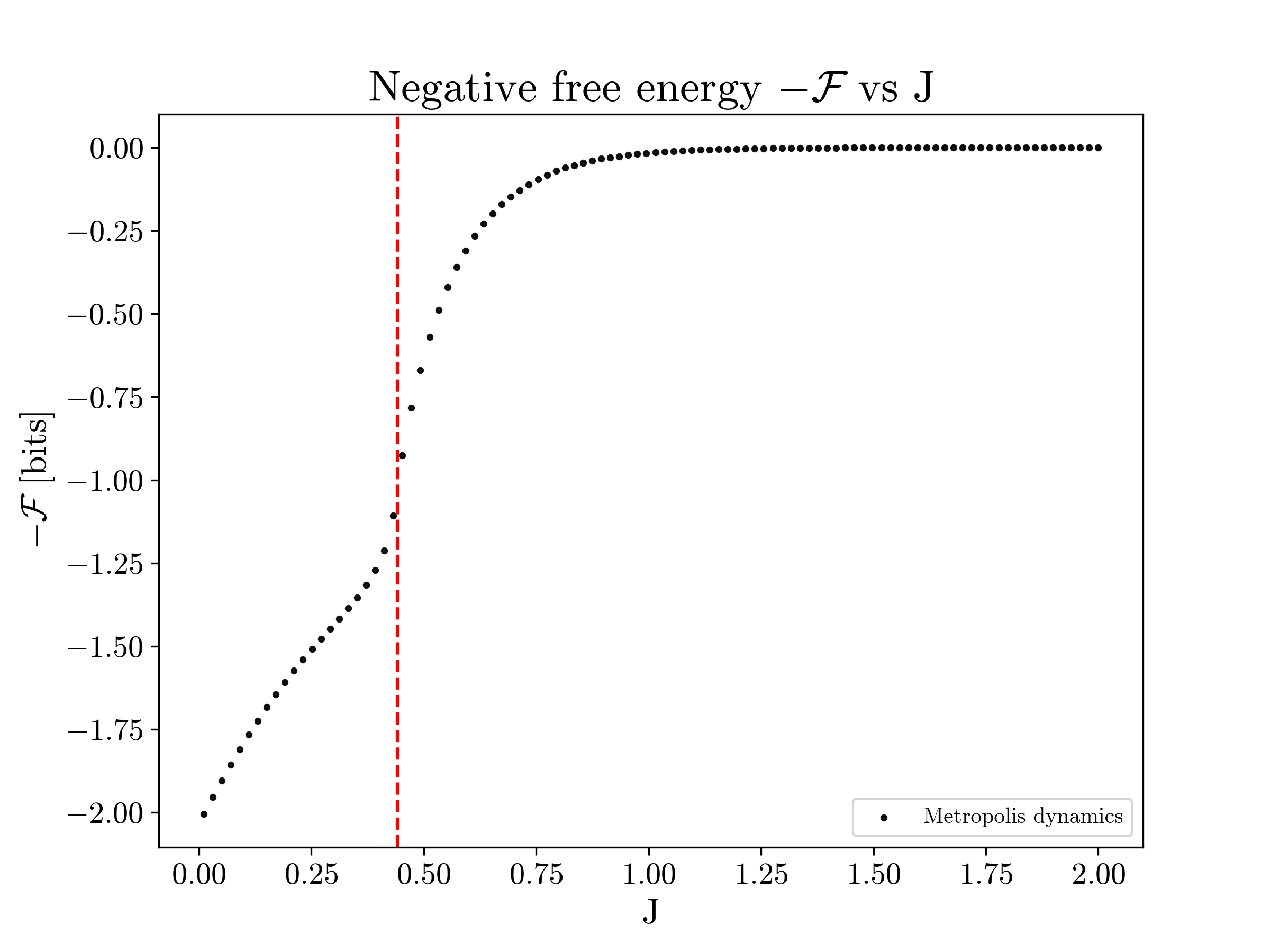}
        \caption{Metropolis dynamics}
        \label{fig:result_nfe_met}
    \end{subfigure}
    \caption{\label{fig:result_fe} Average negative free energy plotted against different values of $J$, computed from 100 simulations, each of 20 million time steps on a $50\times 50$ square lattice with periodic boundary conditions. Negative free energy maximises at strong coupling, where the spins are aligned at equilibrium, and the site experiences minimum surprise comparing its approximate distribution of the world state (based on local sensory information) to the actual distribution that generates the world states (global property).}
\end{figure}

The average negative free energy at equilibrium is computed by weighted average across the proportion of $a_t=-1$ (flip) and $a_t = 1$ (no-flip) actions:
\begin{equation}
    -\mathcal{\bar{F}} = -\sum_{a_t}{p(a_t)\mathcal{F}{(a_t)}} \text{  [bits]}
\end{equation}
For a given coupling strength $J = j$, this measure represents how much discrepancy an average site should expect between its internal model (based on the local sensory history) and the external world (described by the lattice magnetisation); in other words, how well the site's local information aligns with the underlying global situation.

The negative free energy plot (Figure \ref{fig:result_fe}) reveals a similar optimal region for $J$ as empowerment. Maximising negative free energy effectively minimises surprise.  When $J \rightarrow 0$, all the sites are actively flipping their spins. A chosen site's local sensory states can take on all possible values, while the global magnetisation averaged to zero. This mismatch results in large surprise for an average site. On the other hand, more atoms are aligned at large $J$, increasing the likelihood that an average site correctly predicts the overall spin direction, hence reducing surprise. The same trend holds for both Glauber and Metropolis dynamics.

\subsubsection{Thermodynamic efficiency}
Thermodynamic efficiency $\eta$ for each corresponding $J$ is computed as:
\begin{equation}
    \eta = - \frac{d \mathbb{S}(J)/d J}{\int_{J}^{J^*}{\mathbb{I}(J')}dJ'}
\end{equation}
The numerator is the derivative of the configuration entropy $\mathbb{S}$ of the lattice with respect to the control parameter $J$. It represents the reduction of uncertainty in the lattice's configuration as a result of a small variation in the coupling strength $J$. Using the Kikuchi approximation, the configuration entropy is \cite{kikuchi_theory_1951, brandani_quantifying_2013}:
\begin{equation}
    \mathbb{S} = \mathbb{S}_4 - 2\mathbb{S}_2 + \mathbb{S}_1
\end{equation}
where $\mathbb{S}_k$ is the entropy of size k sub-lattices. 

The denominator in this calculation is the integral of Fisher information with respect to the control parameter $J$, representing the work required by the system to instigate the change $\delta J$. The integration limit extends from $J$, the point of evaluation, to $J^*$, the zero-response point. Ideally, $J^* = \infty$, but in this numerical experiment, setting $J^* = 10$ is sufficient, ensuring that the system reaches perfect order at equilibrium and no further work can be done. The method for numerically computing Fisher information is detailed in Appendix \ref{appendix:fisher}.

\begin{figure}[!htpb]
    \begin{subfigure}[b]{0.5\textwidth}
        \includegraphics[width=\textwidth]{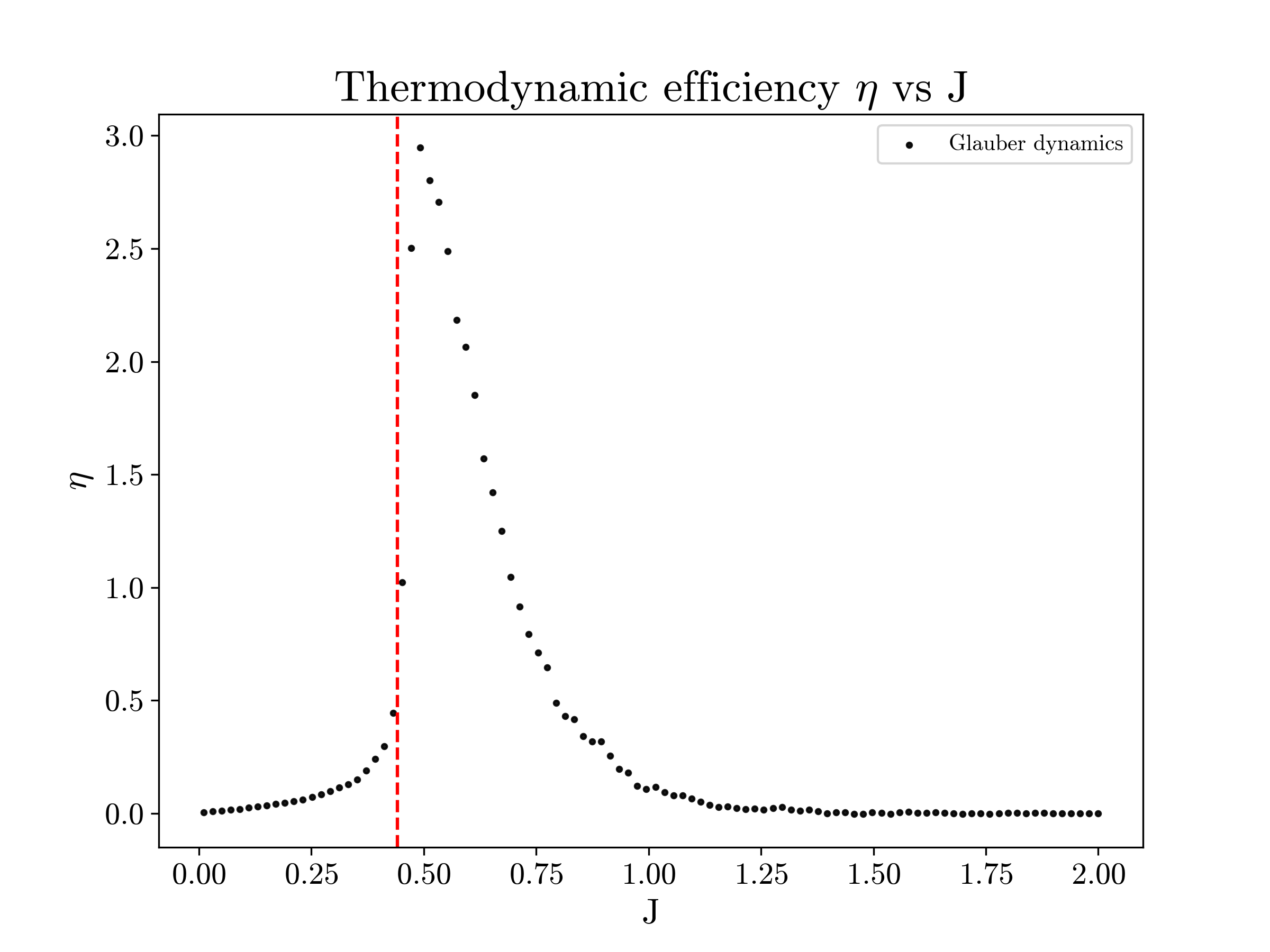}
        \caption{Glauber dynamics}
        \label{fig:result_eta_gla}
    \end{subfigure}  
    \hfill
     \begin{subfigure}[b]{0.5\textwidth}
        \includegraphics[width=\textwidth]{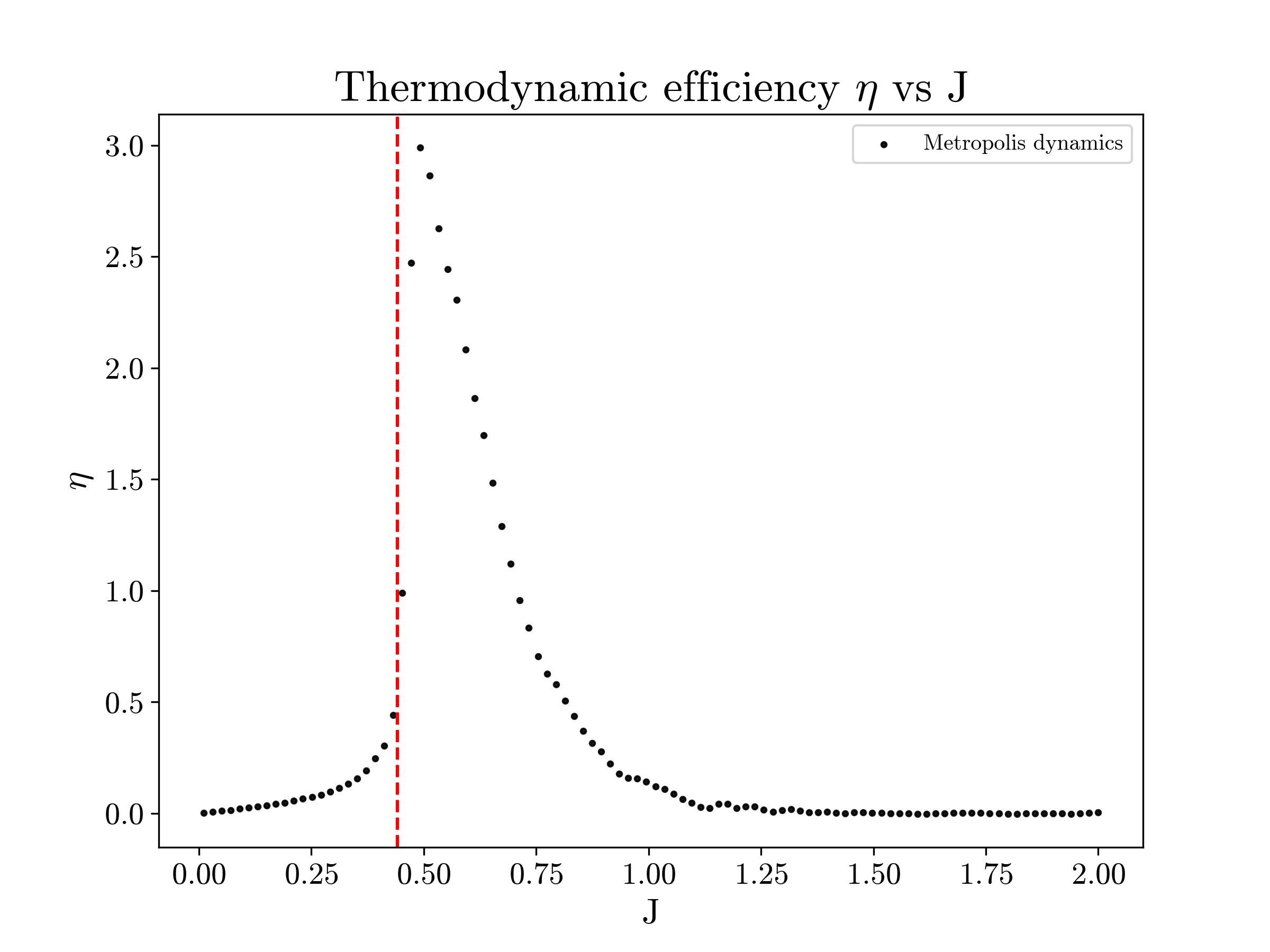}
        \caption{Metropolis dynamics}
        \label{fig:result_eta_met}
    \end{subfigure}
    \caption{\label{fig:result_eta} Average thermodynamic efficiency plotted against different values of $J$, computed from 100 simulations, each of 20 million time steps on a $50\times 50$ square lattice with periodic boundary conditions. Thermodynamic efficiency optimises at the critical regime, where a significant portion of the work expended in tuning the parameter $J$ contributes to reducing configurational entropy; that is, the collective is most energetically efficient in creating internal order.}
\end{figure}

Thermodynamic efficiency reaches optimum when $J$ is near the critical value $J_c \approx 0.4407$ \cite{onsager1944}, as shown in Figure \ref{fig:result_eta}. At the vicinity of the critical point, even a small increase in the control parameter $J$ results in a significant reduction in the system's disorder. Consequently, the work performed to establish order in the system achieves the highest efficiency. This observation indicates that the collective systems that optimise thermodynamic efficiency at the same time operate at the critical regime. The same result holds for both Glauber and Metropolis dynamics.

\begin{figure}[ht]
    \begin{subfigure}[b]{0.5\textwidth}
        \includegraphics[width=\textwidth]{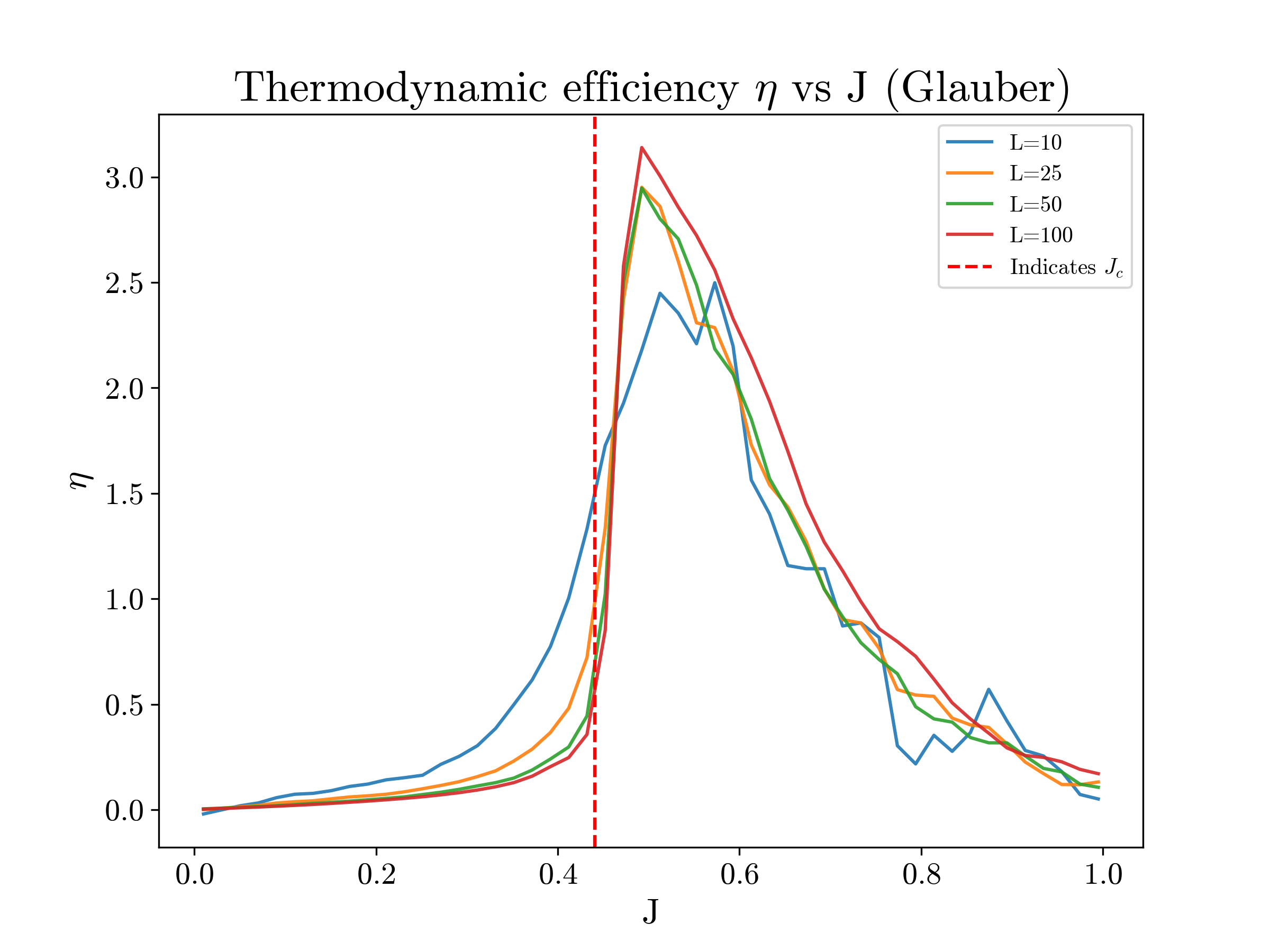}
        \caption{Glauber dynamics}
        \label{fig:result_finite_gla}
    \end{subfigure}  
    \hfill
     \begin{subfigure}[b]{0.5\textwidth}
        \includegraphics[width=\textwidth]{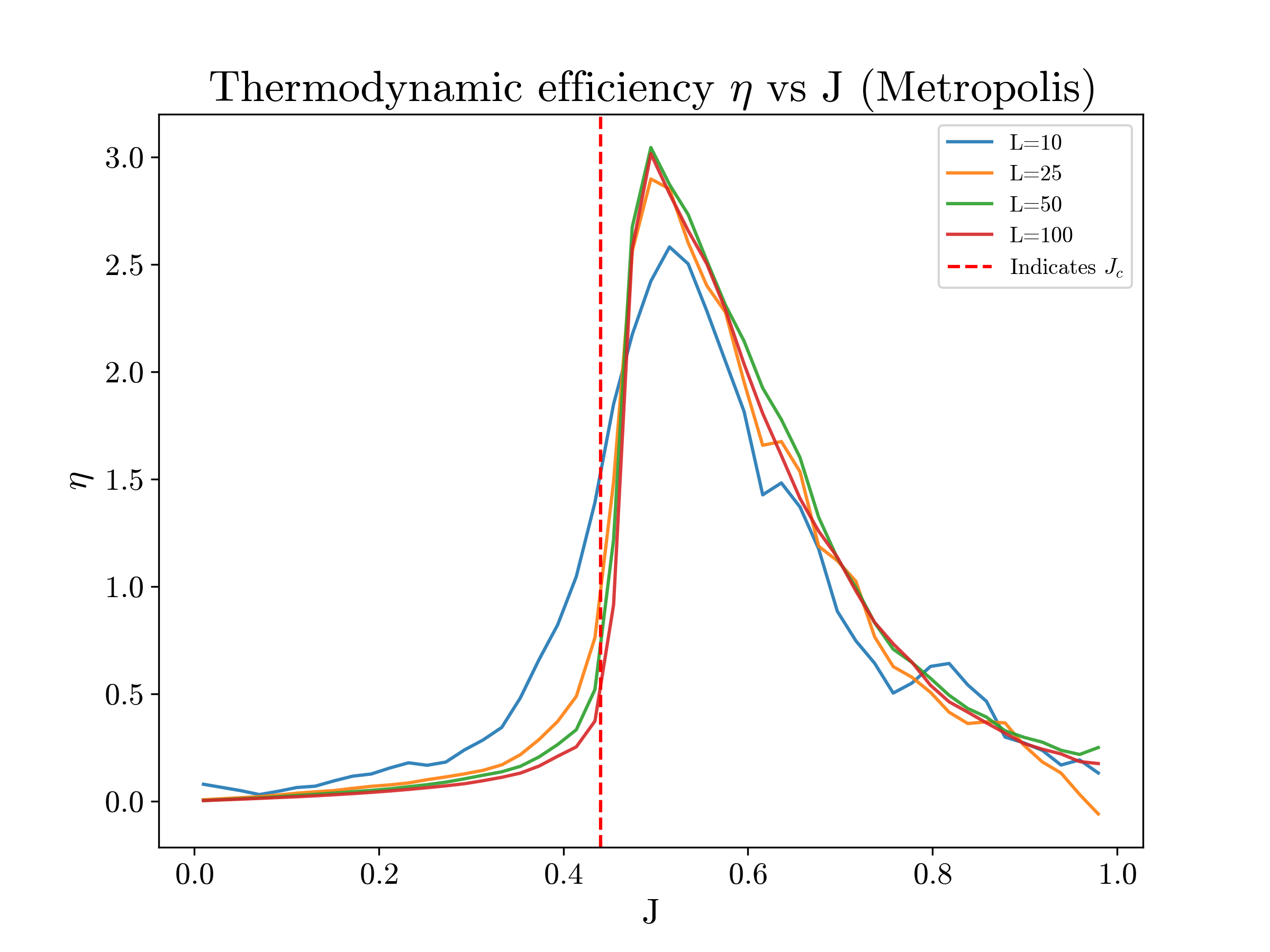}
        \caption{Metropolis dynamics}
        \label{fig:result_finite_met}
    \end{subfigure}
    \caption{\label{fig:result_finitesize} Finite-size analysis near the critical coupling $J_c$ for both Glauber and Metropolis dynamics. Results show that as lattice size increases, the peak of thermodynamic efficiency approaches the critical coupling strength $J_c$.}
\end{figure}

It is important to note that the numerical values of thermodynamic efficiency tend to exhibit more noise than other metrics. This is due to the computation of the entropy derivative and Fisher information. Thermodynamic efficiency peaks slightly to the right of the critical value due to finite-size effect, as shown in figure \ref{fig:result_finitesize}. Despite these numerical nuances, the presented computational results suggest that optimising thermodynamic efficiency within a collective system is achieved at the critical regime. This argument frames the thermodynamic efficiency as an intrinsic utility, and provides an explanation why collective behaviours induced by this utility often exhibit critical phenomena.

\section{\label{sec:principle}Principle of Super-efficiency}

Many natural systems with a large number of interacting components exhibit self-organisation, forming larger structures or coherent collective behaviours without external coordination. Locally interacting neurons collectively perform complex brain functions while processing diverse stimuli \cite{beggs_neuronal_2003, beggs_criticality_2008}. Active matter comprising self-catalytic colloidal particles produces polar collective motion~\cite{ramaswamy2010mechanics,crosato2019irreversibility}. Starlings form flocks that move in intricate patterns in response to environmental changes \cite{cavagna_scale-free_2010, bialek_statistical_2012}.
These self-organising collective behaviours are often observed at critical regimes which seem to balance the fluidity (or adaptability) offered by amorphous, disordered groups (e.g., granular materials or liquids) and the stability (or persistence) provided by rigid, ordered structures (e.g., crystalline materials). 

The ubiquity of collective behaviours that tend to self-organise near or at critical regimes suggests that there is an underlying principle governing such behaviour across different systems. By abstracting from the specific details of each system, we may uncover not only why collective systems self-organise, but also why self-organisation often brings the collective close to the critical regime. A possible underlying principle would need to interpret these behaviours in terms of generic intrinsic utilities. 

Motivated by these observations, we propose the \textit{principle of super-efficiency}:
\begin{quote}
    \textit{Self-organising collective systems strive to maximise the thermodynamic efficiency of interactions by gaining maximal predictability of collective behaviour per unit of the expended work. This efficiency is maximised at the critical regime.}
\end{quote}

In general, the gained predictability enhances coordination within the system, facilitating efficient interactions. Systems with high or maximal thermodynamic efficiency tend to operate at or near the critical regime where long-range correlations and scale invariance bring collective benefits. Thus, thermodynamic efficiency may provide an intrinsic utility to the system by balancing energy costs with group coherence. This would explain the ubiquity of collective animal behaviours, such as swarming, herding, flocking, which self-organise to criticality.

Formally we express the principle of super-efficiency via maximisation of the ratio between the reduction in configuration entropy and the incurred generalised work (equation (\ref{eq:def_eta2})). Given a protocol adjusting the interactions among the constituent components of the system, maximisation of the thermodynamic efficiency occurs through tuning the corresponding control parameter $\theta$\footnote{Note that we use a partial derivative to emphasise that this principle is applicable for settings with multiple control parameters.}:
\begin{equation}
\begin{split}
        \theta^* &= \underset{\theta}{\arg\max} \: \{\eta(\theta)\}\\
        &= \underset{\theta}{\arg\max} \: \frac{-\partial \mathbb{S} / \partial \theta}{\partial \langle \beta\mathbb{W}_{gen}\rangle / \partial \theta}
\end{split}
\end{equation}

The expression above quantifies the abstract notion of super-efficiency: a collective system maximises the ratio of predictability gain to the additional work required to tune the interactions among its components. In the following subsection, we justify this conjecture by taking evolutionary, cognitive, and social perspectives.

\subsection{Maximising thermodynamic efficiency as an intrinsic utility}

Thermodynamic efficiency can be optimised in two ways: either by reducing the additional work (i.e., energy consumption) for a fixed predictability gain or by increasing the predictability gain given a fixed energy budget to carry out the additional work. 

From an evolutionary perspective, minimising energy consumption when establishing structural or functional order preserves resources for survival and reproduction \cite{kaila2008,trenchard2016}. Organisms that achieve goals with less energy per adaptation gain a selective advantage. On the other hand, given energy constraints on behavioural change, increasing predictability enhances group coordination and collective survival \cite{sumpter2006, couzin2009collective,munoz2018}.

In terms of cognitive economy \cite{Simon1956,johnson2024}, individuals use heuristics and mental shortcuts to reduce the energy needed for policy and behavioural changes, freeing resources for other cognitive tasks. Under a fixed energy budget, maximising predictability improves coordination in complex cognitive and social tasks \cite{dunbar1992, harre2016}.

For social dynamics, minimising the energy required for a social transformation conserves resources for the long-term benefit of society, for both current and future generations \cite{hunt2011, li2024advancing}. Alternatively, given energy limitations, maximising predictability enhances strategic planning and policy implementation, particularly in times of crisis \cite{richardson1995,kapucu2008, khialimiab2024}.

\subsection{Super-efficiency in canonical models}

An earlier study by Nigmatullin et al. \cite{nigmatullin_thermodynamic_2021} derived an analytical expression for the thermodynamic efficiency of interactions for the canonical Curie-Weiss model (a fully-connected ferromagnetic model), showing that it diverges at the critical point as the temperature $\theta \rightarrow \theta_c$:
\begin{equation}
  \eta(\theta) =
    \begin{cases}
      -\frac{\theta_c}{2k_B}(\theta - \theta_c)^{-1} & \text{for }\theta < \theta_c\\
       \frac{1}{k_B}(\theta - \theta_c)^{-1} & \text{for }\theta > \theta_c
    \end{cases}       
\end{equation}
that is: 
\begin{equation}
    \eta(\theta) \propto |\theta - \theta_c| ^{-1} \ .
\end{equation}

We can also show that thermodynamic efficiency diverges at the critical point for the canonical Ising model with nearest neighbour interactions. The thermodynamic efficiency $\eta(J)$ as the coupling strength $J \rightarrow J_c$ is given by (see Appendix \ref{appendix:derivation}):
\begin{equation}
    \eta(J) = \frac{\ln(1+\sqrt{2})}{2}|J-J_c|^{-1}
\end{equation}
that is:
\begin{equation}
    \eta(J) \propto |J-J_c|^{-1} \ .
\end{equation}

\subsection{Evidence from simulated systems}
Several previous studies of thermodynamic efficiency~\cite{crosato_critical_2018,crosato_thermodynamics_2018,harding_thermodynamic_2018,nigmatullin_thermodynamic_2021}, as well as the computer simulation presented in Section \ref{sec:example}, exemplify the general principle of super-efficiency.

Crosato et al. \cite{crosato_thermodynamics_2018} explored this relationship near criticality using a model of self-propelled particles. They defined thermodynamic efficiency $\eta$ as the reduction in entropy relative to the work done on the system and demonstrated that as particles undergo a kinetic phase transition from disordered to coherent motion, $\eta$ peaks at the critical regime. In other words, the collective motion becomes coherent at the critical point, where the system is most energetically efficient in reducing its configurational entropy. Hence we can argue that at this point, the system of self-propelled particles is super-efficient (i.e., maximally efficient) in coordinating its collective behaviour, offering a clear utility to the group.

This concept was also applied for urban transformations \cite{crosato_critical_2018}, where the maximum entropy principle coupled with Lotka-Volterra dynamics was used to study shifts in population and income distribution in urban areas. The study considered the thermodynamic efficiency $\eta$ expressed as the increase in predictability of the population income flows relative to the thermodynamic work required to adjust the social disposition (i.e., the factor balancing the suburbs' attractiveness). The study identified a phase transition in urban dynamics, where the number of affluent (i.e., service-abundant) suburbs would change abruptly in response to a small change in the underlying social disposition parameter. Importantly, the thermodynamic efficiency was observed to peak at the phase transition which occurred precisely at the point balancing monocentric and polycentric urban configurations, providing an intrinsic utility for social dynamics.

In the context of epidemic modelling, Harding et al. \cite{harding_thermodynamic_2018} examined the thermodynamic efficiency $\eta$ of contagions diffusing on a network, defining $\eta$ as the ratio of uncertainty reduction in the system to work expenditure required to quasi-statically change the control parameter (e.g. the \textit{infection transmission rate}). Their numerical analyses identified a phase transition between sub-critical (non-epidemic) and super-critical phases (epidemic) as the infection transmission rate increased, with the highest thermodynamic efficiency observed at the critical regime. In this case, the intrinsic utility of an intervention process, considered as social and public health dynamics, is offered by the reduction of pathogen transmission probability: this utility is represented by thermodynamic efficiency maximised at the transition from super-critical to sub-critical phase. Alternatively, the intrinsic utility for the pathogen evolution, considered as a biological phenomenon, would be provided by the increase in the transmission probability: this utility is captured by thermodynamic efficiency maximised at the transition from sub-critical to super-critical phase.

In summary, the principle of super-efficiency is supported by the observed peak in thermodynamic efficiency at the critical regime in various simulated systems, including self-propelled particles, urban transformations, and epidemic modelling. Despite differences in the application settings, the intrinsic utility could be characterised by maximisation of thermodynamic efficiency achieved at the critical regime.

\subsection{Evidence from empirical systems}

The principle of super-efficiency is further supported by empirical studies of collective behaviours in natural systems. While no single empirical study fully justifies the principle, evidence from multiple studies shows that collective systems tend to maximise predictability (through internal order) while minimising energy consumption. 

Empirical studies of European starling show that their flocks exhibit scale-free correlations in velocity and speed fluctuations --- a signature of criticality \cite{cavagna_scale-free_2010, bialek_statistical_2012, bialek_social_2014}. Long-range correlations enhance collective responsiveness to external stimuli (e.g., predators, food resources), allowing flocks to achieve higher coordination of group behaviour (i.e., gain predictability). Friman et al. \cite{friman2024} found that starlings flying in small groups save metabolic energy compared to flying solo, especially when the individuals maintain consistent follower positions behind leaders in V-like formations. These energy savings result from structural coordination, reflecting group-level energy optimisation through positional ordering. Together, these findings suggest that starling flocks operate near a critical regime, balancing predictability gain with energy efficiency. Importantly, these studies contrasted different flocking configurations (ranging between extremes such as solo and coherent groups), so that higher coordination or lower energy consumption can be interpreted as a gained predictability or additional work required to change the configuration. The thermodynamic efficiency is precisely the ratio of these two quantities.

Experiments with ant colonies \cite{beekman_phase_2001, christensen2015, gallotti2018} also show evidence of disorder-to-order phase transition and signatures of criticality (scale-invariant dynamics) in ant foraging activities. These studies suggest that ant colonies benefit from operating at the vicinity of critical point where they maintain a certain level of coordination. These studies show that ants move faster in larger groups, indicating efficiency gains from social coordination. Porfiri et al. \cite{porfiri2024} explained how ant colonies achieve energy savings through social interactions balancing positive and negative feedbacks. Such balance keeps metabolic cost growing sublinearly with colony size, permitting lower energy consumption per individual for larger colonies. Collectively, these findings show that ant colonies self-organise near criticality for maximum efficiency of interactions, where group order is established with low energy consumption. Again, the comparison between different group sizes allows for the interpretation of the coordination increase and energy savings as the predictability gain and additional work respectively. The ratio of these changes is captured by the thermodynamic efficiency.

The organisation and function of brain network also show evidence of balancing predictability gain and energy consumption. Empirical analysis of rat cortex data shows that sensory representation is achieved by reducing entropy (and conditional entropy) of the neuronal responses to stimulus \cite{adibi2013informational}. In other words, sensory adaptation maximises predictability gain in the neuronal activities. Takagi \cite{takagi2021energy} modelled brain network formation by minimising the ratio of activity cost over wiring cost and found structural similarities --- such as hubs and clusters --- between the simulated network and empirical brain networks from various species. By contrasting different network configurations, that is, from completely random networks to small-world networks to fully connected networks, these studies suggest that the brain network evolves to maximise predictability gain while minimising energy consumption per adaptation.

\subsection{Implication: why collective systems self-organise to criticality}

The principle of super-efficiency aims to explain \textit{why} it is beneficial for a group of interacting agents to operate at the critical regime, rather than \textit{how} the system could self-organise to this point --- the latter question is pursued by SOC models. The underlying rationale is that for a self-organising system with many interacting components, being energetically efficient in reducing disorder and creating internal coordination is advantageous, as has been demonstrated by the considered simulation and empirical studies. 

In short, the thermodynamic efficiency offers an intrinsic utility for a collective. When this utility is maximised at a specific configuration (e.g., coupling strength, network connectivity, social disposition), it optimises evolutionary fitness, cognitive economy, or social welfare. In the considered studies the super-efficiency of collective behaviour has occurred at the critical regime, which is expected, as confirmed by the divergence of thermodynamic efficiency at the critical point in canonical models. Thus, super-efficiency may provide a general principle for understanding why collective systems self-organise to the critical regime.

To re-iterate, a super-efficient self-organising system approaches the point where it can gain maximal predictability of its collective behaviour, given the amount of additional work available to change the control parameter.  Alternatively, given a desired predictability gain, a super-efficient system seeks the point where the energy cost of changing the control parameter is minimal. 

\section{\label{sec:conclusion}Discussion}
\textit{Is there an intrinsic utility for self-organising collective systems to operate at the critical regime}? In attempting to explore this question, we overviewed notable intrinsic utility measures, using both information-theoretic and thermodynamic perspectives. The considered measures were directly compared using a common example which we constructed in order to connect the canonical 2D Ising model to the perception-action loop.

The connection is established by conceptualising each site in the Ising lattice as an agent possessing sensory-motor capabilities, thereby linking the model to the perception-action loop framework. The choice of spin-flip dynamic is analogous to the embodiment of the agents. Optimisation of the control parameter --- the coupling strength $J$ --- may be considered analogous to choosing the sensory channels \cite{olsson2004} that maximises specific intrinsic utility given the embodiment.

Optimal $J$ values were computed for different approaches, including predictive information maximisation, empowerment maximisation, free energy minimisation, and thermodynamic efficiency maximisation. For the considered example, each approach exhibited a distinct optimal range of parameter values, offering intuitive insights into the underlying driver shaping collective behaviour:
\begin{itemize}
    \item Predictive information maximises at sub-critical coupling strength for Metropolis dynamics and near-critical regime for Glauber dynamics, balancing sensory richness with predictability;
    \item Empowerment maximises at super-critical coupling strength, where the individuals have maximal influence over the environment;
    \item Free energy minimisation (with intrinsic component only) also leads to super-critical coupling strength, where local observations align most closely with the global configuration hence surprise minimised;
    \item Thermodynamic efficiency maximisation optimises near the critical regime, achieving maximum entropy reduction per unit of work expended.
\end{itemize}
Thus, thermodynamic efficiency, measured by the entropy reduction or predictability gain relative to the associated thermodynamic work carried out, might be a candidate for the intrinsic utility of criticality. 

In the Ising model example, each measure exhibits the same characteristic behaviour reported in previous studies --- balancing diversity and predictability for predictive information \cite{Ay2008, der_predictive_2008}, maximising perceivable influence for empowerment  \cite{klyubin_empowerment_2005, capdepuy_maximization_2007}, aligning the internal model with external generative model for active inference \cite{friston_free_2006, friston_free-energy_2010}, and optimising entropy reduction relative to work for thermodynamic efficiency \cite{crosato_thermodynamics_2018, crosato_critical_2018}. This consistency indicates that the findings we observed in our simulations reflect the measures’ underlying properties rather than artifacts of a specific model. We also tested different Ising model dynamics, both yielding similar results. The considered model represents a broad class of collective systems; this establishes an adequate range of applicability for our comparative analysis.

One limitation of the considered model is its equilibrium dynamics. In contrast, many biological systems operate far from equilibrium, continuously exchanging energy, matter and information with the external environments. Here, we restricted our analysis to equilibrium thermodynamics for two main reasons: (i) previous studies on thermodynamic efficiency are based on equilibrium models, making this a logical starting point for a primer, and (ii) we sought a simplified setting that highlights the characteristics of self-organising behaviours driven by each intrinsic utility without introducing too many additional assumptions. 

Extending the study to non-equilibrium systems is an important direction for future research. For example, a previous study on non-equilibrium flocking \cite{castellana2016} has shown that as velocities of the birds become more aligned, the entropic force can break the flock apart unless it is finely counterbalanced by a cohesive force. Hence, an intrinsic utility such as thermodynamic efficiency might serve as a candidate fitness function for achieving such balance. This would establish whether the principle of super-efficiency is applicable to non-equilibrium systems when significant entropic forces are present.

Informed by this analysis, as well as the relevant studies of thermodynamic efficiency \cite{crosato_thermodynamics_2018, crosato_critical_2018, harding_thermodynamic_2018, nigmatullin_thermodynamic_2021}, we proposed a general principle, the \textit{principle of super-efficiency}, that may explain why collective systems self-organises to the critical point. The principle of super-efficiency states that at the critical point, a self-organising system achieves an optimal entropy reduction relative to the thermodynamic costs. The ability to reduce entropy efficiently grants the collective system an advantage, offering an intrinsic motivation to operate near the critical point. We believe that the principle of super-efficiency has implications for the broader field of guided self-organisation, informing the design of intelligent, adaptive systems that achieve superior coordination, decision-making, and resource management. 

\begin{acknowledgments}
We would like to thank Nihat Ay and Jesse van Oostrum for an insightful discussion on predictive information under different Ising model dynamics. Q.C. would like to thank Joseph Lizier and Jaime Ruiz Serra for discussions on information theory. Sydney Informatics Hub at the University of Sydney provided support for HPC computational resources.
\end{acknowledgments}

\appendix
\section{Information-theoretic quantities}\label{appendix:prelim_info}
We begin by introducing standard notations for information-theoretic quantities that are relevant to the subsequent sections. For readers interested in more details, please refer to \cite{cover_elements_2005}.

For a discrete random variable $X$ that has probability mass function $p(x) = \text{Pr}\{X=x\}$, the \textbf{entropy} $H(X)$ which measures the uncertainty of random variable $X$ is defined as: 
\begin{equation}
    H(X) = - \sum_{x} p(x)\log p(x)
\end{equation}
where by convention, base 2 logarithms are used throughout this paper, and the resulting unit is in \textit{bits}.

If a pair of discrete random variables $(X, Y)$ follows joint probability distribution be $p(x,y)$, the \textbf{conditional entropy} $H(Y|X)$ is defined as:
\begin{equation}
    H(Y|X) = - \sum_{x}\sum_{y} p(x,y)\log p(y|x)
\end{equation}
The conditional entropy $H(Y|X)$ measures the remaining uncertainty in random variable $Y$ given the knowledge of random variable $X$.

The \textbf{mutual information} between two discrete random variables $X$ and $Y$ is defined as:
\begin{equation}
    I(X;Y) = \sum_{x}\sum_{y} p(x,y)\log \frac{p(x,y)}{p(x)p(y)}
\end{equation}
Mutual information is the reduction in surprise about one random variable given the knowledge of the other. It is symmetrical and can be expressed as the difference between entropy and conditional entropy:
\begin{equation}
    I(X;Y) = H(X) - H(X|Y) = H(Y) - H(Y|X)
\end{equation}

The \textbf{relative entropy} or \textbf{Kullback–Leibler divergence} between two probability mass functions $p(x)$ and $q(x)$ is defined as:
\begin{equation}
    D(p||q) = \sum_{x}p(x)\log \frac{p(x)}{q(x)}
\end{equation}
The KL divergence $D(p||q)$ quantifies how much information is lost when an alternative probability distribution $q(x)$ is assumed as a model instead of the actual distribution $p(x)$.

\section{Thermodynamic preliminaries and Fisher Information}\label{appendix:prelim_therm}
\subsection{Thermodynamic preliminaries}
The configuration of a collective system refers to the arrangement of the system's components, usually the geometric or positional arrangement of the components at a specific moment, for example, the up or down orientations of all the spins in a ferromagnetic substance. The configuration entropy represents the amount of uncertainty in the system's arrangement. Lower configuration entropy suggests a limited set of possible configurations, indicating more predictable and coordinated behaviour for the collective system. It is also easier to control or guide the system towards a desired state if there is less uncertainty in the system configurations.

We consider the collective variable $\mathcal{X}_k(x)$ defined as a function of the configuration $x$, which is coupled to the conjugate field or control parameter $\theta_k$. A collective variable characterises the macroscopic state of the system resulting from the specific configuration $x$. An example of the collective variable and its conjugate control parameter could be volume and pressure for an ideal gas system. 

The probability of the system being in configuration $x$ can be expressed by the Gibbs measure \cite{brody_geometrical_1995, prokopenko_relating_2011}:
\begin{equation} \label{eq:eta_px}
    p(x;\Vec{\theta}) = \frac{1}{Z(\Vec{\theta})}e^{-\sum_{k}\theta_k \mathcal{X}_k(x)}
\end{equation}
where $Z(\Vec{\theta})$ is the partition function that normalises this probability over all configurations. For simplicity, we consider single-parameter $\theta$ for the rest of the discussion, but the same framework applies to multi-parameter cases.

In thermodynamic context, the configuration entropy  of the system given by the Gibbs ensemble is defined as:
\begin{equation} \label{eq:eta_entropy}
    \mathbb{S}(\theta) = -k_B \sum_{x}{p(x;\theta)\log \:p(x;\theta)}
\end{equation}
and can be converted to the Shannon entropy $H_{X} = -\sum_{x} p(x)\log p(x)$ by dividing by a factor of $k_B$.

The order parameter $\phi$ conjugate to the control parameter $\theta$ is related to the expected value of the corresponding collective variable $\mathcal{X}$:
\begin{equation}
    \phi = -k_BT\langle \mathcal{X} \rangle
\end{equation}
  The Gibbs free energy $\mathbb{G}$ is defined as:
\begin{equation} \label{eq:G=U-TS}
    \mathbb{G} = \mathbb{U} - T\mathbb{S}- \theta \phi 
\end{equation}
where $\mathbb{U}$ is internal energy. 

Evaluating the work done to or extracted from the system requires the specification of a protocol in which the control parameter varies. Henceforth, we consider a quasi-static protocol, which means that the change of control parameter occurs infinitely slowly so that the system remains in thermal equilibrium with its surroundings at all times. The generalised first law of thermodynamics relates the generalised internal energy $\langle\mathbb{U}_{gen}\rangle = \mathbb{U} - \theta \phi$, the generalised heat flow (from the environment to the system) $\langle\mathbb{Q}_{gen}\rangle$, and the generalised work $\langle\mathbb{W}_{gen}\rangle$:
\begin{equation}
 \Delta\langle\mathbb{U}_{gen}\rangle = \Delta\langle\mathbb{Q}_{gen}\rangle  + \Delta\langle\mathbb{W}_{gen}\rangle
\end{equation} 
Since a change in the configuration entropy
is matched by the heat flow ($\Delta \mathbb{Q}_{gen} = T \Delta\mathbb{S}$), the thermodynamic work equals the change in free energy, that is,  $\Delta\langle\mathbb{W}_{gen}\rangle = \Delta\mathbb{G}$ (the complete argument is presented in Ref.~\cite{crosato_thermodynamics_2018}).  
Taking the first derivative with respect to control parameter $\theta$ yields:
\begin{equation} \label{eq:dW=dG}
    \frac{d \langle\mathbb{W}_{gen}\rangle}{d \theta} = \frac{d \mathbb{G}}{d \theta}
\end{equation}

\subsection{Fisher information}
At this stage we turn our attention to the Fisher Information which is related to Gibbs free energy. The Fisher information measures the amount of information that an observable random variable carries about an unknown parameter $\theta$ which may be influencing the probability of observations. In order words, Fisher information quantifies the sensitivity of observations to the change of parameter $\theta$. Mathematically, Fisher information is defined as the variance of the score function, where the score is the derivative of the log-likelihood function with respect to $\theta$ \cite{cover_elements_2005}:
\begin{eqnarray} \label{eq:def_fisher}
    \mathbb{I}(\theta) &=& E\left[\left(\frac{\partial }{\partial \theta} \log p(x;\theta)\right)^2\right] \notag \\
    &=& \int_{x} \left(\frac{\partial }{\partial \theta} \log p(x;\theta)\right)^2 p(x;\theta) dx
\end{eqnarray}
It has been established that Fisher information is proportional to the rate of change of the order parameter with respect to the change in control parameter, being analogous to susceptibility \cite{prokopenko_relating_2011}:
\begin{equation}
    \mathbb{I}(\theta) = \beta \frac{d\phi}{d \theta} \ 
     \propto \ |\theta - \theta_c|^{-\gamma}
\end{equation}

In explaining the relationship between Fisher information and thermodynamic efficiency, we point out that Fisher information is proportional to the second derivative of Gibbs free energy \cite{brody_geometrical_1995, brody_information_2003, janke_information_2004, crooks_measuring_2007}:
\begin{equation} \label{eq:I(theta)}
    \mathbb{I}(\theta) = -\beta \frac{d^2 \mathbb{G}}{d \theta^2} 
\end{equation}
As established in \cite{crosato_thermodynamics_2018} under a quasi-static protocol, following the first law of thermodynamics, equations (\ref{eq:dW=dG}) and (\ref{eq:I(theta)}) yield:
\begin{equation} \label{eq:eta_integrate2}
    \frac{d \langle\beta \mathbb{W}_{gen}\rangle}{d \theta}= \int_{\theta}^{\theta^{*}}{\mathbb{I}(\theta')}d\theta'
\end{equation}
where the integral is computed from the point of evaluation $\theta$ to the ``zero-response point'' $\theta^*$ defined as the point where a perturbation of control parameter $\theta$ extracts no work from the system. 

\section{Simulation of Ising model}\label{appendix:simulation}

The following settings have been applied to the simulations:
\begin{itemize}
    \item Lattice: a torus-shape lattice of size 50$\times$50;
    \item Coupling strength $J$: values taken from range (0,2) with increments $\delta J$ = 0.02;
    \item Number of simulations: 100 simulations are run for each value of $J$;
    \item Number of time steps: for each simulation, 20.2 million steps are simulated to ensure the system reaches equilibrium. The transient period (first 20 million steps) is excluded from the analysis;
    \item Number of samples (for predictive information, empowerment and variational free energy): in each simulation, the last 200k steps are sampled to compute the corresponding information-theoretic quantities. The quantities are then averaged over 100 simulations;
    \item Number of samples (for thermodynamic efficiency): in each simulation, configuration distributions are sampled from the last 200k time steps with a sub-sampling interval of 2,500 time steps (1 sweep of the lattice). A total of 8,000 sample distributions (80 samples/simulation $\times$ 100 simulations) are collected for each $J$. The average distribution over these 8,000 samples is used for computing the corresponding Fisher information. Configuration entropy is computed using the last snapshot of lattice configuration at each simulation, and then averaged over 100 simulations;
    \item Inverse thermodynamic temperature $\beta$: chosen to be constant 1.
\end{itemize}

\begin{figure*}[!ht]
    \centering
    \begin{subfigure}[b]{0.48\textwidth} 
        \includegraphics[width=\textwidth]{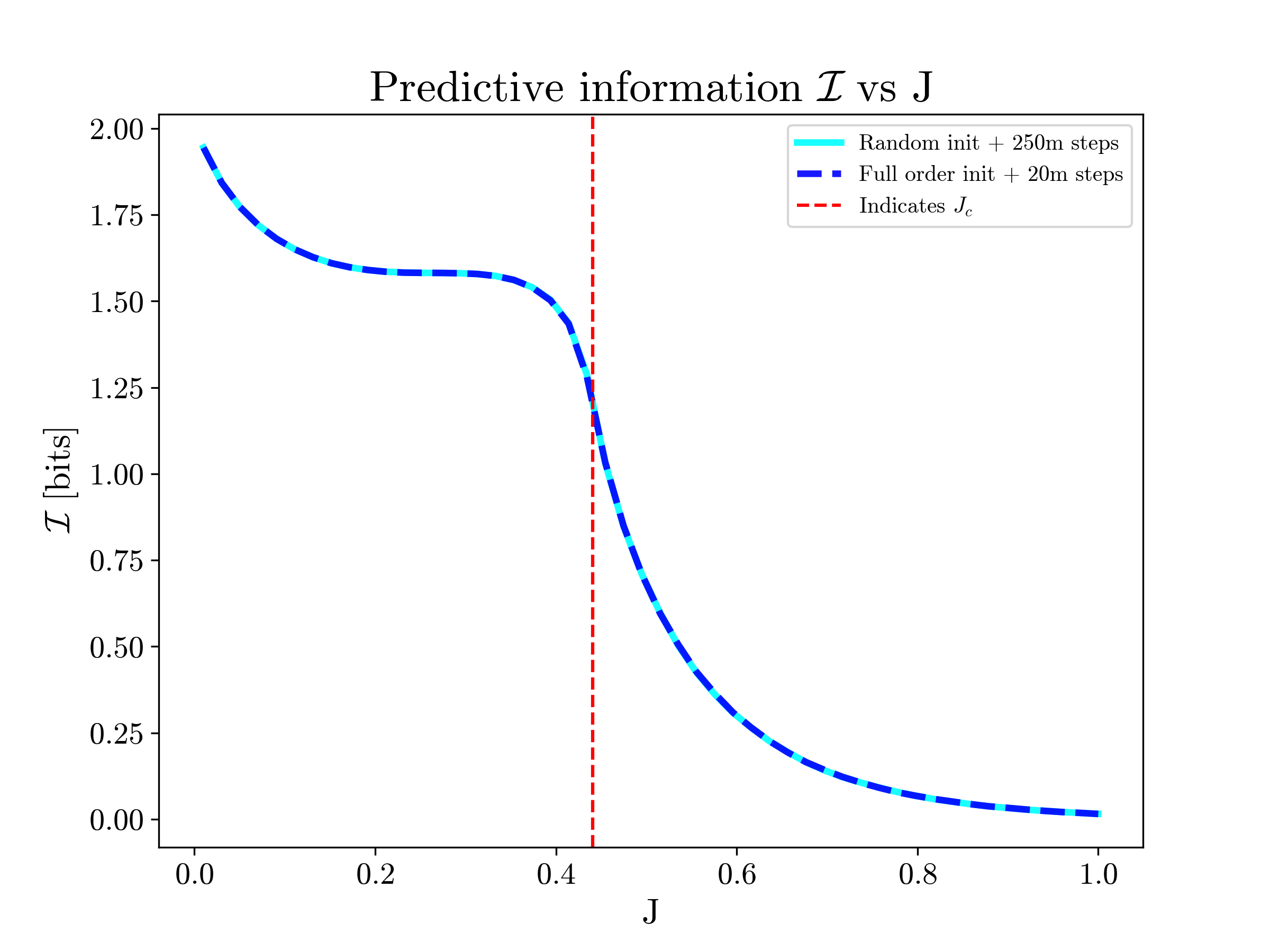}
        \caption{Predictive information}
        \label{fig:pi_v_j}
    \end{subfigure}
    \hspace{0.005\textwidth} 
    \begin{subfigure}[b]{0.48\textwidth}
        \includegraphics[width=\textwidth]{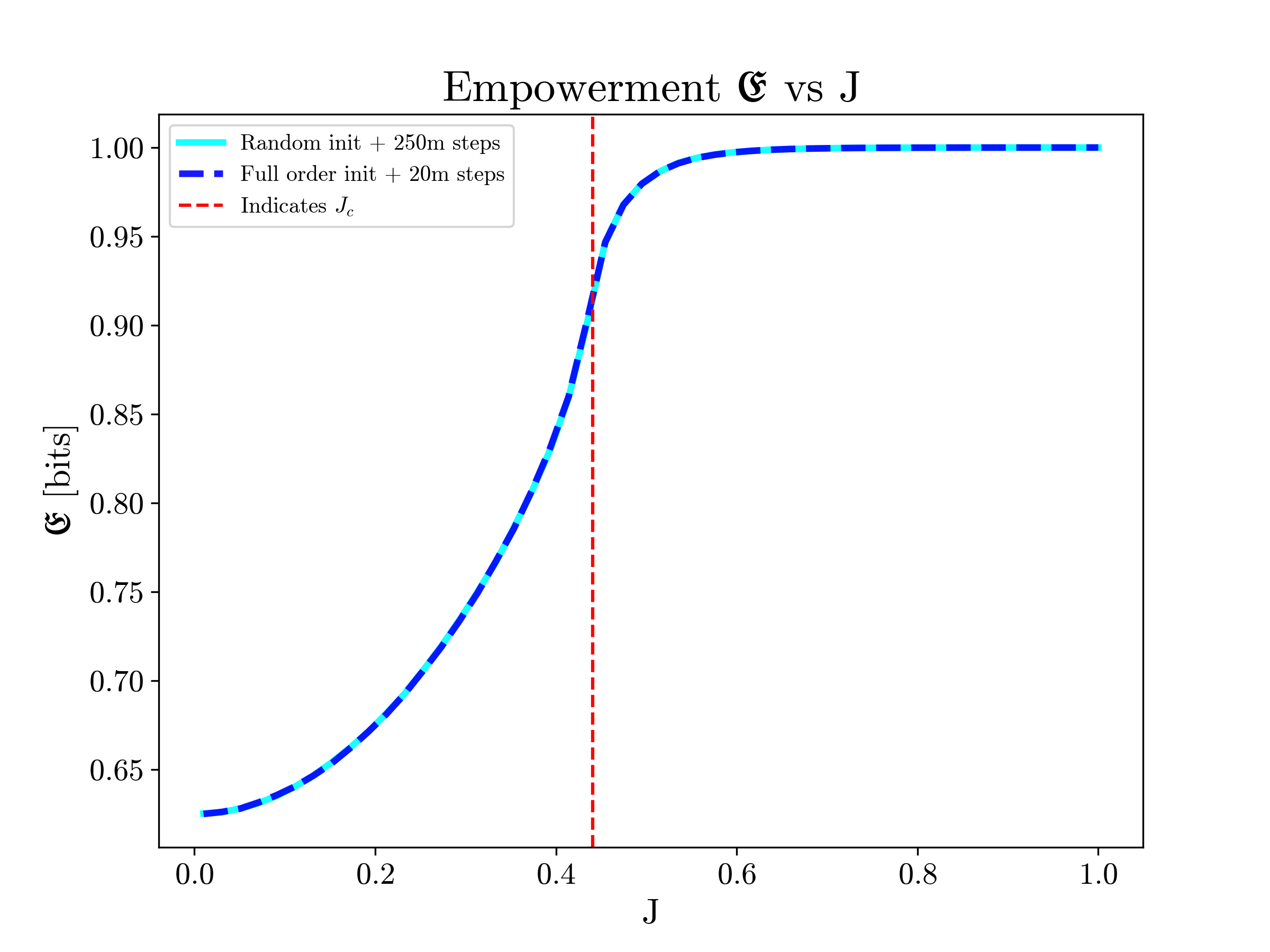}
        \caption{Empowerment}
        \label{fig:emp_v_j}
    \end{subfigure}  
    
    \vspace{0.5em} 
    \begin{subfigure}[b]{0.48\textwidth}
        \includegraphics[width=\textwidth]{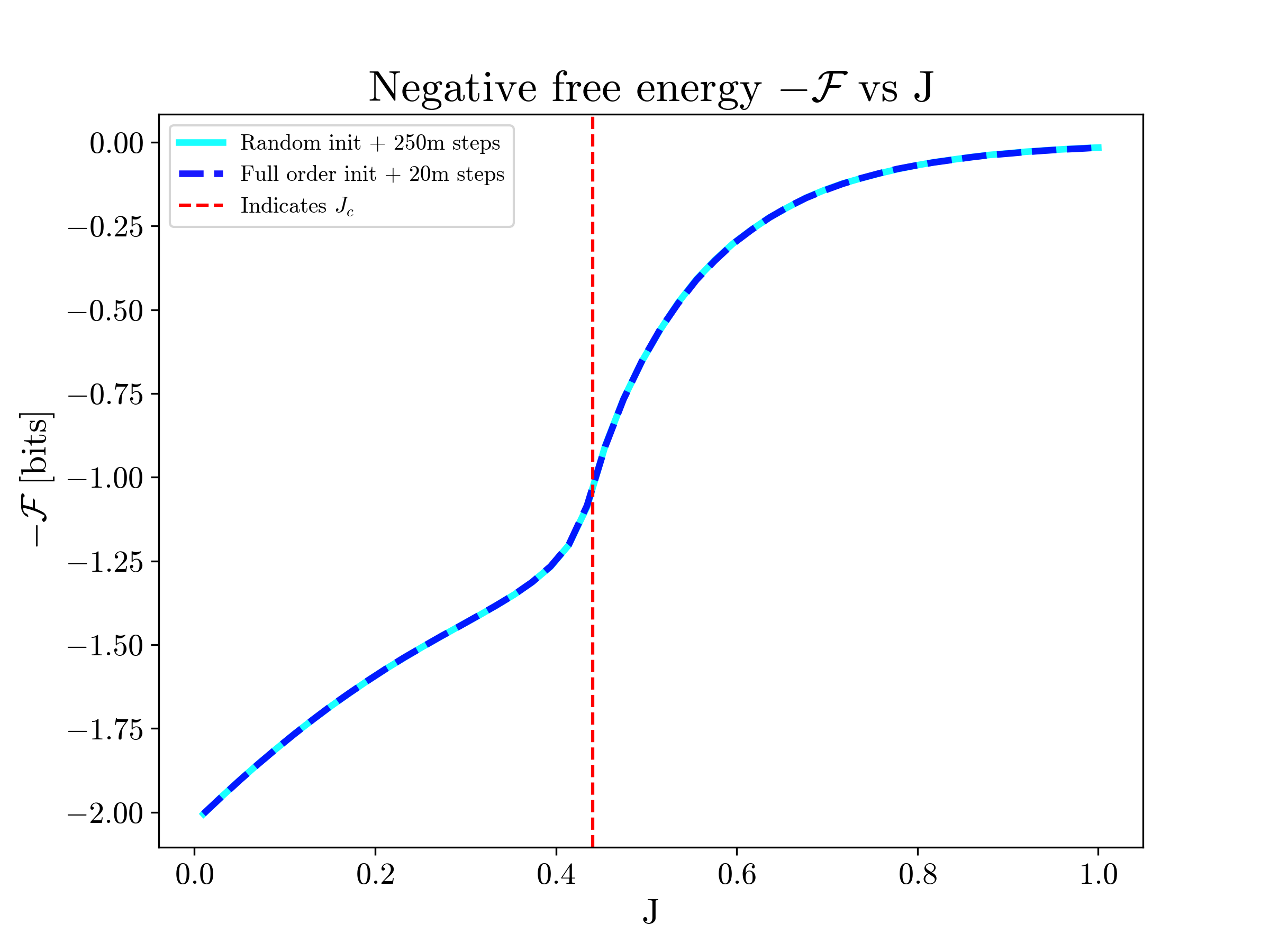}
        \caption{Negative free energy}
        \label{fig:nfe_v_j}
    \end{subfigure} 
    \hspace{0.005\textwidth}
    \begin{subfigure}[b]{0.48\textwidth}
        \includegraphics[width=\textwidth]{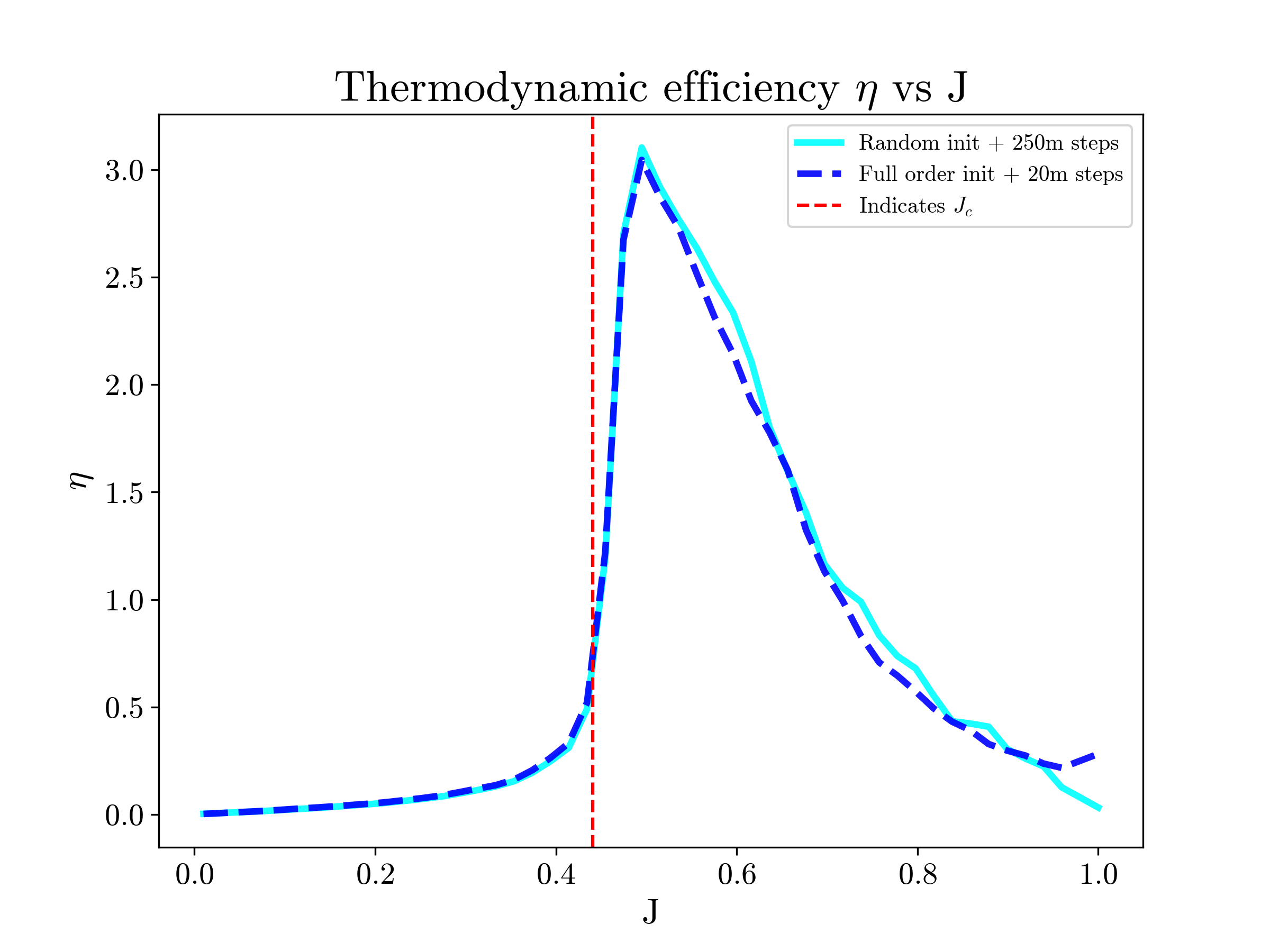}
        \caption{Thermodynamic efficiency}
        \label{fig:eta_v_j}
    \end{subfigure} 
    
    \caption{Compare results for two different simulation settings: random initial configuration with 250 million time steps vs fully-ordered initial configuration with 20 million time steps. The results are almost identical across different initialisation methods when the simulations have converged. Note that the simulations do not fully converge at high coupling strength (low temperature) when using random initialisation with 20 million time steps.}
    \label{fig:init}
\end{figure*}

The lattice is initialised in a fully ordered state. At high coupling strength (low temperatures), single spin-flip algorithms risk trapping the system in a local minimum. Ordered initialisation drives the system to global energy minima at high coupling strength (low temperatures), without affecting outcomes at low coupling strength (high temperature). Since only equilibrium states are analysed and transients are excluded, this method gives the same result as using random initialisation but significantly accelerates convergence (Figure \ref{fig:init}). 


We tested two different spin-flip dynamics -- Gluaber dynamics \cite{Glauber1963} and Metropolis \cite{bhanot_metropolis_1988, janke_coarsening_2019} dynamics -- to assess the impact of model choice on the results. At each time step, a  site was selected uniformly at random, and its spin flipped with the probability:

\textbf{Glauber criterion}:
\begin{equation} \label{eq:glauber_appendix}
  p_G(\text{flip}) = 0.5\left[ 1- \tanh(0.5\beta dE) \right]
\end{equation}

\textbf{Metropolis criterion}:
\begin{equation} \label{eq:metropolis_appendix}
  p_M(\text{flip}) = \min \left[1, e^{-\beta dE}\right]
\end{equation}
where $dE = E_{after} - E_{before}$ is the difference in energy before and after the spin flip of the site. $dE$ depends on the coupling strength $J$ and the number of neighbours aligned with the site's spin.

\begin{figure*}[!ht]
\centering
\includegraphics[scale=0.6]{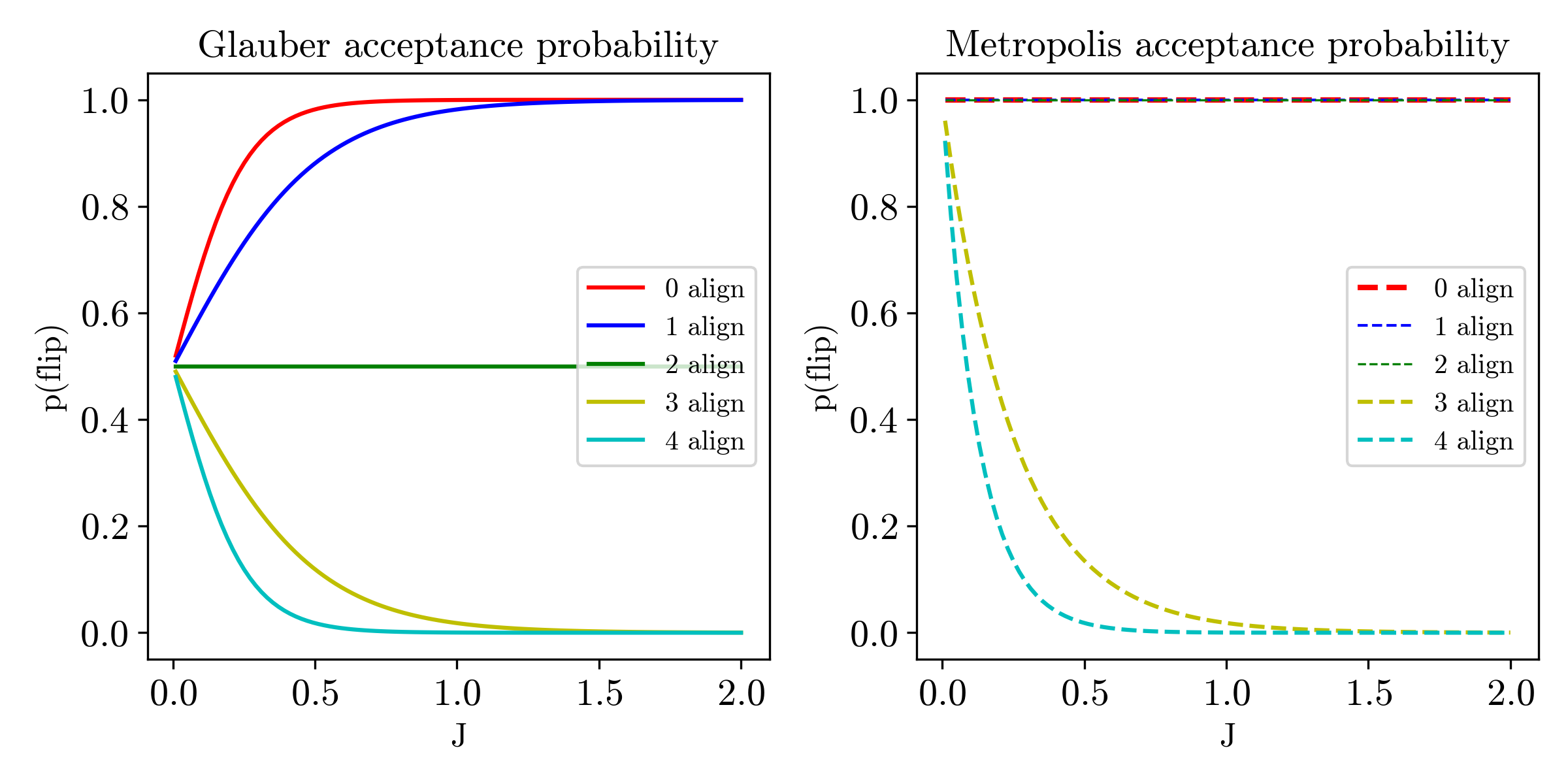}
\caption{\label{fig:met_vs_gla} Compare acceptance probability between Glauber (left) and Metropolis dynamics (right). The acceptance probability of any site depends on the coupling strength J and the number of neighbours that are aligned with it. For example, `0 align' means all four neighbours have opposite spins to the centre site, while `2 align' means the neighbouring sites have equally split spins. Glauber and Metropolis dynamics exhibit distinct behaviours when coupling strength $J$ is small.}
\end{figure*}

Figure \ref{fig:met_vs_gla} illustrates spin-flip probabilities for various coupling strengths under the two different dynamics. As $J \rightarrow +\infty$, both dynamics become deterministic: the site flips its spin in a high energy state, that is, with more than two misaligned neighbours and remains unchanged at a low energy state. At low $J$ values, Glauber dynamics approaches a flipping probability of 0.5 for all different energy states, while Metropolis dynamics approaches a flipping probability of 1. Both criteria satisfy the detailed balance condition at equilibrium; Glauber dynamics is considered more realistic, and Metropolis dynamics result in much faster convergence.

The simulation uses 20 million time steps, equivalent to 8000 lattice sweeps, to ensure the system reaches equilibrium. The transient period is excluded from the analysis, and the final results are obtained by averaging over 20 simulations with different initial conditions.

\begin{figure*}[!ht]
    \begin{subfigure}[b]{0.3\textwidth}
        \includegraphics[width=\textwidth]{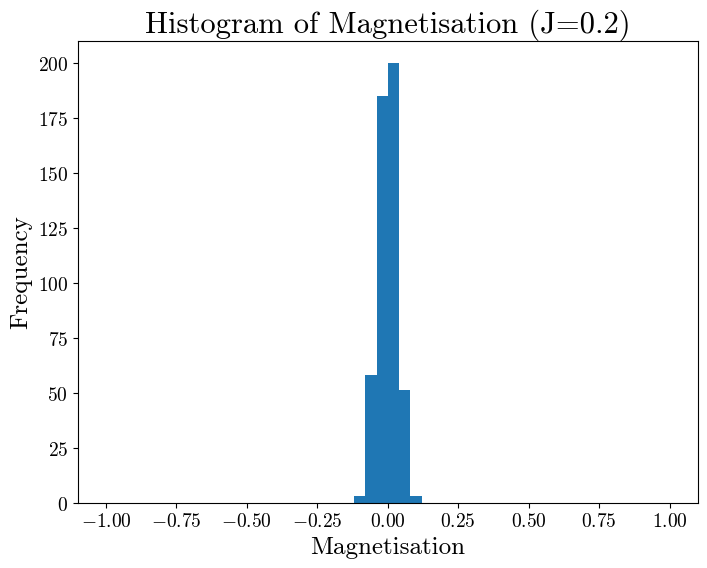}
        \caption{Sub-critical regime}
        \label{fig:distM_subcritical}
    \end{subfigure}
    \hfill
    \begin{subfigure}[b]{0.3\textwidth}
        \includegraphics[width=\textwidth]{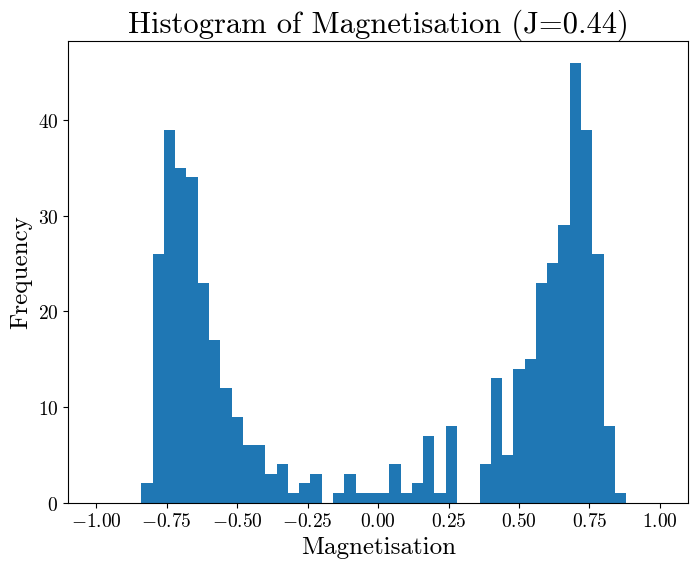}
        \caption{Near-critical regime}
        \label{fig:distM_nearcritical}
    \end{subfigure}  
    \hfill
    \begin{subfigure}[b]{0.3\textwidth}
        \includegraphics[width=\textwidth]{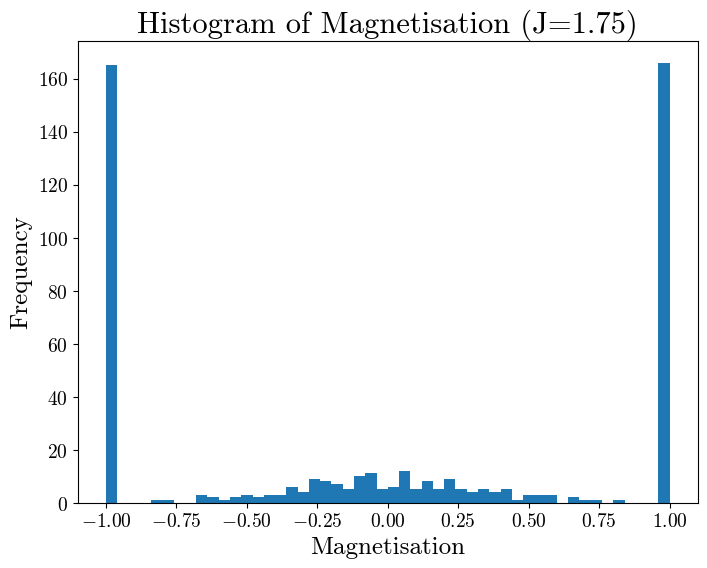}
        \caption{Super-critical regime}
        \label{fig:distM_supcritial}
    \end{subfigure} 
    \caption{Distribution of magnetisation under sub-critical, near-critical and super-critical regimes. 500 simulations run for lattice size $50\times 50$, each simulated for 20 million time steps.}
    \label{fig:distM}
\end{figure*}

The simulation results presented in Section 4 of the main text show increased noise in the super-critical region. This can be attributed to the system's difficulty in reaching equilibrium as it approaches the critical point or enters the super-critical regime. The distribution of magnetisation for various values of $J$ (Figure \ref{fig:distM}) illustrates this behaviour. In the sub-critical regime, the distribution is more centred around zero. In contrast, in the super-critical regime, the magnetisation distribution becomes more polarised at equilibrium, with values closer to -1 or 1, reflecting the dominance of up or down spins. The central mass in Figure \ref{fig:distM_supcritial} represents simulations that have not reached equilibrium after 20 million time steps, contributing to the increased noise observed in this regime.

\section{Computation of Fisher Information} \label{appendix:fisher}
Let $p(x;\theta)$ denote the probability density of random variable $X$, parameterised by $\theta$. The Fisher information is the variance of the score, and a function of the parameter $\theta$ \cite{cover_elements_2005}. The Fisher information can be transformed as follows: 
\begin{equation} \label{eq:appendixFisherContinuous}
    \begin{split}
    \mathbb{I}(\theta) & = \int\left(\frac{\partial \ln p(x;\theta)}{\partial \theta}\right)^2 p(x;\theta)dx \\
    & = \int\left(\frac{\partial p(x;\theta)}{\partial \theta}\right)^2\frac{1}{p(x;\theta)}dx \\
     & = \int\left(\frac{\partial p(x;\theta)}{\partial \theta}\frac{1}{\sqrt{p(x;\theta)}}\right)^2dx \\
     & = 4\int\left(\frac{\partial \sqrt{p(x;\theta)}}{\partial \theta}\right)^2dx \\
    \end{split}
\end{equation}

In this study, the Fisher information is computed numerically using the discretisation method introduced in \cite{sanchez-moreno_discrete_2009}:
\begin{equation} \label{eq:appendixFisherDiscrete}
    \mathbb{I}(\theta) = 4\sum_x\left(\frac{\sqrt{p(x;\theta+\Delta \theta)} - \sqrt{p(x;\theta-\Delta \theta)}}{2\Delta \theta}\right)^2 
\end{equation}

\section{Derivation of thermodynamic efficiency for canonical Ising model}\label{appendix:derivation}
Analytically, thermodynamic efficiency can be shown to diverge at the critical point. Consider a 2D square-lattice Ising model with nearest neighbour interactions as described in Section 4 of the main text. For isotropic
coupling strength, the dominant singular part of free energy density is given by \cite{baxter1982critical}:
\begin{equation} \label{eq:fs}
    -\frac{f_s}{k_B T} = \frac{(1 + k)(1 - k)^2}{2\pi k \cosh^2 \left( \frac{2J}{k_B T} \right)} \ln \left| \frac{1 + k}{1 - k} \right|
\end{equation}
where $k_B$ is the Boltzmann constant, $T$ is the temperature, and $k = (\sinh\frac{2J}{k_BT})^{-2}$.
The criticality condition is given by \cite{onsager1944}:
\begin{equation}
    \left(\sinh \frac{2J}{k_B T} \right)^{2} = 1
\end{equation}
Hence, given a constant temperature $T$, we have the following expressions for the critical coupling strength $J_c$ and the corresponding value of $k$:
\begin{gather}
        J_c(T) = \frac{k_B \ln \left( 1 + \sqrt{2} \right)}{2}T = C_0T \label{eq:Jc_T} \\
        k(J_c) = 1 \label{eq:k_Jc}
\end{gather}
where $C_0 = \frac{k_B \ln \left( 1 + \sqrt{2} \right)}{2}$ is a constant.
Expanding $k(J)=(\sinh\frac{2J}{k_BT})^{-2}$ around $J_c$ gives:
\begin{equation} \label{eq:k_J_expand}
    k(J) = k(J_c) + \frac{\partial k}{\partial J}\bigg|_{J=J_c} (J - J_c) + O((J - J_c)^2)
\end{equation}
The partial derivative term $\frac{\partial k}{\partial J}$ is:
\begin{equation} 
    \frac{\partial k}{\partial J} = \frac{-4}{k_B T}\cosh\frac{2J}{k_B T}\bigl(\sinh\frac{2J}{k_B T}\bigr)^{-3}
\end{equation}
Substituting $J=J_c$ into the above equation and noting that $\cosh\frac{2J_c}{k_B T} = \sqrt{2}$ and $\sinh\frac{2J_c}{k_B T} = 1$, yields:
\begin{equation} \label{eq:partial_k}
    \frac{\partial k}{\partial J}\bigg|_{J=J_c} = -\frac{4\sqrt{2}}{k_B T}
\end{equation}
Substituting (\ref{eq:k_Jc}) and (\ref{eq:partial_k}) into the expression (\ref{eq:k_J_expand}) gives:
\begin{align} 
    k(J) &= 1 - \frac{4\sqrt{2}}{k_B T}(J-J_c)+ O((J - J_c)^2) \notag \\
    &\approx 1 - \frac{4\sqrt{2}}{k_B T}(J-J_c)
\end{align}
Therefore, near the critical point:
\begin{equation} \label{eq:prox_k}
    k - 1 \approx -\frac{4\sqrt{2}}{k_B T} (J - J_c)
\end{equation}
When $J > J_c$, $k < 1$, equation (\ref{eq:fs}) becomes:

\begin{align} \label{eq:fs_J>Jc}
    f_s &= -k_B T \frac{(1 + k)(1 - k)^2}{2\pi k \cosh^2 \left( \frac{2J}{k_B T} \right)} \ln \left( \frac{1+k}{ 1-k} \right) \notag \\
    &= -k_B T \frac{(1 + k)(1 - k)^2}{2\pi k \cosh^2 \left( \frac{2J}{k_B T} \right)} \ln (1+k) \notag \\
    & + k_B T \frac{(1 + k)(1 - k)^2}{2\pi k \cosh^2 \left( \frac{2J}{k_B T} \right)} \ln (1-k) 
\end{align}

Substituting equation (\ref{eq:prox_k}) into (\ref{eq:fs_J>Jc}) and noting that $\cosh^2 \left( \frac{2J_c}{k_B T} \right) = 2$ and $k_c = 1$, we approximate the singular part of free energy density as $J \rightarrow J_c^{+}$:
\begin{equation} \label{eq:fs_J>Jc_approx}
    f_s \approx C_1(J-J_c)^2\ln(J-J_c) 
\end{equation}
where $C_1$ is a constant. Equation (\ref{eq:fs_J>Jc_approx}) has the same form as the expression of $f_s$ near the critical temperature point \cite{baxter1982critical}. 

Taking the derivative of $f_s$ with respect to $J$ and keeping only the leading order term gives:
\begin{align}
    \frac{\partial f_s}{\partial J} &= C_1\bigl[2(J-J_c)\ln(J-J_c) + (J-J_c)^2\frac{1}{J-J_c}\bigr] \notag\\ 
    &= C_1(J-J_c)\bigl[2\ln(J-J_c)+1\bigr] \notag\\
    &\approx 2C_1(J-J_c)\ln(J-J_c) \label{eq:df_dJ}
\end{align}
Entropy $S$ and free energy $F$ are related by $S = -\frac{\partial F}{\partial T}$. Together with equation (\ref{eq:Jc_T}), the derivative of entropy density with respect to $J$ can be expressed as:
\begin{align}
    \frac{\partial s}{\partial J} &= \frac{\partial}{\partial J} \left(\frac{-\partial f_s}{\partial T}\right) = \frac{\partial}{\partial T} \left(\frac{-\partial f_s}{\partial J}\right) \notag \\
    &= -\frac{\partial}{\partial T} \bigl[2C_1(J-J_c)\ln(J-J_c)\bigr] \notag \\
    &= -2C_1\frac{\partial (J-J_c)}{\partial T}\left[\ln(J-J_c) + 1\right] \notag \\
    &\approx 2C_1C_0\ln(J-J_c) \label{eq:ds_dJ}
\end{align}
Replacing the free energy density $f$ by its dominant singular part $f_s$ and using equations (\ref{eq:df_dJ}) and (\ref{eq:ds_dJ}), the thermodynamic efficiency of interactions $\eta$ can be computed as $J \rightarrow J_c^{+}$. Following the sign convention that work extracted from the system is positive, the change in free energy satisfies $\Delta f= - \Delta \mathbb{W}$:
\begin{align}
    \eta(J) &= \frac{1}{k_B} \; \frac{\partial s}{\partial J} \Big/ \frac{\partial f}{\partial J} \notag \\
    &\approx \frac{1}{k_B} \; \frac{2C_1C_0\ln(J-J_c)}{2C_1(J-J_c)\ln(J-J_c)} \notag \\
    &= \frac{\ln(1+\sqrt{2})}{2} \; \frac{1}{J-J_c} \label{eq:eta_J+}
\end{align}

Similarly, when $J < J_c$, $k > 1$, the dominant part of the free energy density is given by:
\begin{equation}
    f_s = -k_B T \frac{(1 + k)(1 - k)^2}{2\pi k \cosh^2 \left( \frac{2J}{k_B T} \right)} \ln \left( \frac{k+ 1}{k - 1} \right)
\end{equation}
Using the same procedure as above, we can show for $J \rightarrow J_c^{-}$:
\begin{align}
    \frac{\partial f_s}{\partial J} &\approx 2C_1(J_c-J)\ln(J_c-J) \label{eq:df_dJ-} \\
    \frac{\partial s}{\partial J} &\approx 2C_1C_0\ln(J_c-J) \label{eq:ds_dJ-}
\end{align}
Hence, the thermodynamic efficiency of interactions $\eta$ can be computed as $J \rightarrow J_c^{-}$:
\begin{align}
    \eta(J) &= \frac{1}{k_B} \; \frac{\partial s}{\partial J} \Big/ \frac{\partial f}{\partial J} \notag \\
    &\approx \frac{1}{k_B} \; \frac{2C_1C_0\ln(J_c-J)}{2C_1(J_c-J)\ln(J_c-J)} \notag \\
    &= \frac{\ln(1+\sqrt{2})}{2} \; \frac{1}{J_c-J} \label{eq:eta_J-}
\end{align}
Combining equations (\ref{eq:eta_J+}) and (\ref{eq:eta_J-}), we have:
\begin{equation}
    \eta(J) \propto |J-J_c|^{-1}
\end{equation}
This shows that the thermodynamic efficiency diverges as $J$ approaches $J_c$ from either side.

\clearpage

\end{document}